\documentclass[final,3p,times]{elsarticle}
\usepackage{amssymb}
\usepackage{amsthm}
\usepackage{amsmath}
\usepackage{mdframed}
\usepackage{tcolorbox}
\usepackage{subcaption}
\usepackage[table,xcdraw]{xcolor}
\usepackage{commath}
\newcommand{\iu}{\mathrm{i}\mkern1mu}

\biboptions{sort&compress}  % <-- this is the key line

\journal{Physics Reports}
\usepackage{hyperref}
\hypersetup{
    colorlinks=true,
    linkcolor=blue,
    filecolor=magenta,      
    urlcolor=cyan,
    pdftitle={Overleaf Example},
    pdfpagemode=FullScreen,
    }
\begin{document}

\begin{frontmatter}



\title{Interplay of sync and swarm: Theory and application of swarmalators}
\author[label1]{Gourab Kumar Sar}
\affiliation[label1]{organization={Physics and Applied Mathematics Unit, Indian Statistical Institute, 203 B. T. Road}, state={Kolkata}, postcode={700108}, country={India}}
\ead{mr.gksar@gmail.com}

\author[label2]{Kevin O'Keeffe}
\affiliation[label2]{organization={Senseable City Lab, Massachusetts Institute of Technology, Cambridge}, state={Massachusetts}, postcode={02139}, country={USA}}
\ead{kevin.p.okeeffe@gmail.com}

\author[label3]{Joao U. F. Lizárraga}
\affiliation[label3]{organization={Instituto de Fisica Gleb Wataghin, Universidade Estadual de Campinas}, state={Sao Paulo}, postcode={13083-970}, country={Brazil}}
\ead{jaoa3333@ifi.unicamp.br}

\author[label3]{Marcus A. M. de Aguiar}
\ead{aguiar@ifi.unicamp.br}

\author[label4]{Christian Bettstetter}
\affiliation[label4]{organization={Institute of Networked and Embedded Systems, University of Klagenfurt}, state={9020 Klagenfurt am Wörthersee},  country={Austria}}
\ead{christian.bettstetter@aau.at}

\author[label1]{Dibakar Ghosh}
%\correspondingauthor\email{dibakar@isical.ac.in}
\ead{Corresponding author:dibakar@isical.ac.in}

\begin{abstract}
%\textcolor{red}{General comment: Double-check acronyms. Are we consistently using ``periodic boundary conditions'' throughout the text?}
\par Swarmalators, entities that combine the properties of swarming particles with synchronized oscillations, represent a novel and growing area of research in the study of collective behavior. This review provides a comprehensive overview of the current state of swarmalator research, focusing on the interplay between spatial organization and temporal synchronization. After a brief introduction to synchronization and swarming as separate phenomena, we discuss the various mathematical models that have been developed to describe swarmalator systems, highlighting the key parameters that govern their dynamics. The review also discusses the emergence of complex patterns, such as clustering, phase waves, and synchronized states, and how these patterns are influenced by factors such as interaction range, coupling strength, and frequency distribution. Recently, some minimal models were proposed that are solvable and mimic real-world phenomena. The effect of predators in the swarmalator dynamics is also discussed. Finally, we explore potential applications in fields ranging from robotics to biological systems, where understanding the dual nature of swarming and synchronization could lead to innovative solutions. By synthesizing recent advances and identifying open challenges, this review aims to provide a foundation for future research in this interdisciplinary field.
\end{abstract}

%\begin{keyword}
%{\bf Keywords: }{Complex networks, coupled oscillators, dynamical robustness, aging transition}



%\end{keyword}

\end{frontmatter}

%\newpage
\tableofcontents

%\newpage
\section{Introduction}\label{sec.1}
Many natural and engineered systems exhibit striking patterns that arise from the collective behavior of interacting units. Two well-studied manifestations of such collective dynamics are swarming -- where agents self-organize in physical space, and synchronization -- where they adjust their internal rhythms in time. Traditionally, these phenomena have been studied independently, with models such as the Vicsek model~\cite{vicsek1995novel} capturing alignment and spatial coherence in swarming, and the Kuramoto model~\cite{kuramoto1975self} describing the temporal synchronization of coupled oscillators. However, in many real-world systems--ranging from biological cells and fireflies to robotic collectives -- the spatial and temporal dynamics are intricately intertwined. This interplay between motion and phase gives rise to a new class of systems known as swarmalators: agents that both swarm and synchronize. The simultaneous exploration of swarming and synchronization is driven by the need to understand complex systems where both spatial positioning and internal states play critical roles in collective behavior. We begin by briefly reviewing each of these two phenomena before turning to the rationale for studying their interplay.

Synchronization~\cite{boccaletti2018synchronization} is a ubiquitous phenomenon in complex systems, where coupled units with internal dynamics spontaneously adjust their rhythms to operate in unison. From the synchronized flashing of fireflies~\cite{buck1978toward} and chorusing frogs~\cite{aihara2008mathematical} to firing neurons~\cite{montbrio2015macroscopic,laing2014derivation} and circadian rhythms in living organisms~\cite{gonze2005spontaneous}, synchronization appears across a wide range of natural systems. It is also crucial in many technological settings, such as power grids~\cite{motter2013spontaneous,dorfler2012synchronization}, Josephson junction arrays~\cite{wiesenfeld1996synchronization}, and wireless sensor networks~\cite{rhee2009clock}. Theoretical efforts to understand synchronization have led to a rich mathematical framework including mean-field approximations, renormalization group analyses and finite-size scaling, most notably centered around the Kuramoto model and its many variants~\cite{acebron2005kuramoto,ott2008low,daido1988lower,hong2007entrainment}. These models capture the onset of phase coherence among interacting oscillators through coupling schemes that vary with frequency, topology, interaction range etc. Extensive research has explored on synchronization transitions, chimera states, partial synchronization, and noise-induced effects, offering deep insights into both the universality and diversity of synchronous behavior in complex networks~\cite{rodrigues2016kuramoto,ghosh2022synchronized,majhi2019chimera}.

Swarming, in contrast, focuses on the spatial organization of mobile agents governed by simple interaction rules~\cite{sumpter2010collective,couzin2007collective,buhl2006disorder}. These agents -- such as birds, fish, bacteria, or drones—adjust their positions and velocities based on local cues, giving rise to emergent collective motion like flocking, milling, or aggregation~\cite{ballerini2008interaction,herbert2016understanding}. Swarming has been studied extensively using mathematically tractable models~\cite{vicsek1995novel,cucker2007emergent}, maximum entropy methods~\cite{bialek2012statistical}, attraction-repulsion models~\cite{sar2023flocking,chen2014minimal}, agent-based simulations~\cite{reynolds1987flocks}, and in the continuum limit~\cite{fetecau2011swarm,bernoff2013nonlocal}. These models reveal how coherent structures in space can emerge from purely local interactions, often without any centralized control. The field has branched into various domains, including biology, physics, and robotics, addressing questions related to pattern formation, robustness, and the role of heterogeneity or noise. Swarming models also serve as design blueprints for distributed robotic systems and autonomous agents engaged in exploration, mapping, or collective transport~\cite{ducatelle2014cooperative,brambilla2013swarm,rubenstein2013collective}.

Although synchronization and swarming have traditionally been treated as distinct phenomena -- one concerned with phase coordination and the other with spatial organization -- there has been growing interest in studying systems where these two aspects are intertwined. Over the past two decades, research on mobile oscillators and moving agents has, to some extent, bridged the domains of synchronization and swarming. In many of these studies, an oscillator's spatial position influences its phase dynamics, but the reverse coupling (from phase to motion) is typically absent or neglected~\cite{frasca2008synchronization, uriu2013dynamics}. The agents’ movement is often modeled as a random walk or governed by external forces, remaining independent of their internal phase. However, numerous examples in both natural and engineered systems feature mobile oscillators whose position and phase mutually influence one another~\cite{yan2012linking, nguyen2014emergent}. An early effort to model these systems was made by Tanaka and colleagues~\cite{tanaka2007general, iwasa2010hierarchical}, who studied chemotactic oscillators whose spatial dynamics are mediated by the diffusion of a background chemical field. The term swarmalator, introduced by O’Keeffe et al. in 2017, formalizes these systems exhibiting bidirectional interplay between spatial and internal dynamics in a minimal yet expressive framework. The study of swarmalators is motivated by the desire to unravel the intricacies of complex systems where the dynamics of individual agents are governed not only by their spatial interactions with neighbors but also by their internal states. In natural and artificial systems alike, the coupling of these internal states with spatial positioning often leads to emergent behavior that cannot be fully understood through traditional models of swarming or synchronization in isolation. For instance, in biological systems, the activity of certain cells or organisms depends on both their location and their phase in a biological cycle, such as circadian rhythms or metabolic cycles~\cite{riedl2023synchronization}. Similarly, in robotic systems, the effectiveness of decentralized operations can depend on both the physical proximity of agents and their synchronization in executing tasks~\cite{barcis2019robots}. In such systems, the position of an agent and its internal state are intrinsically linked, leading to emergent behaviors that are richer and more intricate than those predicted by models of swarming or synchronization alone. Swarmalators provide a framework to capture these bidirectional dependencies, allowing for a more nuanced exploration of how coordinated behavior arises in such systems.

In this review article, we present a comprehensive overview of existing research, offering a roadmap for advancing the study of swarmalators and enhancing their relevance across various scientific and engineering disciplines. The rapid evolution of the field, the growing interdisciplinary interest, and the potential for real-world applications gives us the perfect opportunity to conduct a comprehensive review of swarmalator systems. The rest of the article is organized as follows. In Sec.~\ref{sec.2}, we outline the conceptual foundations of swarmalators by tracing their origins in the literature on swarming and synchronization. Section~\ref{sec.3} explores the phenomenology of swarmalators, focusing on their dynamic behaviors across various configurations and spatial dimensions. In Sec.~\ref{sec.4}, we develop the mathematical framework underlying swarmalator models. Section~\ref{sec.5} examines the influence of an external factor, specifically the presence of a predator-like agent, on swarmalator dynamics. In Sec.~\ref{sec.6}, we review emerging applications of swarmalators across diverse scientific and technological domains. Finally, Sec.~\ref{sec.7} concludes with a summary and a discussion of open questions and future research directions.

%----------------------------------------------------------
\section{Foundations of swarmalators}\label{sec.2}

%\subsection{Swarming and synchronization}
% \subsection{The Kuramoto model and the Vicsek model}
% \subsection{Mobile oscillators or moving agents}
% \subsection{Oscillators that synchronize and swarm}
Before starting to discuss swarmalators, we first briefly discuss synchronization and then swarming.

\subsection{Synchronization of coupled oscillators: Pioneering work by Huygens and Winfree}

Although synchronization (sync) has been observed by Huygens' experiments with pendulum clocks already in the 17th century \cite{oliveira2015huygens},
it was not until the 1960s that the first theoretical understanding of the phenomenon was developed by Winfree~\cite{winfree1967biological}. Impressed by the large number of physical and biological systems exhibiting synchronization, he realized that a general problem formulation should be possible. 

Winfree's approach was based on the idea that, no matter how complicated an oscillatory system is, if it exhibits sustained periodic motion, there must be a stable limit cycle in its phase space. On the limit cycle, the important variable is the phase along the orbit, which can be parameterized by $\theta=2\pi t/T_0 = \omega_0 t$, where $T_0$ is the period of the orbit and $\omega_0$ is the angular frequency. Winfree argued that if such a system is perturbed by a small external periodic force of frequency $\omega_f$ and phase $\phi$, it should remain very close to the limit cycle, suggesting that the phase is still the only relevant variable for the perturbed system. The frequency around the periodic orbit, on the other hand, would suffer an instantaneous change proportional to the intensity of the perturbation. This change can be expressed as $\Delta \omega = S(\phi) \, Z(\theta)$, where $S$ is the small stimulus produced by the perturbation and $Z$ is the sensitivity of the oscillator, also known as phase response function. If the oscillator synchronizes with the external force, their phases will be locked and $\theta = \phi + \psi$ with a constant  $\psi$. Moreover,  the average frequency variation over one cycle must be just enough to achieve the matching with the external force, that~is,  
\begin{equation}
	\frac{1}{2\pi} \int_0^{2\pi} Z(\phi+\psi) \, S(\phi) \, d\phi \equiv M(\psi) = \omega_f-\omega_0.
	\label{found1}
\end{equation}
Here, $M(\psi)$ is a periodic function that depends on the stimulus and sensitivity of the system. Eq.~\eqref{found1} implies that synchronization is possible only if $|\omega_f-\omega_0|$ is smaller than the amplitude of $M$. For example, if $M(\psi)= A \cos\psi$, two solutions exist if $|\omega_f-\omega_0| < A$ and the stable solution is the one satisfying $dM/d\psi < 0$: if $\psi$ increases a little, $M$ decreases, pushing the solution back to its equilibrium.  Using this basic idea, Winfree studied populations of oscillators perturbing each other according to the equations
\begin{equation}
    \dot{\theta}_i = \omega_i + Z(\theta_i) \:\frac{K}{N}  \sum_{j=1}^N S(\theta_j)\:,
\end{equation}
where $K$ controls the coupling strength. He concluded, using simulations, that synchronization would occur only if a threshold in the coupling intensity was crossed. 

\subsubsection{The Kuramoto model}

In a series of papers, Kuramoto formalized the ideas developed by Winfree, arriving at the famous model bearing his name  \cite{kuramoto1975self,kuramoto1981rhythms,kuramoto1984cooperative}. %Here we give a brief account of his formulation. 
An important step to put Winfree's insights on firmer grounds was the extension of the concept of phase, originally defined only on the limit cycle, to points in its vicinity. This is achieved as follows: let the system be described by the differential equations 
\begin{equation}
	\frac{d \mathbf{X}}{dt} = \mathbf{F}(\mathbf{X}),
\end{equation}
where $\mathbf{X}$ and $\mathbf{F}$ are vectors with $n$ components each. We assume that the system admits a stable limit cycle ${\cal L}$, consisting of a periodic orbit $\mathbf{X}_0(t)$ with period $T_0=2\pi/\omega_0$. Following Winfree, we parameterize the orbit by the phase $\theta=\omega_0 t$. %The limit cycle solution can be written as $\mathbf{X}_0(\theta)$ with $\dot{\theta} = \omega_0$.  

We now consider the stroboscopic map $\mathbf{X}(t) \rightarrow \mathbf{X}(t+T_0)$. All points on ${\cal L}$ are fixed points of the map. The orbit of a point $\mathbf{X}$ in the vicinity of ${\cal L}$, on the other hand, asymptotically approaches a point on the limit cycle: $\lim_{n\rightarrow \infty} \mathbf{X}(t+nT_0) = \mathbf{X}_0^*$ for some $\mathbf{X}_0^* \in {\cal L}$, since ${\cal L}$ is stable. We assign the phase of  $\mathbf{X}_0^*$ to $\mathbf{X}$, creating a correspondence $\theta=\theta(\mathbf{X})$. The set of points in the vicinity of ${\cal L}$ that converge to the same $\mathbf{X}_0^*$ is called its iso\-chron \cite{pikovsky2001universal}. The dynamics can now be described in terms of $\theta$ for all points in the neighborhood of ${\cal L}$. Using the chain rule, we obtain
\begin{equation}
	\frac{d \theta}{d t} = \nabla_{\mathbf{X}} \theta \cdot \frac{d \mathbf{X}}{d t} = \nabla_{\mathbf{X}} \theta \cdot \mathbf{F} = \omega_0.
\end{equation}
If a perturbation is added to the system, so that $ \dot{\mathbf{X}}= \mathbf{F}(\mathbf{X}) + \epsilon \mathbf{S(\mathbf{X},t)}$, the equation for $\theta$ becomes
\begin{equation}
	\dot{\theta} =  \nabla_{\mathbf{X}} \theta \cdot (\mathbf{F} + \epsilon \mathbf{S(\mathbf{X},t)}) = \omega_0 + \epsilon \mathbf{Z}(\theta) \cdot \mathbf{S(\mathbf{X}(\theta),t)}),
\end{equation}
where $\mathbf{Z}(\theta) = \nabla_{\mathbf{X}} \theta$ is computed at $\mathbf{X}=\mathbf{X}_0(\theta)$ to keep the theory to first order in the small parameter $\epsilon$. 

We now take the next step and consider that the perturbation $\mathbf{S}$ is caused by interactions with other similar oscillators. Let us first deal with only two nearly identical and symmetrically coupled systems described by 
\begin{equation}
		\frac{d \mathbf{X}_i}{dt} = [ \mathbf{F}(\mathbf{X}) + \epsilon \mathbf{G}_i(\mathbf{X})] + \epsilon \mathbf{V}(\mathbf{X}_i,\mathbf{X}_j)
\end{equation}
with $i \neq j=1,2$. We assume that the vector fields $ \mathbf{F}(\mathbf{X}) + \epsilon \mathbf{G}_i(\mathbf{X})$ have stable limit cycles ${\cal L}_i$ with frequencies $\omega_i = \omega_0 + \delta \omega_i$. As before, the phase along the cycle is extended to its vicinity through the map $\theta_i=\theta_i(\mathbf{X}_i)$. Then, the equation for the phase becomes
\begin{equation}
	\dot{\theta}_i =  \nabla_{\mathbf{X}} \theta_i \cdot (\mathbf{F} + \epsilon \mathbf{G}_i+ \epsilon  \mathbf{V}(\mathbf{X}_i,\mathbf{X}_j)) = \omega_i + \epsilon \mathbf{Z}(\theta_i) \cdot \mathbf{V}(\theta_i,\theta_j),
	\label{found2}
\end{equation}
where  $\mathbf{Z}(\theta_i) = \nabla_{\mathbf{X}} \theta_ i$ calculated at  ${\cal L}_i$. Since the effect of the interaction on the angular frequency is small compared to $\omega_i$, it can be approximated by its average over one period, as suggested by Winfree. Defining the disturbances $\psi_i$ caused by the mutual interaction by $\theta_i = \omega_i t + \psi_i$, the last term of Eq.(\ref{found2}) can be approximated by
\begin{equation}
	\frac{\omega_i}{2\pi} \int_0^{2\pi/\omega_i}  \mathbf{Z}(\omega_i t + \psi_i) \cdot \mathbf{V}(\omega_i t + \psi_i,\omega_j t + \psi_j) dt.
\end{equation}
Finally, using that $\omega_i = \omega_j = \omega_0$ to first order in $\epsilon$ and changing the integration variable to $\phi = \omega_0 t + \psi_i$ results in 
\begin{equation}
	 \frac{1}{2\pi} \int_0^{2\pi}  \mathbf{Z}(\phi) \cdot \mathbf{V}(\phi, \phi+\psi_j - \psi_i) d \phi
		\approx \frac{1}{2\pi} \int_0^{2\pi}  \mathbf{Z}(\phi) \cdot \mathbf{V}(\phi, \phi+\theta_j - \theta_i) d \theta \equiv \Gamma(\theta_j-\theta_i)
\end{equation}
to first order in $\epsilon$. The equation for the phases becomes
\begin{equation}
    \dot{\theta}_i =   \omega_i + \epsilon 	\Gamma(\theta_j-\theta_i).
\end{equation}
Extending the dynamics to the case of $N$ oscillators and choosing the simplest case of equally weighted, all-to-all sinusoidal interactions leads to the famous Kuramoto model
\begin{equation}
	\dot{\theta}_i =   \omega_i + \frac{K}{N} \sum_{j=1}^N	\sin(\theta_j-\theta_i),
	\label{found3}
\end{equation}
where we replaced $\epsilon$ by $K/N$.

\subsubsection{Behavior of the Kuramoto model for large $N$}

\begin{figure}
	\centering
	\includegraphics[width =0.8 \columnwidth]{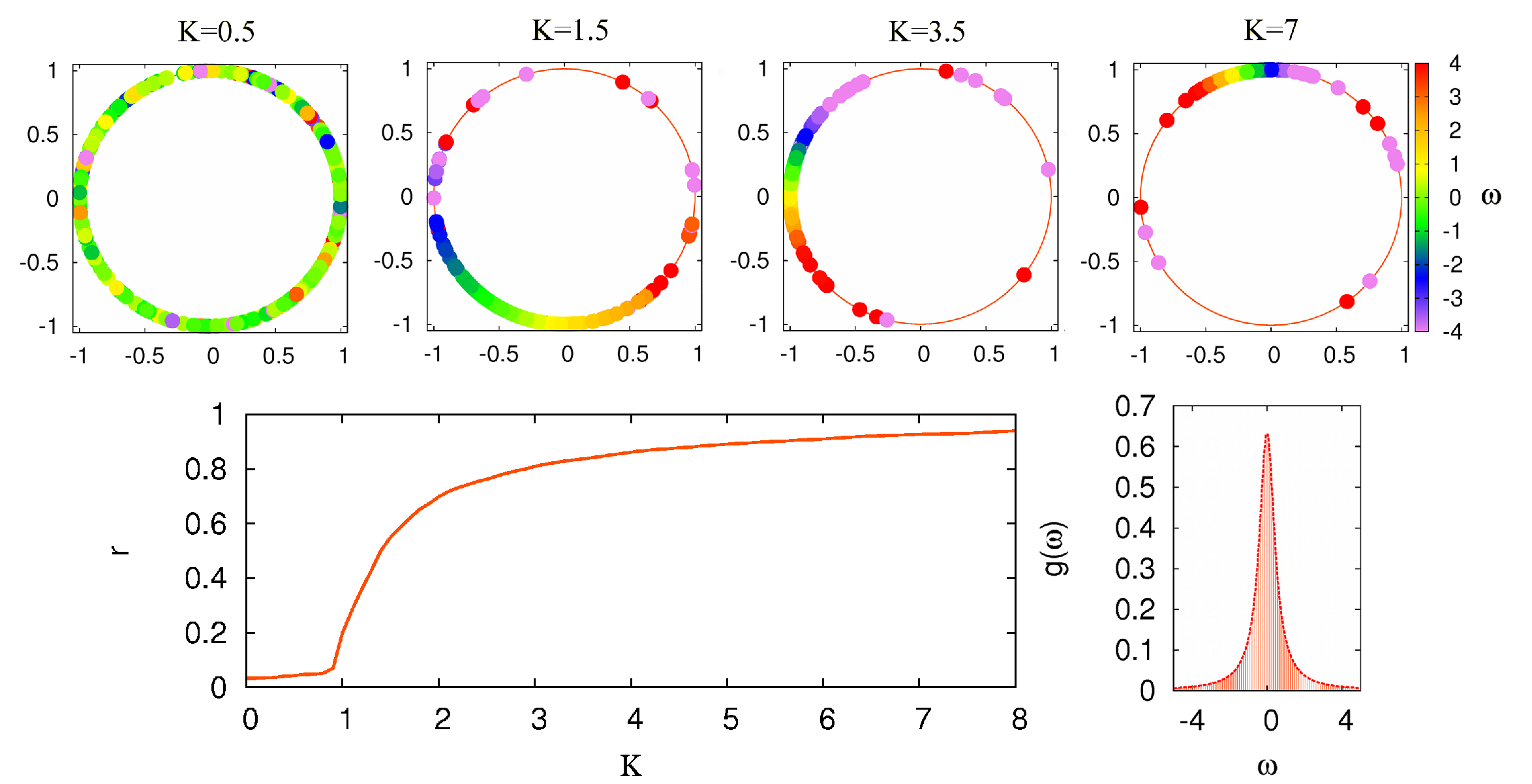}
	\caption{Each panel on the top shows the collection of oscillators situated in the unit circle (when each oscillator $j$ is represented as $e^{i\theta_j}$). The color of each oscillator represents its natural frequency. From left to right we observe how oscillators start to concentrate as the coupling $K$ increases. In the panels below we show the synchronization diagram, i.e., the Kuramoto order parameter $r$ as a function of $K$. It is clear that $K_c=1$ as obtained by using the distribution $g(\omega)$ shown in the right panel. Eq.~\eqref{found3} is used to perform simulations and obtain the results.\\ Source: Reprinted figure with permission from Ref. \cite{hermoso2014synchronization}.}
    \label{figfound1}
\end{figure}

A useful way to visualize the oscillators in the Kuramoto model is to picture them as particles moving on the unit circle. In this representation, disordered motion corresponds to particles evenly distributed along the circle at all times, whereas during phase synchronization the particles group together (see Fig.~\ref{figfound1} top). The degree of synchronization can be measured by the complex number
\begin{equation}
	z = \frac{1}{N} \sum_{j=1}^N e^{\mbox{i} \theta_j} \equiv r e^{\mbox{i} \psi}, ~~ \mbox{i}=\sqrt{-1}.
	\label{found4}
\end{equation}
In this way, $r$ works as an order parameter: it tends to zero when the oscillators are uniformly distributed and to one when all oscillators have exactly the same phase.  The real vector $\vec{r} = (r\cos\psi,r\sin\psi)$ can also be interpreted as the center of mass of the  particles.

We now assume that  the `natural' frequencies $\omega_i$ are  drawn from a symmetric unimodal distribution $g(\omega)$ centered at $\omega_0$. Summing both sides of Eq.~(\ref{found3}) over the index $i$,  the last term on the right vanishes and the average angular velocity of the group of oscillators is exactly $\omega_0$. Moreover, because Eq.~(\ref{found3}) is invariant under rotations we might take $\omega_0=0$ without loss of generality, which amounts to change to a frame of reference rotating with $\omega_0$. In this frame of reference, the synchronized oscillators are static, with $\dot{\theta}_i=0$. 

Numerical simulations of Eq.~\eqref{found3} for finite $N$ show that, for small values of $K$, the oscillators behave as if they were independent, thus no synchronization occurs. The order parameter $r$ fluctuates around zero with amplitude of the order of $1/\sqrt{N}$ \cite{kuramoto1984cooperative,daido1989intrinsic}. As $K$ crosses a threshold $K_c$, $r(t)$ increases exponentially reaching a plateau that tends to 1 as $K \rightarrow \infty$. Figure \ref{figfound1} illustrates the distribution of oscillators on the unit circle and the typical behavior of the model by varying the coupling strength $K$. For a detailed analysis of the model's behavior for finite $N$, we refer to  \cite{strogatz2000kuramoto,bronski2012fully,gottwald2017finite}. Here, instead, we review the main properties of the model for large $N$. For a detailed review see \cite{acebron2005kuramoto}.

Before we describe the mean field limit $N \rightarrow \infty$, we use Eq.~(\ref{found4}) to rewrite Eq.~(\ref{found3}) as
\begin{equation}
	\dot{\theta}_i =   \omega_i + K r	\sin(\psi-\theta_i).
	\label{found5}
\end{equation}
This equation can be solved for each $\theta_i$, treating $r$ and $\psi$ as parameters and imposing that the solutions self-consistently return, via Eq.~(\ref{found4}), the same $r$ and $\psi$ we started with. Synchronization requires $\dot{\theta}_i=0$, which is possible only if $|\omega_i| < K r$.  Next, we replace individual oscillators by $\rho(\theta,t,\omega)$, representing the density of oscillators with natural frequency $\omega$ at position $\theta$ in time $t$. In equilibrium, $\rho$ becomes stationary and can be divided into a synchronized part, $\rho_{sync}$, and a non\-synchronized part, $\rho_{async}$, which is assumed to be uniform and does not contribute to $r$ (see, however, \cite{strogatz2000kuramoto}). With these considerations, the behavior of the order parameter $r$ as a function of $K$ can be obtained as follows: First, we choose the origin of our reference frame in such a way that $\psi=0$. Second, we set $\dot{\theta}_i$ to zero in Eq.~(\ref{found5}) to get $\rho_{sync}(\theta,\omega) = g(\omega) \delta(\theta-\theta^*)$ with $\sin\theta^*=\omega/Kr$. Finally, write Eq.~(\ref{found4}) as
\begin{equation}
	r = \int_0^{2\pi} d\theta \int_{-Kr}^{Kr} d\omega g(\omega) \delta(\theta-\theta^*)   e^{i\theta} =  \int_{-Kr}^{Kr} d\omega g(\omega) \cos\theta^*,
\end{equation}
where we used $e^{\mbox{i} \theta} = \cos\theta+\mbox{i} \sin\theta$ and noted that the integral with $\sin\theta$ vanishes because $g(\omega)$ is symmetric. Changing the integration variable to $\theta^*$, we obtain
\begin{equation}
	r = K r \int_{-\pi/2}^{\pi/2}  g(K r\sin\theta^*) \cos^2\theta^* d\theta^*.
	\label{found6}
\end{equation}
The trivial solution is $r=0$. Canceling $r$ on both sides gives an implicit equation for $K=K(r)$. This equation can be solved analytically in at least two cases: \\
\noindent (i) Lorenz distribution 
\begin{equation}
	g(\omega)= \frac{\Delta}{\pi} \frac{1}{\omega^2+\Delta^2},
    \label{found7}
\end{equation}
which results in 
\begin{equation}
	r=\sqrt{1-\frac{2\Delta}{K}},
\end{equation}
for $K > K_c = 2\Delta$ \cite{kuramoto1984cooperative}, and

\noindent (ii) Gaussian distribution
\begin{equation}
	g(\omega)= \frac{1}{\sqrt{2\pi\Delta^2}} e^{-\frac{\omega^2}{2\Delta^2}}.
\end{equation}
In this case, a parametric relation is obtained as,
\begin{equation}
	r = \sqrt{\frac{\pi x}{2}} e^{-x}\left[I_0(x)+I_1(x)\right] = F(x)\:,
\end{equation}
where $x=K^2r^2/4\Delta^2$ and $I_n(x)$ is the modified Bessel function of the first kind \cite{barioni2021complexity}.
For each $x$, we find $(r,K) = (F(x), 2 \Delta\sqrt{x}/ F(x))$ with $K_c = \lim_{x\rightarrow 0} 2 \Delta \sqrt{x}/F(x) = \sqrt{8/\pi} \Delta$.

For general distributions, we can expand the argument of $g$ in Eq.~(\ref{found6}) to second order around the critical point $r=0$ to obtain
\begin{equation}
	r = \sqrt{\frac{16}{\pi g''(0) K_c^3}\left(1- \frac{K_c}{K}\right)}
\end{equation}
for $K > K_c$ where
\begin{equation}
	K_c = \frac{2}{\pi g(0)}.
\end{equation}

The Kuramoto model has became a paradigm in the study of synchronization and has been extended in different directions \cite{rodrigues2016kuramoto}. Examples include frustration \cite{sakaguchi1986soluble}, networks of connections \cite{arenas2008synchronization}, different distributions of natural frequencies \cite{gomez2011explosive,ji2013cluster,climaco2019optimal}, multi-dimensional models \cite{olfati2006swarms,Strogatz2019higher,barioni2021ott,fariello2024exploring} and periodically forced oscillators \cite{childs2008stability,moreira2019global,moreira2019modular}. However, not all biological or artificial systems, the coupled oscillators can be described by the Kuramoto model.

\subsubsection{Pulse-coupled oscillators}

Several biological and artificial systems of coupled oscillators do not interact continuously as described by the Winfree and Kuramoto models. Instead, the interactions occur at discrete points in time --- they are pulsatile, occurring only when the oscillators {\it fire}. A population is considered as synchronized if these pulse-coupled oscillators fire in unison. Examples in nature include fireflies, which ``communicate" through flashes of luminescent signals~\cite{buck1978toward, takatsu2012spontaneous}; heart pacemaker cells, which emit electric signals when the cell's voltage reaches a threshold~\cite{peskin1975mathematical, baruscotti2010cardiac}; and neural networks, where neurons interact by sending and receiving electric impulses called \textit{spikes}~\cite{strong1998entropy, brown2004multiple}.   %and certain types of telecommunication. 

Peskin~\cite{peskin1975mathematical} proposed one of the earliest models of pulse-coupled oscillator synchronization to describe the cardiac pacemaker. It consists of a network of $N$ ``integrate-and-fire" oscillators, each characterized by the variable $x_i$ representing its internal voltage and subject to the dynamics \cite{mirollo1990synchronization}
\begin{equation}
    \frac{d x_i}{d t} = S_0 - \gamma x_i,
    \label{foundpeskin}
\end{equation}
where the saturation $S_0$ and dissipation rate $\gamma$ are constants. When $x_i=1$, the oscillator fires and its state resets to $x_i=0$. The other oscillators in the network, on the other hand, have their state variables $x_j$ incremented by a small fixed amount $\epsilon$. Then, if $x_j + \epsilon > 1$, the $j$-th oscillator also fires and resets to $x_j=0$, in synchrony with $x_i$. This update rule can be stated as
\begin{equation}
    x_i(t)=1 \quad \Longrightarrow \quad x_j(t^+) = {\min}(0,x_j(t)+\epsilon).
\end{equation}
Peskin conjectured that a set of pulse-coupled oscillators would always synchronize, independent of their initial conditions, even if they were not quite identical, and proved this for the case of two identical oscillators with small coupling $\epsilon$ and dissipation $\gamma$ constants.

\begin{figure}
	\centering
	\includegraphics[width =1\columnwidth]{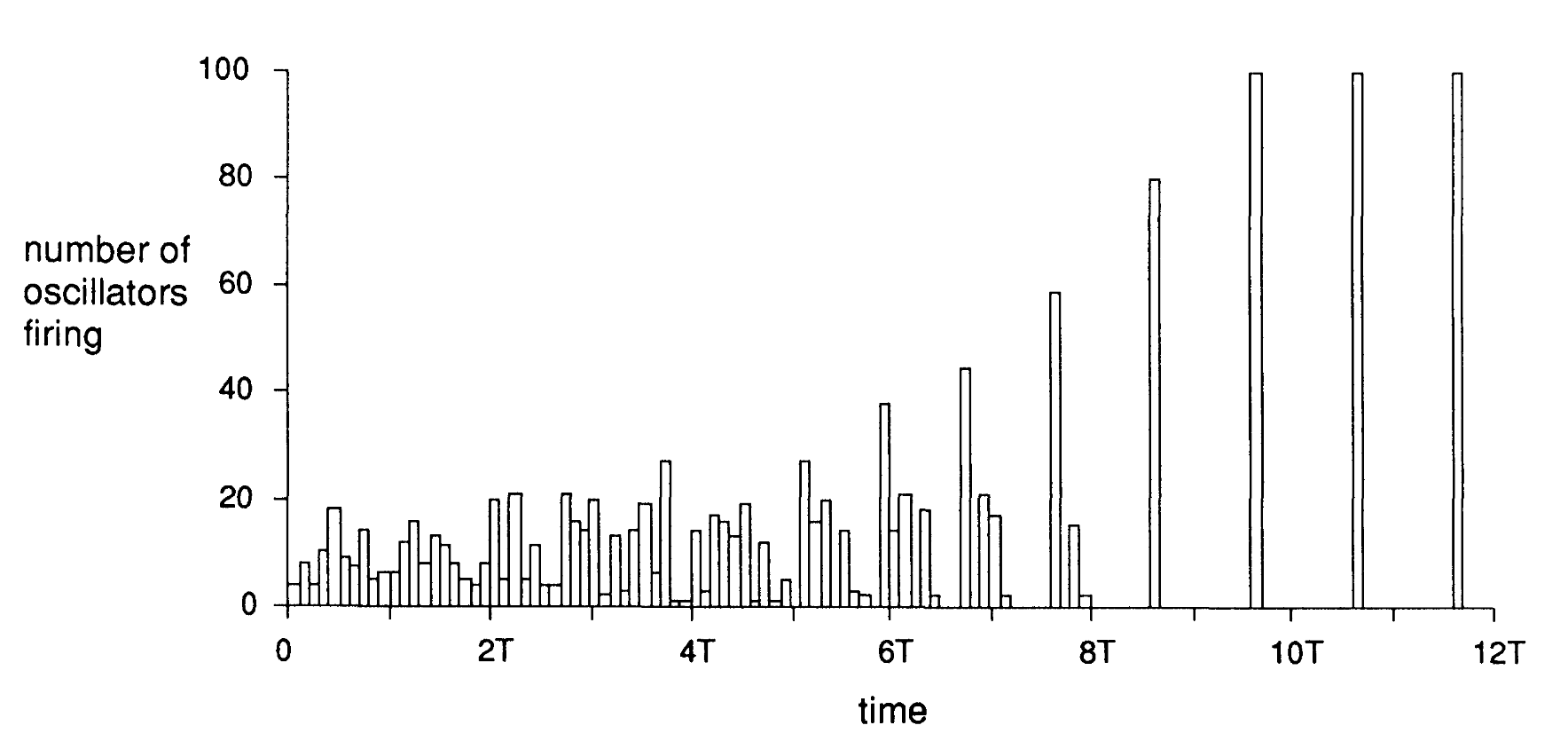}
	\caption{Number of oscillators firing as a function of time for $S_0=2$, $\gamma=1$ and $\epsilon=0.3$ and random initial conditions.  Time is plotted in multiples of the natural period $T$ of the oscillators. Each period is divided into 10 equal intervals, and the number of oscillators firing during each interval is plotted vertically. For simulation, Eq.~\eqref{foundpeskin} is used with $N=100$ for $S_0=2$, $\gamma=1$ and $\epsilon=0.3$. \\ Source: Reprinted figure with permission from Ref. \cite{mirollo1990synchronization}. }
    \label{fig:mirollo}
\end{figure}

Mirollo and Strogatz \cite{mirollo1990synchronization} extended the work of Peskin to the case of $N$ identical oscillators, whose state variables $x_i$ increase monotonically with time and are concave down. This generalizes Eq.~(\ref{foundpeskin}), whose solution $x_i(t) = S_0 \gamma^{-1}(1-e^{-\gamma t})$ has these two properties. Coupling between the oscillators was considered to be all-to-all as in Peskin's model. They showed that, for all $N$ and for almost all initial conditions, the system eventually becomes synchronized. As a result, they also proved that Peskin’s conjecture for identical oscillators is true for all $N$ and for all $\epsilon, \gamma > 0$. Figure ~\ref{fig:mirollo} shows the results of simulations for $N=100$ pulse-coupled oscillators using Eq.~(\ref{foundpeskin}) with $S_0=2$, $\gamma=1$ and $\epsilon=0.3$. Full synchronization occurs after about ten oscillation periods. For an application of the model to robots, see ref.~\cite{berlinger2021implicit}.

In the next subsection we review some more recent results that are relevant for the theory of swarmalators.

\subsubsection{Recent advances}

\hspace{0.5cm}

%\paragraph*{Ott-Antonsen (OA) ansatz} % could be used but is in italics not bold

\paragraph*{Ott-Antonsen (OA) ansatz} --  A breakthrough in the solution of the Kuramoto model was achieved by Ott and \mbox{Antonsen} in 2008 \cite{ott2008low}. Expanding the density of oscillators in Fourier modes as
\begin{equation}
 	\rho(\theta,t,\omega) = \frac{g(\omega)}{2 \pi} \Big[1 + \sum_{m=1}^\infty \alpha_m(\omega,t) e^{\mbox{i} m\theta} + c.c.\Big]
\end{equation}
and substituting into the continuity equation $\partial \rho / \partial t + \partial (\rho v)/\partial \theta=0$ would result in infinite series of coupled equations for the coefficients $\alpha_m$. According to Eq.~(\ref{found5}), the velocity field is given by $v =  \omega + K r	\sin(\psi-\theta)$, and c.c. stands for complex conjugate. Ott and Antonsen realized that the {\it ansatz} $\alpha_m = \alpha^m$ would make $\rho$ satisfy the continuity equation~if
\begin{equation}
    \dot{\alpha} + \frac{K}{2}(z |\alpha|^2-z^*) + \mbox{i} \omega \alpha = 0.
    \label{found8}
\end{equation}
The complex order parameter in Eq.~(\ref{found4}) becomes
\begin{equation}
    z^* = \int_0^{2\pi} d\theta \int_{-\infty}^{\infty} d\omega \rho(\theta,t,\omega) e^{-\mbox{i}\theta} = \int_{-\infty}^\infty d\omega g(\omega) \alpha(\omega,t),
\end{equation}
which can be solved analytically for the Lorenz distribution Eq.~(\ref{found7}), resulting in $z^* = \alpha(\omega_0-\mbox{i} \Delta)$. Calculating (\ref{found8}) at $\omega = \omega_0 - \mbox{i} \Delta$ gives an equation for $z$, and separating the real and imaginary parts leads to
\begin{equation}
    \dot{r} = -\Delta r + \frac{K}{2}r(1-r^2)
\end{equation}
and $\dot{\psi} = \omega_0$. This equation not only recovers the equilibrium solutions obtained by Kuramoto but it provides the exact dynamical behavior of the order parameter. It reveals, in particular, that the bifurcation at $K=2\Delta$ is pitchfork supercritical~\cite{skardal2020higher}.\\

%\noindent {\it {\bf Forced Kuramoto model}}
\paragraph*{Forced Kuramoto model} -- External driving plays a key role in many biological systems, such as circadian rhythms and artificial pacemakers for the heart. The forced Kuramoto model was considered by Ott and Antonsen \cite{ott2008low} but was studied in detail by Child and Strogatz \cite{childs2008stability}. The equations are given by
\begin{eqnarray}
	\dot{\theta}_i &=& \omega_i + \frac{K_1}{N} \sum_{j=1}^N \sin{(\theta_j-\theta_i)} + F \sin(\sigma t - \theta_i).
	\label{found9}
\end{eqnarray}
Changing to a reference frame rotating with $\sigma$ and applying the OA ansatz for a Lorenz distribution of natural frequencies leads to the following equations for the order parameter:
\begin{equation}
	\dot{r} = -r + \frac{K}{2}r(1-r^2)  + \frac{F}{2} (1-r^2) \cos \psi 
    \label{found10}
\end{equation}
and
\begin{equation}
	\dot{\psi} = -\Omega -  \frac{F}{2} \left(r+\frac{1}{r} \right) \sin \psi, 
    \label{found11}
\end{equation}
where $\Omega = \sigma-\omega_0$.

\begin{figure}
	\centering
	\includegraphics[width =1.0\columnwidth]{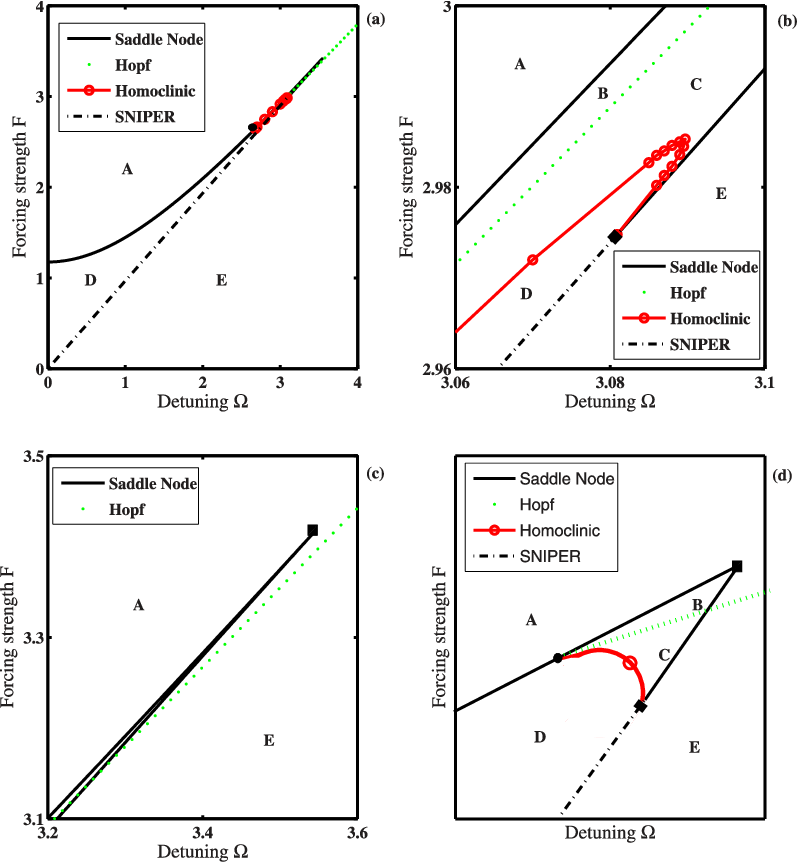}
	\caption{Stability diagram for the forced Kuramoto model (Eq.~\eqref{found9}) for $K=5$. (a) Regions A-E correspond to qualitatively different phase portraits. Four types of bifurcations occur: supercritical Hopf bifurcation; homoclinic bifurcation; and two types of saddle-node bifurcations. The filled circle marks a codimension-2 Takens-Bogdanov point, at which the Hopf, homoclinic, and upper saddle-node curve intersect tangentially. (b) Enlargement of the crossover region, where all four bifurcation curves run nearly parallel to one another. (c) Enlargement of the region near the codimension-2 cusp point (filled square), where the upper and lower branches of saddle-node bifurcations meet tangentially. (d) Schematic version of the stability diagram, intended to show how the bifurcation curves connect in the confusing crossover region. Tangential intersections have been opened up for clarity. \\ Source: Reprinted figure with permission from Ref. \cite{childs2008stability}. }
    \label{figfound2}
\end{figure}

The system has three parameters: $K$, $F$, and $\Omega$. Setting Eqs.~(\ref{found10}) and (\ref{found11}) to zero gives the fixed points $(r^*,\psi^*)$ while the eigenvalues of the Jacobian matrix $J$ determine their stability. Bifurcation manifolds can be computed, for instance, by setting $\det{(J)}=0$ for saddle-node points and $tr(J)=0$ for Hopf bifurcations. Figure \ref{figfound2} shows the bifurcation curves in the $(\Omega,F)$ plane for $K=5$.\\

%\noindent {\it {\bf Generalizations}} 

\paragraph*{Generalizations} -- A large number of modifications and generalizations of the Kuramoto model has been proposed since its publication. Kuramoto and Sakaguchi, in particular, introduced a parameter to describe phase alignment frustration and showed that it makes synchronization harder and induces rotation of the macroscopic group of synchronized oscillators even if the average natural frequency of the group is zero \cite{sakaguchi1986soluble}. Other generalizations include different types of coupling functions \cite{hong2011kuramoto,yeung1999time,breakspear2010generative}, different distributions of the oscillator's natural frequencies \cite{martens2009exact}, inertial terms \cite{acebron2005kuramoto,dorfler2011critical,olmi2014hysteretic}, and higher dimensions \cite{chandra2019continuous,barioni2021complexity,de2023generalized}.

One particularly important generalization of the Kuramoto model was the introduction of pairwise interactions based on networks \cite{strogatz2001exploring,moreno2004synchronization,rodrigues2016kuramoto}. In this case, Eq.~(\ref{found3}) becomes
\begin{equation}
	\dot{\theta}_i =   \omega_i + \frac{K}{k_{av}} \sum_{j=1}^N A_{ij}	\sin(\theta_j-\theta_i),
\end{equation}
where the elements of the adjacency matrix $A_{ij}$ are one if oscillators $i$ and $j$ interact and zero otherwise. It was shown that when the network represented by $A$ is scale-free and the natural frequencies are proportional to the degree of the corresponding node, $\omega_i = k_i$ where $k_i=\sum_j A_{ij}$, the system exhibits explosive synchronization and hysteresis \cite{gomez2011explosive,ji2013cluster}.\\

%\noindent {\it {\bf Higher order interactions}} 

\paragraph*{Higher order interactions} -- In many complex systems, interactions involve the simultaneous action of three or more agents, requiring the inclusion of higher order interactions in the corresponding mathematical models \cite{battiston2020networks}. Examples can be found in neuroscience \cite{ganmor2011sparse,petri2014homological,giusti2015clique,reimann2017cliques,sizemore2018cliques,majhi2024patterns}, 
ecology \cite{grilli2017higher,ghosh2024chimeric}, biology \cite{sanchez2019high}, the social sciences \cite{benson2016higher,de2020social}, propagation of epidemics \cite{iacopini2019simplicial,jhun2019simplicial,vega2004fitness}, and synchronization \cite{berec2016chimera,skardal2019abrupt,skardal2020higher,sayeed2024global,muolo2024phase}. 

For the Kuramoto model, higher order terms can be derived from phase reduction methods \cite{ashwin2016hopf,leon2019phase}. Three-body interactions, for example, proportional to $\sin{(2\theta_j-\theta_k -\theta_i)}$ (asymmetric coupling) or $\sin{(\theta_j+\theta_k -2\theta_i)}$ (symmetric coupling) can be added to Eq.~(\ref{found3}) \cite{battiston2020networks}. In both cases the interaction goes to zero when oscillators $i$, $j$, and $k$ have the same phases. Symmetric coupling has been studied in \cite{tanaka2011multistable,skardal2019abrupt,dai2021d,leon2024higher} whereas asymmetric interactions were considered in \cite{skardal2020higher,dutta2023impact,fariello2024third,suman2024finite}. The important feature of asymmetric interactions is the possibility of applying the OA ansatz and derive exact equations for the order parameter. Skardal and Arenas, in particular, have shown that 3 and 4-body asymmetric interactions can lead to bi-stability and to the appearance of a first-order phase transition~\cite{skardal2020higher}. 

Higher dimensions have also been studied in the context of multi-dimensional Kuramoto models \cite{dai2021d,sarika2024,biswas2024symmetry} and externally forced systems \cite{costa2024bifurcations}.

\subsection{Swarming}

Swarming refers to self-organized collective motion --- occuring in physical space, as opposed to synchronization that occurs in the time domain. Very much like synchronization, a large number of different physical and biological systems can exhibit complex collective dynamical spatial patterns, such as schools of fish~\cite{parrish2002self}, flocks of birds~\cite{cavagna2010scale}, hopper bands of locusts~\cite{buhl2006disorder, weinburd2024anisotropic}, and bacterial aggregates~\cite{allison1991bacterial}, suggesting that a general theory could describe their basic properties.

%While synchronization refers to collective behavior in the time domain, swarming describes a similar phenomenon that occurs in physical space. Once again, a large number of different physical and biological systems can exhibit complex collective dynamical patterns, such as schools of fish, flocks of birds,  self-propelled particles and bacteria. 

One of the first successful models of collective motion was proposed by Aoki \cite{Aoki1982simulation} to simulate the behavior of schools of fish. The model considered stochastic interactions with close neighbors within a finite angular view. The direction of movement of each individual was calculated as a weighted average over the directions of neighbors within sight. Although simulations were carried out only for very small numbers of individuals (typically 8 and up to 32), the results demonstrated conditions for pattern formation and flocking without the need for a group leader. The same conclusion was presented by Patridge~\cite{partridge1982structure}. His model, however, was based on the hypothesis that schooling relied on information acquired trough the lateral line, an organ sensitive to water displacement, complemented by visual cues. Thus, attraction between individuals was attributed to their ability to see, while repulsion and spacing within the flock depended on the lateral line.

Reynolds \cite{reynolds1987flocks}, who was interested in simulating a small number of flying birds, published a similar work a few years later. The trajectories of individuals were described by differential equations that would: (i) avoid collisions; (ii) head in the direction of the neighbors and; (iii) stay close to the center of mass of the flock.

\subsubsection{The Vicsek model}

\begin{figure}
	\centering
	\includegraphics[width =0.8 \columnwidth]{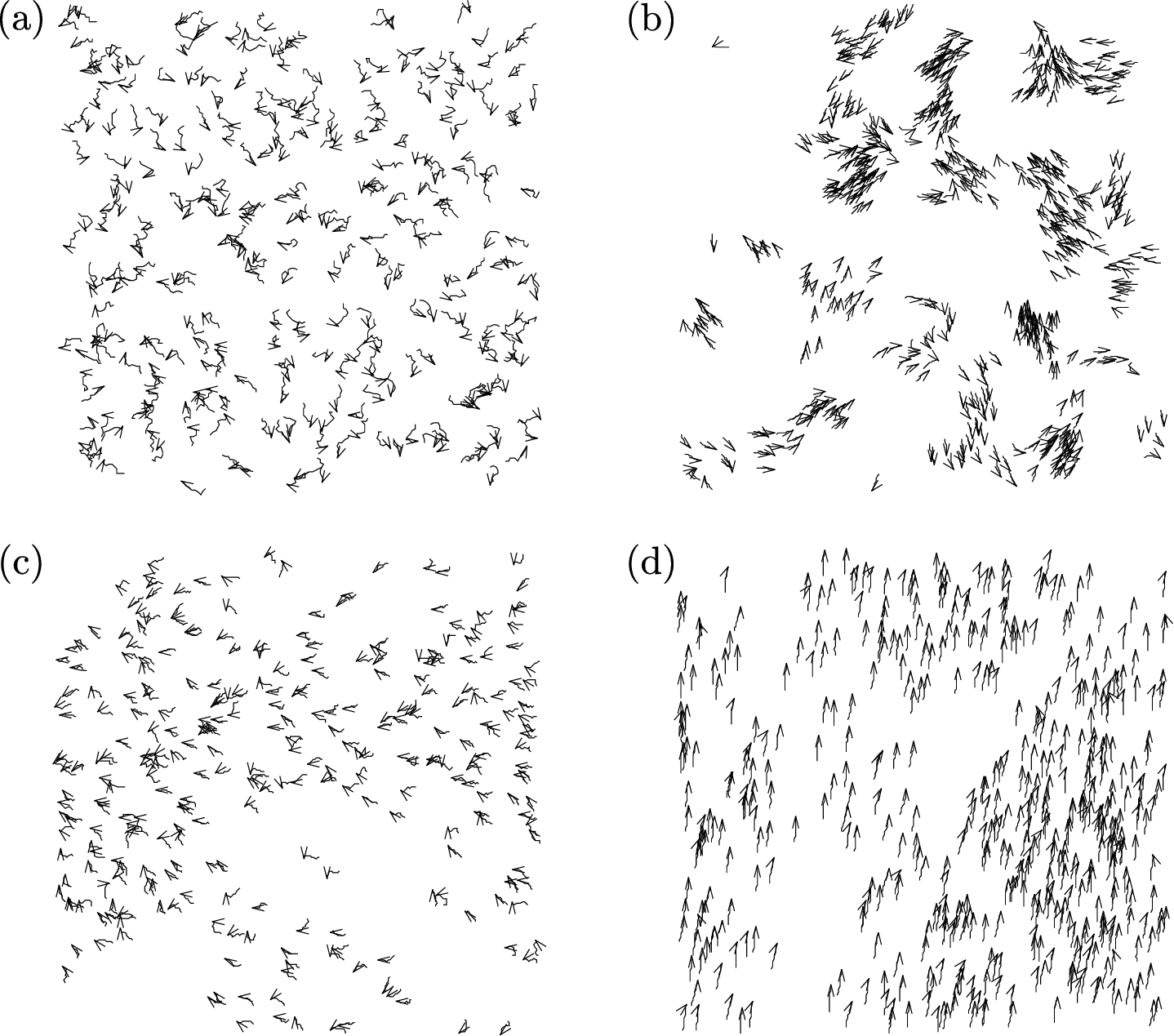}
	\caption{Simulation of the Vicsek model given by Eqs.~\eqref{vicsek1}-\eqref{vicesk2}. Velocities of the particles are indicated by small arrows, while their trajectories for the last 20 time steps are shown by a short
    continuous curve. The number of particles is $N = 300$ and $v=0.03$. (a) initial configuration at $t = 0$. (b) $L = 25$, $\eta = 0.1$; for small densities and
    noise the particles tend to form groups moving coherently in random directions. (c) $L = 7$, $\eta = 2.0$; at higher densities and noise the particles move randomly with some correlation. (d) $L = 5$, $\eta = 0.1$; for higher
    density and small noise the motion becomes ordered. \\ Source: Reprinted figure with permission from Ref. \cite{vicsek1995novel}.}
    \label{figfound3}
\end{figure}

The first model capable of dealing with large collectives and provide quantitative statistical results was proposed by Vicsek et al.~in 1995~\cite{vicsek1995novel,vicsek2012collective}. They considered a square cell of size $L$ with periodic boundary conditions where $N$ self-propelled particles could move and interact with each other. The particle $i$ was characterized by its position $\mathbf{x}_i(t)$ $\in \mathbb{R}^2$ and velocity $\mathbf{v}_i(t)$ $\in \mathbb{R}^2$, which were updated at discrete time steps. All particles were assumed to have the same absolute velocity $v$ and only their directions $\theta_i$ were updated. Initially, the $N$ particles were randomly distributed over the cell in random directions (see Fig. \ref{figfound3}(a)). Positions and velocities were then updated according to
\begin{equation}
    \mathbf{x}_i(t+h) = \mathbf{x}_i(t) + h \, \mathbf{v}_i(t), \label{vicsek1}
\end{equation}
and
\begin{equation}
    \theta_i(t+h) = \langle \theta_i \rangle_r + \Delta \theta_i, \label{vicesk2}
\end{equation}
where $h$ is the time step and $\langle \theta_i \rangle_r$ denotes the average direction of the velocities of particles that are within a circle of radius $r$ around particle $i$, including itself. The average direction was given by the angle $\arctan{[\langle \sin (\theta(t) \rangle_r/(\cos \langle \theta(t)) \rangle_r]}$
and $\Delta \theta_i$ is a random number chosen with a uniform probability from the interval $(-\eta/2,\eta/2)$, representing noise and playing the role of a temperature-like variable.

There are three free parameters in the Vicsek model for fixed cell size $L$: noise intensity $\eta$, speed $v$, and particle density $\rho = N/L^2$. In the limit $v \rightarrow 0$, the particles do not move and the model becomes an analog of the well-known XY model \cite{kosterlitz2018ordering,toner1995long}. For $v \rightarrow \infty$, the particles become completely mixed between updates, corresponding to the mean-field behavior of a ferromagnet. Results are especially interesting for intermediate values, particularly for $0.03 < v < 0.3$~\cite{vicsek1995novel}. Figure \ref{figfound3} shows typical configurations: panel (a) displays a random configuration at $t=0$ whereas (b) shows the formation of coherent groups at low density and noise. For high density and noise, the particles move randomly but keep some correlation among them (panel (c)), and for high density and small noise the motion becomes ordered (panel (d)). For fixed density there is a critical noise $\eta_c$ above which the motion becomes disordered. Similarly, for fixed noise, there is a critical density $\rho_c$ below which no order is observed. Both critical values were determined in the limit $L \rightarrow \infty$.

The order parameter measuring the transition from disordered to ordered motion is the average speed
\begin{equation}
    v_a = \frac{1}{Nv} \left| \sum_{j=1}^N \mathbf{v}_i \right|,
\end{equation}
which goes from zero, when the particles are randomly oriented, to 1, when they move in the same direction. The phase transition can be characterized by measuring how $v_a$ changes with noise and particle density. Writing 
\begin{equation}
    v_a \propto (\eta_c(\rho) - \eta)^\beta \qquad;\qquad v_a \propto (\rho - \rho_c(\eta))^\delta
\end{equation}
it was found, through simulations, that $\beta = 0.45 \pm 0.07$ and $\delta = 0.35 \pm 0.06$, indicating a continuous phase transition. The critical values $\eta_c(\rho)$ and $\rho_c(\eta)$ are computed in the limit $L\rightarrow\infty$. It has been argued that the phase transition might actually be discontinuous for very large $L$ \cite{chate2008collective} and for $v > 0.5$, but it remains continuous in the limit where $v \rightarrow 0$~\cite{baglietto2009computer}. These results inspired Toner and Tu~\cite{toner1995long} to formulate a field theory of flocking, which also gained notoriety~thereafter.

\subsubsection{The Cucker-Smale model}

The model proposed by Cucker and Smale \cite{cucker2007emergent} differs from the Vicsek model in three important aspects: motion occurs in 3D, it is deterministic (no noise), and interactions are long-range (controlled by a parameter $\beta$).

Let $\mathbf{x}_i$ and $\mathbf{v}_i$ $\in \mathbb{R}^3$ represent the position and velocity of the $i$-th particle, respectively. The dynamics in the Cucker-Smale model is governed by the rules:
\begin{eqnarray}
    \mathbf{x}_i(t+h) &=& \mathbf{x}_i(t) + h \, \mathbf{v}_i \nonumber\\
    \mathbf{v}_i(t+h) &=& \mathbf{v}_i(t) + h \, \sum_{j=1}^N A_{ij} (\mathbf{v}_j - \mathbf{v}_i)
    = \mathbf{v}_i(t) - h \, \sum_{j=1}^N L_{ij} \mathbf{v}_j,
\end{eqnarray}
where $h$ is the time step, the time-dependent adjacency matrix $A_{ij}$ is defined as
\begin{equation}
    A_{ij} = {(1+||\mathbf{x}_i - \mathbf{x}_j||^2)^{-\beta}}
\end{equation}
and the Laplacian matrix by
\begin{equation}
    L_{ij} =  k_i \delta_{ij} - A_{ij},
\end{equation}
where $k_i = \sum_k A_{ik} - A_{ij}$ is the degree of particle $i$. In the limit $h\to 0$, the system equations become
\begin{eqnarray}
    \dot{\mathbf{x}}_i &=& \mathbf{v}_i \nonumber \\
    \dot{\mathbf{v}}_i &=& -\sum_{j=1}^N L_{ij} \mathbf{v}_j.
    \label{found12}
\end{eqnarray}
The Vicsek model is recovered if motion is restricted to 2D and the adjacency matrix is replaced by $A_{ij}=1$ if $||\mathbf{x}_i - \mathbf{x}_j||< r$ and $A_{ij}=0$ otherwise. In the Cucker-Smale model, the sharp cutoff in the interaction range is replaced by a smooth function of the distance between particles. This allowed to prove the following theorem \cite{cucker2007emergent,vicsek2012collective}:\\

\noindent {\bf Theorem}: For the equations of flocking (\ref{found12}) there exists a unique solution for all $t \in \mathbb{R}$. If $\beta < 1/2$, the velocities $\mathbf{v}_i(t)$ tend to a common limit $\mathbf{v}^* \in \mathbb{R}^3$ as $t \rightarrow \infty$, where $\mathbf{v}^*$ is independent of $i$, and the vectors $\mathbf{x}_i - \mathbf{x}_j$ tend to a limit-vector $\mathbf{x}^{ij}$ for all $i,j \leq N$, as $t \rightarrow \infty$ that is, the relative positions remain bounded. If $\beta \geq 1/2$ dispersal, the split-up of the flock, is possible. However, provided that some certain initial conditions are satisfied, flocking will still occur.

\subsubsection{The Couzin model}

Couzin et al.~\cite{couzin2002collective} introduced a more realistic model taking into account local repulsion, alignment, and attraction. Individual dynamics were based on the definition of three behavioral zones, dubbed `zone of repulsion' ({\it zor}), `zone of orientation' ({\it zoo}), and `zone of attraction' ({\it zoa}). 

Let $\mathbf{r}_i$ and  $\mathbf{v}_i \in \mathbb{R}^3$ denote the position and velocity of the individual $i$, respectively. Velocities are assumed to be of unit module and only their directions change over time. The maximum rotation allowed per time step $h$ is $\theta h$. The {\it zor} is a sphere of radius $r_r$ centered on the individual. If the number of individuals in the {\it zor} is $n_r \neq 0$, the direction of individual $i$ is updated according to
\begin{equation}
    \mathbf{d}_{ir}(t+h) = - \sum_{j\neq i}^{n_r} \frac{\mathbf{r}_{ij}(t)}{|\mathbf{r}_{ij}(t)|}
\end{equation}
where $\mathbf{r}_{ij} = (\mathbf{r}_j - \mathbf{r}_i)/|\mathbf{r}_j - \mathbf{r}_i|$. This guarantees that the individuals maintain their personal space, or avoid collisions. If $n_r \neq 0$, the individual turns towards $\mathbf{d}_{ir}(t+h)$ up to the maximum angle $\theta h$.

If $n_r=0$ individuals respond to the orientation zone, which is spherical but excludes a volume behind the individual (blind zone, similar to Aoki's model \cite{Aoki1982simulation}). Individuals try to align themselves with neighbors within the {\it zoo}, resulting in
\begin{equation}
    \mathbf{d}_{io}(t+h) = \sum_{j\neq i}^{n_o} \frac{\mathbf{v}_j(t)}{|\mathbf{v}_j(t)|}
\end{equation}
or towards the positions of individuals within the {\it zoa},
\begin{equation}
    \mathbf{d}_{ia}(t+h) = \sum_{j\neq i}^{n_a} \frac{\mathbf{r}_{ij}(t)}{|\mathbf{r}_{ij}(t)|}.
\end{equation}
If neighbors are found in both zones, $\mathbf{d}_{ioa}(t+h) = [\mathbf{d}_{io}(t+h)+ \mathbf{d}_{ia}(t+h)]/2$. The velocity of particle $i$ is then rotated towards $\mathbf{d}_{io}$, $\mathbf{d}_{ia}$, or $\mathbf{d}_{ioa}$ up to the maximum angle $\theta h$. Positions are also updated according to $\mathbf{r}_i(t+h) = \mathbf{r}_i(t) + h \mathbf{v}_i(t)$.

The model predicts several types of collective behaviors, with sharp transitions between them. It shows that small changes in individual responses result in large changes in group properties and organization, such as observed fish schools \cite{krause1993relationship}. 

\subsection{Swarmalators}
A `swarmalator' is an individual with an internal phase and the ability to move in space. The term is a neologism for `swarming oscillator'. The phase is one-dimensional, similar to that defined by Winfree and Kuramoto, and its motion space can be one-, two-, or three-dimensional. The main feature of swarmalators is that spatial and phase dynamics are \textit{coupled}, meaning that the phases affect a swarmalator's spatial dynamics and the locations in space affect its phase dynamics. 

A model describing the behavior of interacting swarmalator collectives was proposed by O'Keeffe et al.~\cite{o2017oscillators} in 2017. That paper also introduced the term `swarmalator'. Before that, inspired by natural and artificial systems, many efforts were made to describe and explain the nature of mobile agents with internal degrees of freedom, where the two-way coupling observed in swarmalators was not necessarily taken into account.

\subsubsection{Early models}
We can start by considering extreme case models: those where swarming or synchronization dominates, while the other states are barely defined, yet they still exist. For instance, back in 1987, Sakaguchi and Kuramoto modified the model named after the latter to distribute a population of phase oscillators on a $d$-dimensional hyper\-cubic lattice~\cite{sakaguchi1987local}. As a result, synchronization patterns were unveiled and found to be related to both the dimensionality of the lattice and the interaction ranges. Despite being distributed on static sites, a sense of spatial distribution for oscillators was already suggested. Much work exists along the same line, where oscillators are connected through networks of static nodes, and discussions consistently highlight the relevance of the distance between these in synchronization~\cite{arenas2008synchronization, rodrigues2016kuramoto}. However, since our focus is on oscillators with spatial dynamics, we narrow our scope to mobile oscillators.

To begin with, the model presented in~\cite{uriu2013dynamics} closely follows the ideas of Sakaguchi and Kuramoto~\cite{sakaguchi1987local}. The key difference, evidently, is that the oscillators are distributed in space and can switch positions between neighbors following a Poisson process. Then, despite being confined to a one-dimensional space, the \textit{information spreading}, as synchronization is described by the authors, occurs faster as translation velocities increase. More elaborate systems, in which oscillators are distributed on evolving networks representing their spatial states, have also yielded interesting results. In~\cite{stilwell2006sufficient}, a \textit{moving neighborhood network} of oscillators is considered, representing a group of moving individuals who share information only when proximity conditions are met. The authors then derive conditions for synchronization, which depend on the network's topology evolution speed. A model based on the same framework is presented in~\cite{frasca2008synchronization} for the synchronization of chaotic oscillators moving in a two-dimensional space. However, using this model, it is shown that synchronization conditions depend mainly on the density of oscillators in the plane. A complementary work for a similar setup of chaotic oscillators is presented in~\cite{fujiwara2016synchronization}, where simplifications related to time-scales considered in~\cite{stilwell2006sufficient} and~\cite{frasca2008synchronization} are left out. Additionally, more recent applications such as the synchronization of reservoir computing oscillators~\cite{weng2023synchronization} and multilayer mobile networks~\cite{meli2023mobile} have also been studied.

Beyond systems where the spatial states of oscillators are determined by a network topology, much work has also been done on systems where oscillators move freely and interact with those within a proximity radius. The spatial dynamics of these individuals follow the Vicsek model~\cite{vicsek1995novel} or slight variations of it. For instance, in~\cite{degond2014hydrodynamics}, the authors present a system of self-propelled phase oscillators whose dynamics are described by what they refer to as the Kuramoto-Vicsek model. In this system, particles can align spatially and synchronize their internal phases, but their phases do not influence their positions. The spatial effects on the phases are indirect: oscillators are considered coupled only when they are close to each other. A similar approach is presented in~\cite{kruk2018self}, where, however, the internal phases represent the individuals' orientations, as their spatial dynamics are described by a polar parametrization of the Vicsek model. Hence, the spatial dynamics depend on the phase states. In addition to this one-way dependency, the model uses a modification from the Kuramoto framework: phase frustration~\cite{sakaguchi1986soluble}. Both considerations --- the one-way dependency of states and the phase frustration --- enable the emergence of chimera states, previously observed in coupled phase oscillator systems~\cite{kuramoto2002coexistence, abrams2004chimera} and Vicsek-like models with stochastic forcing~\cite{chate2008modeling}. It is clear, from the last example, that even a one-way coupling between phases and positions enrich the dynamics of a system of mobile oscillators significantly.

One-way coupling is also an inherent property of biologically inspired synchronization techniques for computer and communication networks. The fact that the nodes' physical locations influence their phase dynamics is particularly evident in wireless multi-hop networks, such as sensor networks and mobile ad hoc networks \cite{werner2005firefly,hong2005scalable,tyrrell2010emergent,pagliari2010scalable}. In these systems, nodes communicate over channels subject to signal attenuation (path loss), meaning that a node with limited transmit power can only synchronize with neighbors within a certain vicinity. Additionally, environmental factors like multi\-path propagation and shadowing obstacles cause variations in neighbor reachability over time (fading), introducing a dynamic and random element to the coupling.

\subsubsection{Two-way coupled models}
Prior to the seminal model of swarmalators~\cite{o2017oscillators}, a few other models were proposed that incorporated a two-way coupling between internal degrees of freedom and the positions of mobile individuals. In~\cite{shibata2003coupled}, individuals are described as elements with time-dependent internal states and spatial positions. These elements behave as self-sustained oscillators in the sense that, in isolation, their internal states exhibit oscillatory dynamics. The evolution of their internal states is influenced by local interactions with their neighbors, and their spatial movement also depends on forces generated by interactions between their internal states. This is verified by their equations of motion:
\begin{align}
\begin{split}
x_{i, n+1} &= (1-\epsilon) f(x_{i, n}) + \frac{\epsilon}{N_{i, n}}\sum_{j\in \mathcal{N}_i}f(x_{j, n}),\\
\vec{r}_{i, n+1} &= \vec{r}_{i, n} + \sum_{j\in \mathcal{N}_i}\frac{\vec{r}_{j, n} - \vec{r}_{i, n}}{|\vec{r}_{j,n} - \vec{r}_{i, n}|}\mathcal{F}(x_{i, n+1}, x_{j, n+1}),
\end{split}
\label{eq:shibata}
\end{align}
where at time-step $n$, $x_{i, n}$ and $\vec{r}_{i, n}$ represent, respectively the internal state and position of the $i$-th element. The evolution of the internal states is governed by $f(x) = 1-ax^2$ and, similarly, the spatial states depend on the force $\mathcal{F}(x_i, x_j) = Fx_ix_j$. In both, internal and spatial dynamics, the elements affecting the focal one are its neighbors, positioned within a radius $R$ as imposed by $\mathcal{N}_i = \{j: |\vec{r}_{j, n} - \vec{r}_{i, n}| \leq R\}$. The effect of these is averaged by their number $N_{i, n}$. The control parameters are $a$, $\epsilon$, the space dimensionality $d$, $F$, and $R$.

\begin{figure}
    \centering
    \includegraphics[width=0.95\textwidth]{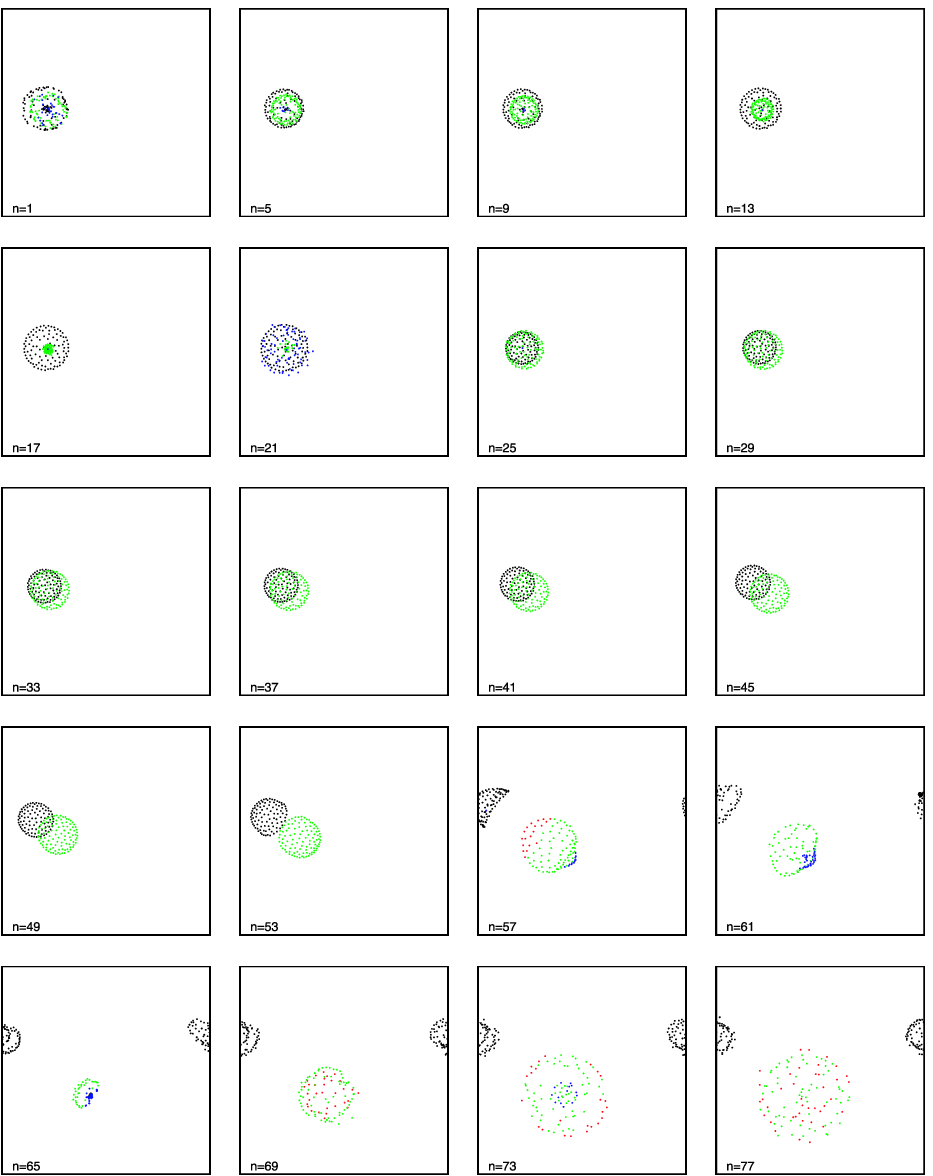}
    \caption{Patterns formed in a two-dimensional space ($d = 2$) of length $L = 50$, with control parameters $a = 1.8$, $\epsilon = 0.3$, $N = 200$, $R = 15$, and $F = 0.2$. Results are obtained using Eq.~\eqref{eq:shibata}. Panels show the collective states at each four time-steps, from left to right and from top to bottom. Internal states are represented by colors. \\ Source: Reprinted figure with permission from Ref.~\cite{shibata2003coupled}.}
    \label{fig:shibata}
\end{figure}

The two-way coupling introduced in~\cite{shibata2003coupled}, as depicted by Eqs.~\eqref{eq:shibata}, leads to the emergence of five collective states where coherence between spatial and internal states may or may not exist. One example of successive pattern formation over time is shown in Figure~\ref{fig:shibata}, where individuals, represented by small colored circles, form clusters in a two-dimensional periodic space. As shown, individuals initially gather to form a circularly shaped structure. However, as time passes, it splits and shape-shifts, while still retaining a hint of its former structure.

Another example of this two-way coupling is presented in~\cite{tanaka2007general}, where the model is inspired by the alignment of organisms in a chemotactic process. In addition to spatial organization, the model considers that each individual possesses an internal oscillatory degree of freedom. Then, motion is driven by density gradients of a chemical that mediates the interaction between individuals, and the internal degree of freedom controls the production or consumption of this chemical by each of them. This two-way dependency gives rise to numerous collective states where individuals cluster in space and move in various ways, while either synchronizing or not. The dynamics of this system are described by
\begin{align}
    \begin{split}
        \dot{\vec{x}}_i(t) &= \vec{f}(\vec{x}_i) + k\vec{g}(S(\vec{r}_i, t)),\\
        m\ddot{\vec{r}}_i(t) &= - \gamma\dot{\vec{r}_i} - \Gamma(\vec{x}_i)\vec{\nabla}S|_{\vec{r} = \vec{r}_i},
    \end{split}
    \label{eq:tanaka_coupled}
\end{align}
where, for the $i$-th individual, $\vec{x}_i$ and $\vec{r}_i$ represent the $N$- and $D$- dimensional internal and position states, respectively. The evolution of the internal state is driven by $\vec{f}$, representing its intrinsic behavior, and $\vec{g}$, representing the influence of the chemical concentration $S$, scaled by a factor $k$. The position dynamics appear affected by the individual's mass, a viscosity term related to the coefficient $\gamma$, and the spatial gradient of the chemical concentration, given by $\Gamma\vec{\nabla}S$, where $\Gamma$ is a matrix. The concentration, moreover, follows its own dynamics, given by 
\begin{equation}
        \tau \partial_tS(\vec{r}, t) = -S + d\nabla^2 S+ \sum_jh(\vec{x}_j)\delta(\vec{r} - \vec{r}_j),
        \label{eq:tanaka_conc}
\end{equation}
where $\tau$ is a time constant, $d$ is the coefficient of a spatial diffusion term, and the last addend represents the local consumption and production carried out by the elements.

\begin{figure}
    \centering
    \includegraphics[width=\textwidth]{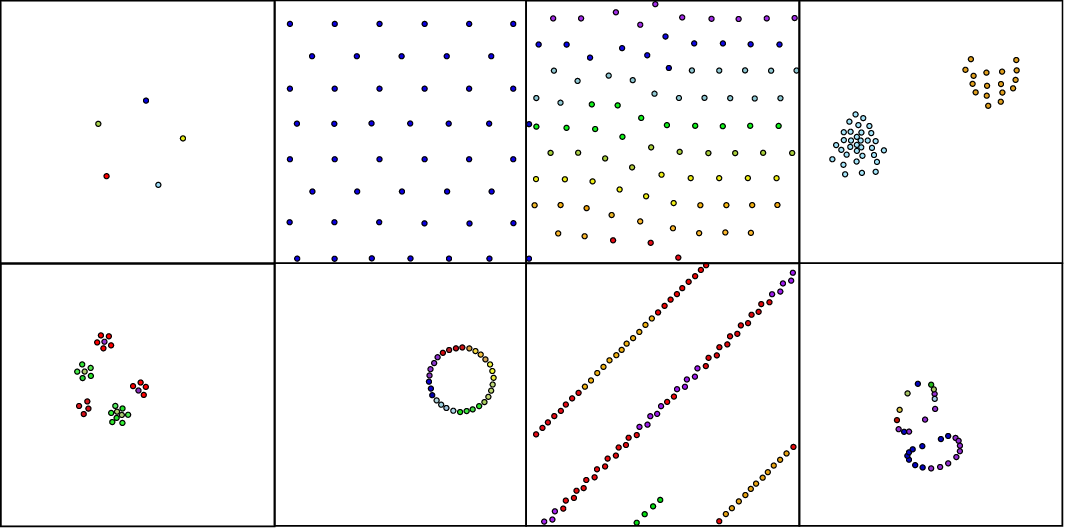}
    \caption{Emergent collective patterns in a simplified model of chemotaxis given by Eq.~\eqref{eq:tanaka_reduced}. Individuals are distributed in a two-dimensional space, and colors represent their internal phases. \\ Source: Reprinted figure with permission from Ref. \cite{iwasa2010hierarchical}.}
    \label{fig:iwasa}
\end{figure}

The system for a chemotactic process, introduced in Eqs.~\eqref{eq:tanaka_coupled} and~\eqref{eq:tanaka_conc}, can be greatly reduced by making considerations specific to the dynamical process (see~\cite{tanaka2007general} for details). The simplified model is described by
\begin{align}
    \begin{split}
    \dot{\psi}_i(t) &= \sum_{j\neq i}e^{|\vec{r}_{ji}|}\sin\left(\psi_{ji} + \alpha |\vec{r}_{ji}| - c_1\right),\\
    \dot{\vec{r}}_i(t) &= c_3\sum_{j\neq i}\frac{\vec{r}_{ji}}{|\vec{r}_{ji}|}e^{|\vec{r}_{ji}|}\sin\left(\psi_{ji} + \alpha |\vec{r}_{ji}| - c_2\right),\\
    \end{split}
    \label{eq:tanaka_reduced}
\end{align}
where $\vec{r}_{ji} = \vec{r}_j - \vec{r}_i$ and $\psi_{ji} = \psi_j - \psi_i$. The variable $\psi$, dubbed phase, represents the internal state of the individuals, and $c_1$, $c_2$, $c_3$, and $\alpha$ are control parameters. The individuals are now distributed in a two-dimensional space while the phases are one-dimensional. Notice that, despite the simplifications, the two-way coupling shown in Eqs.~\eqref{eq:tanaka_coupled} is still present in the reduced model. Additional analytical work on this system is presented in~\cite{iwasa2010hierarchical}, where individuals are confined to move on a single dimension. Then, stability conditions are derived for states of spatial clustering. Snapshots of some patterns that emerge after simulating the reduced model are shown in Figure~\ref{fig:iwasa}. These states can be static or active.

The model introduced in~\cite{ito2001spontaneous} is also related, however, individuals are not oscillators but dynamical elements capable of synchronizing. These are distributed across networks whose connections have time-dependent strengths, and their dynamics, as well as those of the elements, are coupled. In this vein, aiming to close this section, we shift our perspective to discuss specific active matter models in which coherence between self-propelled individuals is achieved through an alignment mechanism. Even though internal phases in these models are not defined according to a Kuramoto-like criteria, their formulation presents a two-way coupling between relevant degrees of freedom. We find that the simplest way to elaborate on this is by relying on the Vicsek model; specifically, on a continuous variation of it~\cite{peruani2008mean}. Here, the dynamics of the system are governed by
\begin{align*}
\begin{split}
    \dot{\vec{x}_i} &= v_0\vec{v}(\theta_i),\\
    \dot{\theta}_i &= -\gamma\frac{\partial U}{\partial \theta_i}(\vec{x}_i, \theta_i) + \eta_i(t),
\end{split}
\end{align*}
where $\gamma$ is a constant, $v_0$ represents the speed of individuals, $\vec{v}(\theta_i) = (\cos(\theta_i), \sin(\theta_i))$  is the velocity direction, $\eta(t)$ is white noise, and the alignment interactions are given by the potential $U = -\sum_{|\vec{x_j} - \vec{x_i}|}\cos(\theta_i - \theta_j)$ (we limit our description to this potential for the sake of brevity, though the authors also address results for a Lebwohl and Lasher potential). After a trivial calculation, it is evident that the dynamics of $\theta_i$ follow the Kuramoto model. However, in this case, the internal phases correspond to the individual orientations. Moreover, the interactions between these phases are constrained by the spatial proximity of their respective individuals, given by $|\vec{x}_j - \vec{x}_i|$. The individual spatial dynamics, on the other hand, are directly influenced by their orientations, thereby sealing the two-way coupling. This mechanism has been replicated and tailored to address the behavior of active individuals showing dependence on density~\cite{farrell2012pattern}, shape~\cite{sese2021phase}, limited spatial interactions~\cite{barberis2016large}, among others. However, the main difference between these and the definition of swarmalators is that their internal phases are always restricted to being the individuals' orientations.

%==============================================================
\section{Phenomenology} \label{sec.3}
%\section{Collective behaviors emerging from swarming and synchronization} \label{3}
%\color{blue}
Swarmalators capture the interplay between the spatial and phase dynamics and their mutual influence. In the following, we review the behavior of swarmalators in various spatial dimensions, incorporating multiple interaction mechanisms. A rich spectrum of collective states emerges, ranging from static patterns to dynamic spatiotemporal structures. \textcolor{black}{To aid readability, we briefly outline the modeling roadmap at the start of this part of the review. We treat the two-dimensional (2D) swarmalator model as the primary and more realistic formulation, since it is the natural setting in which the canonical collective states are defined. The one-dimensional (1D) model is then introduced as a derived, analytically tractable reduction, used to obtain solvable results and build intuition that can be carried back to the full two-dimensional dynamics. For completeness, we also discuss extensions of swarmalator models to three and higher dimensions.}

\subsection{Swarmalators in two dimensions (2D)}

%\subsubsection{Global coupling}
A 2D swarmalator model for global interaction among the units is governed by the pair of equations,
\begin{align}
	\dot{\textbf{x}}_{i} &= \textbf{v}_{i}+\frac{1}{N} \sum_{\substack{j = 1\\j \neq i}}^{N}\bigg[\text{I}_{\text{att}}(\textbf{x}_{j}-\textbf{x}_{i}) \text{F}_{\text{att}}(\theta_{j}-\theta_{i}) - \text{I}_{\text{rep}}(\textbf{x}_{j}-\textbf{x}_{i}) \text{F}_{\text{rep}}(\theta_{j}-\theta_{i})\bigg], \label{eq.2.9}\\
	\dot{\theta}_{i} &= \omega_{i}+\frac{K}{N}\sum_{\substack{j = 1\\j \neq i}}^{N} \text{H}(\theta_{j}-\theta_{i})\text{G}(\textbf{x}_{j}-\textbf{x}_{i}), \label{eq.2.10}
\end{align}
for $i = 1,2,\ldots,N$, where $N$ is the total number of swarmalators, $\textbf{x}_i = (x_i, y_i) \in \mathbb{R}^2$ is the position vector of the $i$th swarmalator, $\theta_i \in \mathbb{S}^1$ is its internal phase. $\textbf{v}_i$ and $\omega_i$ denote its self-propulsion velocity and natural frequency, respectively. It is clearly noted from the above pair of equations that the spatial dynamics (Eq.~\eqref{eq.2.9}) and the phase dynamics (Eq.~\eqref{eq.2.10}) affect each other. There are spatial attraction and repulsion among the swarmalators that are governed by the functions $\text{I}_\text{att}$ and $\text{I}_\text{rep}$, respectively. Attractive force is needed in the system so that the swarmalators remain bounded, whereas, the repulsive force ensures collision avoidance. $\text{F}_{\text{att}}$ and $\text{F}_\text{rep}$ represent the influence of phase dynamics on spatial attraction and repulsion, respectively. In Eq.\ \eqref{eq.2.10}, phase interaction between the swarmalators is given by the function $\text{H}$ and the influence of spatial proximity on phase dynamics is given by $\text{G}$. Swarmalators' phases are coupled with strength $K \in \mathbb{R}$. When $K$ is positive (attractive coupling), swarmalators try to minimize their phase differences while a negative value (repulsive coupling) of $K$ increases the incoherence of their phases. 

\sloppy

The alignment component of the swarming dynamics in Eq.\ \eqref{eq.2.9} is reflected through the velocity $\textbf{v}_{i} = (v\cos \eta_i, v\sin \eta_i)$, where $\eta_i$ is the orientation of the $i$th swarmalator and $v$ is the speed.

\subsubsection{Coupling functions}
\label{sec.3.1.1}
We start by reviewing a particular instance of the model with the following choices,
\begin{equation}
	\text{I}_{\text{att}}(\textbf{x})=\frac{\textbf{x}}{|\textbf{x}|^{\alpha}}, \, \text{I}_{\text{rep}}(\textbf{x})=\frac{\textbf{x}}{|\textbf{x}|^{\beta}}, \, \text{G}(\textbf{x}) = \frac{1}{|\textbf{x}|^{\gamma}},
	\label{eq.2.11}
\end{equation}
where $\text{I}_{\text{att}}$, $\text{I}_{\text{rep}}$, and $\text{G}$ are chosen as power laws with positive exponents $\alpha$, $\beta$, and $\gamma$, respectively, and where $|\cdot|$ represents the Euclidean norm. The exponents $\alpha$ and $\beta$ play a crucial role in determining the collective motion in the system. One wants to ensure short-range repulsion and long-range attraction, so that collisions are avoided and the solution remains bounded \cite{fetecau2011swarm}. It is required that the exponents $\alpha$ and $\beta$ satisfy the following condition 
\begin{equation}
	1 \le \alpha < \beta.
	\label{eq.2.12}
\end{equation}
It is further assumed that $\text{F}_{\text{att}}$ and $\text{F}_{\text{rep}}$ are even and bounded functions of their arguments.
%Now, we have sufficient tools in our hand to proceed with the theoretical analysis of the model.
For notational simplicity, the following are set, $\textbf{X} := (\textbf{x}_1,\textbf{x}_2, \ldots,\textbf{x}_{N})$, $\textbf{V} := (\textbf{v}_1,\textbf{v}_2,\ldots,\textbf{v}_{N})$, $\Theta := (\theta_{1},\theta_{2},\ldots,\theta_{N})$, $W := (\omega_{1},\omega_{2},\ldots,\omega_{N})$, $\mathcal{N} := \{1,2,\ldots,N\}$. One defines the following functionals
\begin{gather}
	\mathcal{C}_1(\textbf{X},\textbf{V},t) := \sum_{i \in \mathcal{N}} \textbf{x}_i - t \sum_{i \in \mathcal{N}} \textbf{v}_i, \label{eq.2.13}\\
	\mathcal{C}_2(\Theta,W,t) := \sum_{i \in \mathcal{N}} \theta_i - t \sum_{i \in \mathcal{N}} \omega_i. \label{eq.2.14}
\end{gather}
Now, taking the sum over Eqs.\ \eqref{eq.2.9} and \eqref{eq.2.10}, it is elementary to see that $\dfrac{d}{dt}\mathcal{C}_{1,2} = 0$. Therefore, these two quantities $\mathcal{C}_1$ and $\mathcal{C}_2$ are conserved along the swarmalators dynamics. However, the presence of singular functions like $\frac{1}{|\textbf{x}_{j}-\textbf{x}_{i}|^{\mu}}$ raises the question whether the model is well-posed or not. It is imperative to rule out collisions, which violates the well-posedness of the model. Inter-swarmalator collision can be avoided if one assumes the initial data \((\textbf{X}(0), \Theta(0))\) to be non-collisional, i.e., $\min_{1\le i,j \le N} |\mathbf{x}_{i}(0)-\mathbf{x}_{j}(0)| > 0$ for all $i,j \in \mathcal{N}$ and $i \ne j$, along with $\alpha, \beta$ satisfying Eq.\ \eqref{eq.2.12}. These conditions further guarantee the existence of a minimal inter-particle distance among the swarmalators which eliminates the possibility of unbounded solution in case of swarmalators being asymptotically close. The detailed discussions of these results can be found in \cite{ha2019emergent}.

\begin{figure}[hpt]
	\centerline{
		\includegraphics[scale = 0.4]{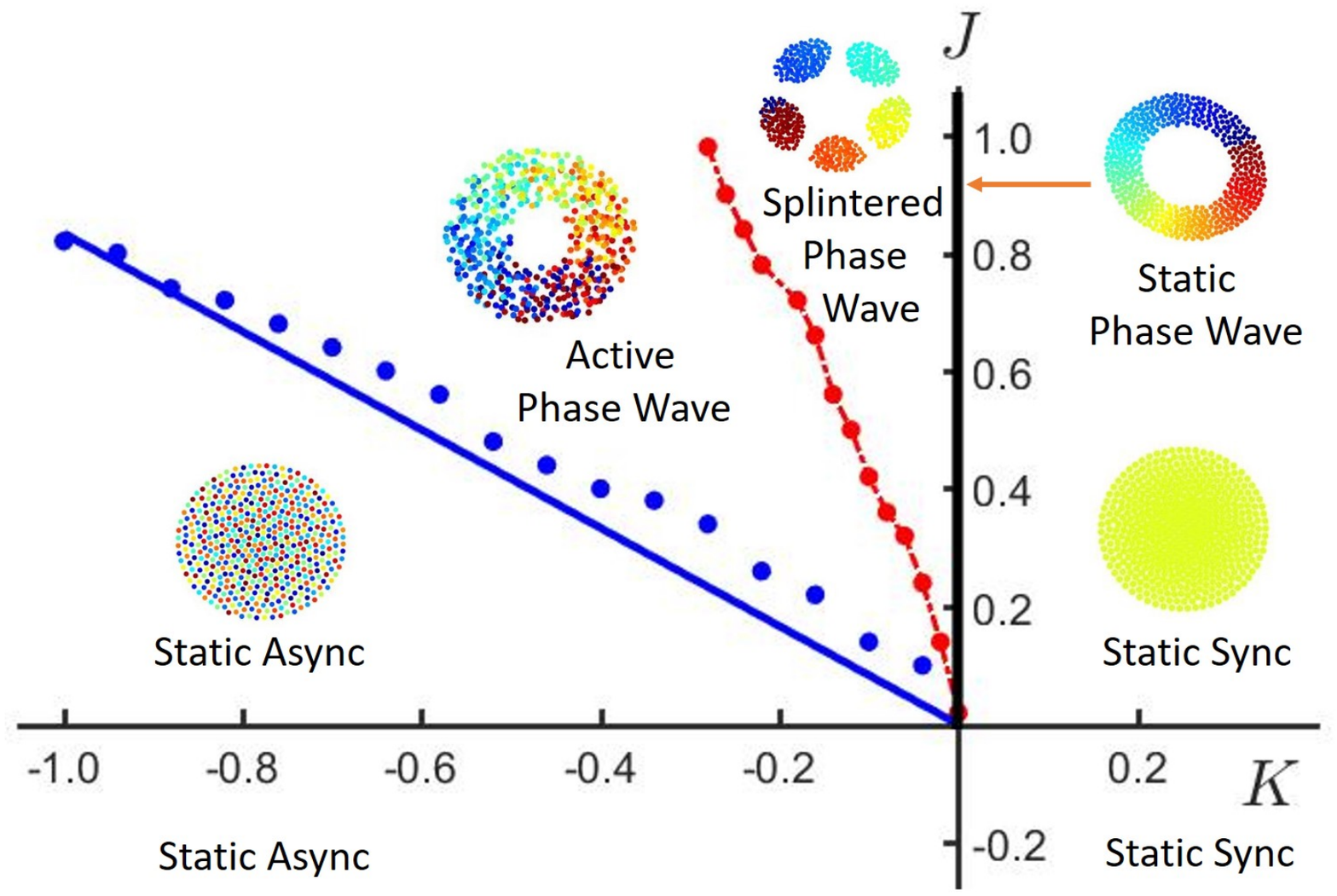}}
	\caption{$K$-$J$ parameter region for emerging collective states of the swarmalator system given by Eqs.~\eqref{eq.2.15}-\eqref{eq.2.16}. Red dots are the points where $T$ bifurcates from zero to non-zero. They are joined with dashed line to separate the regions between splintered phase wave and active phase wave states. Blue dots are numerically calculated points where $S$ bifurcates from non-zero to zero. \\ Source: Reprinted figure with permission from Ref. \cite{sar2022dynamics}.}
	\label{fig.2.3}
\end{figure}

Note that, the above results are independent of the choices of the functions $\text{F}_{\text{att}}$, $\text{F}_{\text{rep}}$, and $\text{H}$, the interaction functions that depend on the phases of the swarmalators. Now, a more specific instance of the model is considered where the spatial functions $\text{I}_{\text{att}}$, $\text{I}_{\text{rep}}$, and $\text{G}$ are chosen following Eq.\ \eqref{eq.2.11} with $\alpha=1$, $\beta = 2$, and $\gamma=1$ that satisfy Eq.\ \eqref{eq.2.12}. The influence of phase similarity on spatial attraction is considered via the function $\text{F}_{\text{att}}(\theta_j-\theta_i) = 1 + J \cos(\theta_j-\theta_i)$. For $J>0$, {\it like attracts like}, i.e., swarmalators in nearby phases attract each other. When $J$ is negative, swarmalators are preferentially attached to the ones in opposite phases. $-1<J<1$ is considered so that $\text{F}_{\text{att}}$ is always strictly positive. The influence of phase dynamics on the spatial repulsion is not considered here, i.e., $\text{F}_{\text{rep}}(\theta_j-\theta_i) = 1$. The phase interaction among the swarmalators is taken in the spirit of Kuramoto model \cite{kuramoto1975self} by taking $\text{H}(\theta_j-\theta_i) = \sin(\theta_j-\theta_i)$. Here identical swarmalators are chosen with same velocity $\textbf{v}_i = \textbf{v}$ and frequency $\omega_{i} = \omega$, and by choice of reference frame both of them are set to zero. Finally, the dynamics of the swarmalators are governed by the following pair of equations~\cite{o2017oscillators},
\begin{align}
	\dot{\textbf{x}}_{i} &= \frac{1}{N} \sum_{\substack{j = 1\\j \neq i}}^{N}\bigg[\frac{\textbf{x}_{j}-\textbf{x}_{i}}{|\textbf{x}_{j}-\textbf{x}_{i}|} (1+J \cos(\theta_{j}-\theta_{i})) - \frac{\textbf{x}_{j}-\textbf{x}_{i}}{|\textbf{x}_{j}-\textbf{x}_{i}|^2}\bigg], \label{eq.2.15}\\
	\dot{\theta}_{i} &= \frac{K}{N}\sum_{\substack{j = 1\\j \neq i}}^{N} \frac{\sin(\theta_{j}-\theta_{i})}{|\textbf{x}_{j}-\textbf{x}_{i}|} 
	\label{eq.2.16}.
\end{align}
Each swarmalator is connected to all the other swarmalators in the system, and its spatial and phase dynamics are influenced by the dynamics of other swarmalators present in the system. The two controllable parameters $J$ and $K$ determine the system's long-term dynamical state. Five different asymptotic states are found depending on the values of $J$ and $K$ which is contemplated in Fig.\ \ref{fig.2.3}. In the snapshots, the swarmalators are colored according to their phases. These states are {\it static sync}, {\it static async}, {\it static phase wave}, {\it splintered phase wave}, and {\it active phase wave}. In the static states, the movement of the swarmalators is ceased and the phases become stationary, while in the active states, the swarmalators move. However in all states, the density of swarmalators $\rho(x,\theta,t)$ is stationary or time-independent, where $\rho(x,\theta,t) dx d\theta$ gives the fraction of
swarmalators with positions between $\textbf{x}$ and $\textbf{x} + d\textbf{x}$, and phases between $\theta$ and $\theta + d\theta$ at time $t$. Initial positions of the swarmalators are chosen from $[-1,1] \times [-1,1]$ and the initial phases are drawn from $[0,2\pi]$, both uniformly and randomly. From Eq.~\eqref{eq.2.13} and the subsequent discussions, it is understood that the center of positions of the system is conserved throughout the system. The description and characteristics of the five emerging are provided below:

{\it Static sync}: The swarmalators arrange themselves in a crystal-like pattern with circular symmetry, and exhibit complete phase synchronization, as reflected by their uniform coloring in Fig.~\ref{fig.2.3}. Given that their spatial positions become stationary and their phases converge to a common value, this configuration is referred to as the "static sync" state. This state emerges when $K>0$, irrespective of the value of $J$, as illustrated in Fig.~\ref{fig.2.3}. The radius of the disc structure of the static sync state is calculated to be $R_{\text{sync}} = (1+J)^{-1/2}$.

{\it Static async}: Swarmalators can also settle into a state where all phases $\theta$ are equally likely to occur at any spatial location $\textbf{x}$ inside a circular disc, resulting in a full spectrum of colors throughout the disc. Such a fully asynchronous state emerges when $ J < 0, K < 0 $, and also for $J > 0$ within the wedge $J < |K_c|$, as shown in the phase diagram in Fig.~\ref{fig.2.3}. Similar to the static sync case, the radius of the circular distribution can be analytically determined under a linear attraction kernel, yielding $R_{\text{async}} = 1$.

{\it Static phase wave}: The third stationary configuration arises in the special case where $K=0$ and $J>0$. In this regime, the phases of the swarmalators remain fixed at their initial values, while their spatial arrangement continues to evolve. The positive $J$ promotes spatial attraction among swarmalators with similar phases, resulting in a self-organized annular structure. In this state, the spatial angle $\phi$ of each swarmalator becomes perfectly aligned with its phase $\theta$. Since the phases are distributed continuously around the annulus, completing a full cycle, this configuration is referred to as the static phase wave state. The inner and outer radii of this state can be analytically calculated with a linear attraction kernel.

{\it Splintered phase wave}: As the coupling parameter $K$ decreases from zero into the negative domain ($K<0$), the system transitions into its first non-stationary state, illustrated in Fig.~\ref{fig.2.3}. In this regime, the previously continuous static phase wave breaks apart into separate clusters, each composed of swarmalators with similar phases. This configuration is referred to as the splintered phase wave. The precise mechanism that determines the number of such clusters remains unclear. A tendency toward fewer clusters is observed when the interaction functions $\text{I}_{\text{att}}$, $\text{I}_{\text{rep}}$, and $\text{G}$ operate at smaller length scales. Parameters $J$ and $K$ also influence clustering, although the relationship has yet to be fully characterized. Within each cluster, swarmalators exhibit low-amplitude oscillations in both space and phase, fluctuating around their mean values.

{\it Active phase wave}: As the coupling strength $K$ is further reduced, the oscillations grow in amplitude, eventually leading the swarmalators to perform regular, cyclical movements in both spatial angle and phase. Such motion arises from a conserved quantity in the system: $\langle \dot{\phi} \rangle = \langle \dot{\theta} \rangle = 0$, which becomes evident upon averaging Eqs.\eqref{eq.2.15} and~\eqref{eq.2.16} across the population, where $\phi = \tan^{-1}({y}/{x})$ denotes the spatial angle of the swarmalators. In addition to angular motion, the swarmalators exhibit radial oscillations, traveling from the inner to the outer edge of the annulus and back during each orbit.

{\it Order parameters:} After characterizing the five distinct states of the systems \eqref{eq.2.15} and \eqref{eq.2.16}, we now turn to the question of how to differentiate between them. To distinguish them, the following order parameters are introduced,
\begin{equation}
	W_{\pm}=S_{\pm} e^{\mbox{i} \Psi_{\pm}} = \frac{1}{N} \sum_{\substack{j = 1}}^{N} e^{\mbox{i}(\phi_j \pm \theta_j)}.
	\label{eq.2.17}
\end{equation}
In the static phase wave state, the spatial angle and phase of each swarmalator are perfectly correlated, such that $\phi_j = \pm \theta_j + C$, where the sign and the constant $C$ are determined by the initial conditions. This implies that either $S_+$ or $S_-$ is 1 and we define
\begin{equation}\label{smaxmin}
    S = \max \{ S_+, S_- \},
\end{equation}
so that $S = 1$ at $K=0$, where the static phase wave state occurs. As we move into the $K<0$ regime, the system enters the splintered phase wave state. Here, the correlation between $\phi_i$ and $\theta_i$ weakens, so $S < 1$. Further decreasing $K$ results in a non-monotonic decay of this correlation. Once the system transitions into the active phase wave state, this non-monotonicity disappears. Consequently, $S$ decreases steadily and ultimately approaches zero in the static async state, where $\phi_j$ and $\theta_j$ become completely uncorrelated. Since $S$ is non-zero in both the splintered and active phase wave states, it is insufficient for distinguishing between them. To resolve this, an additional order parameter $U$ is introduced. This parameter represents the fraction of swarmalators that have completed at least one full cycle in both phase and position, after transients are excluded. Accordingly, $U=0$ corresponds to the splintered phase wave, while $U>0$ indicates the presence of the active phase wave. A later study has revealed that the motion of each swarmalator in the 2D plane is chaotic both in the splintered phase wave and in the active phase wave states~\cite{ansarinasab2024spatial}. \textcolor{black}{In a very recent study, the clustering conditions and the influence of the number of clusters on cluster existence are established using a mean-field approximation and a self-consistency argument, and estimates are obtained for the cluster size in the splintered phase wave state and for the phase transition threshold between the splintered phase wave and active phase wave states~\cite{gong2024approximating}.}

{\it Effect of phase similarity on spatial repulsion}: Consider $\text{F}_{\text{att}}(\theta) = 1 + J_1\cos(\theta)$ and $\text{F}_{\text{rep}}(\theta) = 1 - J_2\cos(\theta)$ to explore the behavior of swarmalators when phase similarity influences both spatial attraction and repulsion~\cite{o2018ring}. A different parameter set is selected: $\alpha = 0$, $\beta = 2$, and $\gamma = 2$. Note that Eq.~\eqref{eq.2.12} does not hold for these values, and thus collision avoidance cannot be guaranteed. However, numerical simulations indicate that collisions are avoided as long as $J_2 \leq 1$. These parameter choices also simplify the mathematical analysis of the resulting dynamical states, assuming the model remains well-posed.

Depending on the parameters $J_1$, $J_2$, $K$, and $N$, swarmalators exhibit various dynamic behaviors. For certain combinations, a stationary configuration emerges where swarmalators arrange themselves evenly on a ring centered at the origin. In this configuration, each phase is perfectly correlated with its spatial angle, such that $\theta_i = \phi_i + C$ for some constant $C$ determined by the initial conditions. This configuration is referred to as the \textit{ring phase wave}. The position and phase of the $k$th swarmalator in this state are given by,
\begin{align}
	\textbf{x}_k &= R\cos(2\pi k / N)\,\hat{i} + R\sin(2\pi k / N)\,\hat{j},
	\label{eq.21}\\
	\theta_k &= \frac{2\pi k}{N} + C,
	\label{eq.22}
\end{align}
where $R$ is the ring radius, and $\hat{i}$, $\hat{j}$ are unit vectors in the $x$ and $y$ directions, respectively.

\begin{figure}[hpt]
	\centerline{
		\includegraphics[width = 0.75\columnwidth]{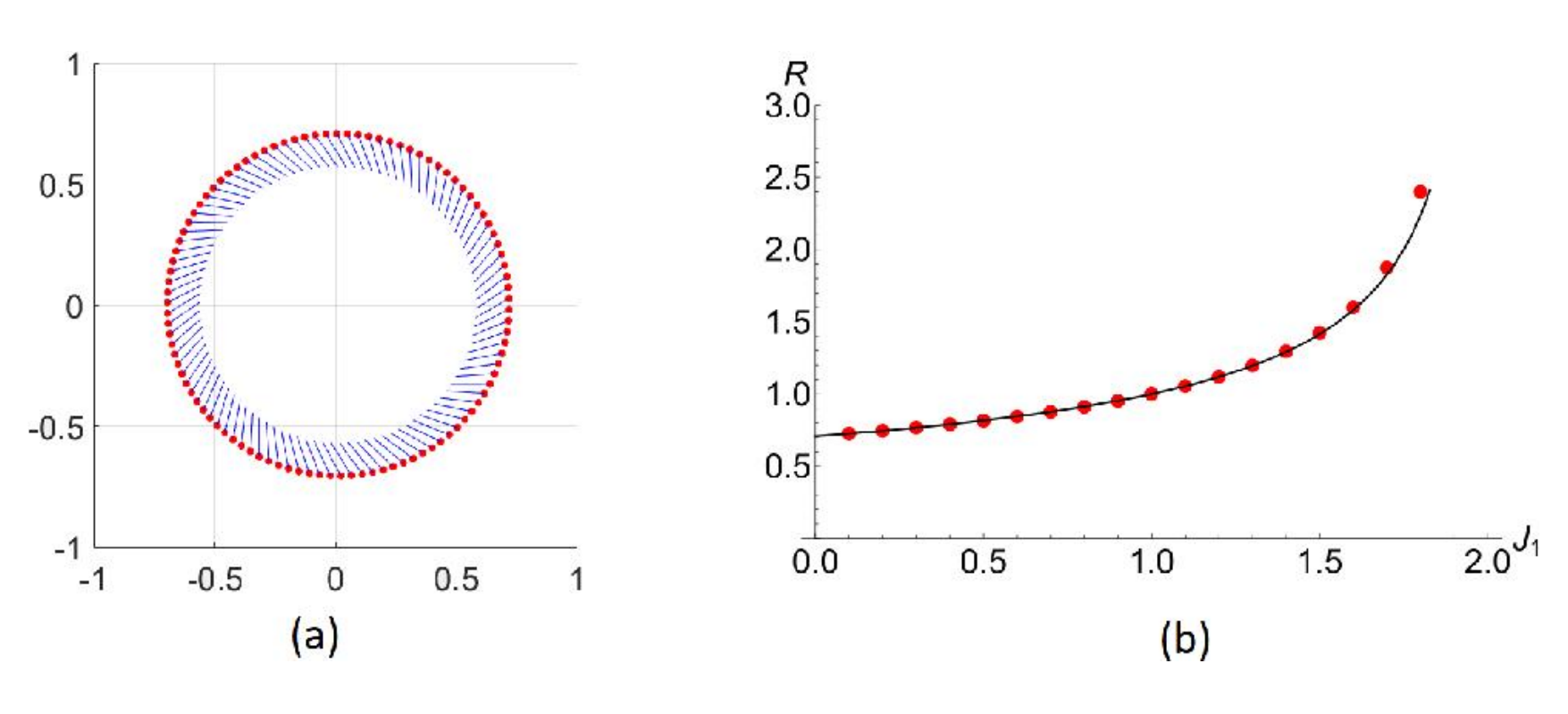}}
	\caption{(a) Scatter plot of the ring phase wave state. Each swarmalator’s phase is indicated by a blue ray, corresponding to the angle it forms with the positive $x$-axis. The configuration is shown for parameters $J_1 = 0$, $J_2 = 1$, $K = -0.003$, and $N = 100$. (b) Radius of the ring state as a function of $J_1$. Red dots represent numerically computed values for $J_2 = 1$ and $N = 100$, while the black curve denotes the theoretical prediction. \\ Source: Reprinted figure with permission from Ref.~\cite{o2018ring}.}
	\label{Fig.502}
\end{figure}

To simplify the analysis, it is helpful to adopt a complex representation, identifying the position vector $\textbf{x}_k = (x_k, y_k) \in \mathbb{R}^2$ as a complex number $z_k = x_k + \mbox{i} y_k$ in the complex plane. This facilitates the use of complex identities, particularly given the choice $\alpha = 0$, $\beta = 2$, and $\gamma = 2$. Substituting Eqs.~\eqref{eq.21} and~\eqref{eq.22} into the governing equations yields the radius of the ring,
\[
R = \sqrt{\frac{N - 1 + J_2}{N(2 - J_1)}}.
\]
Thus, the ring state exists only within the parameter region $\{J_1 < 2,\; J_2 > 1 - N\} \cup \{J_1 > 2,\; J_2 < 1 - N\}$. Figure~\ref{Fig.502} illustrates the spatial arrangement and radius of the ring phase wave state. A detailed stability analysis is provided in Ref.~\cite{o2018ring}. When $K = 0$, this state becomes unstable beyond a critical value $N_{\text{max}} = \frac{8}{(2 - J_1)(1 - J_2)}$, beyond which the static phase wave state appears. For $K < 0$, the system undergoes a bifurcation into the splintered phase wave state.

\begin{figure}[hpt]
	\centerline{
		\includegraphics[width = 0.6\columnwidth]{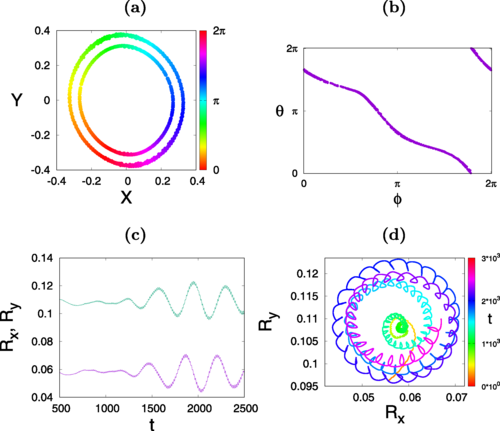}}
	\caption{Swarmalator exhibiting collective oscillations for parameters $K = -0.1$, $\sigma = 0.4$. (a) Spatial distribution in the $X$-$Y$ plane. (b) Scatter plot in the $(\phi, \theta)$ phase space. (c) Time series of the center of mass coordinates $R_x(t)$ and $R_y(t)$. (d) Trajectory of the center of mass. The color of the points indicates the progression of time. \\ Source: Reprinted figure with permission from Ref. \cite{jimenez2020oscillatory}.}
	\label{Fig.503}
\end{figure}

{\it Short-range repulsive function}: Although long-range interaction functions such as those discussed until this point are used frequently in the study of physical systems, most biological systems such as bacterial colonies~\cite{chen2011statistical}, flocks of birds~\cite{cavagna2015short}, and flying midges~\cite{puckett2014searching} are governed by both long- and short-range interactions. To capture this kind of short-range interaction in the swarmalator model, in Ref.~\cite{jimenez2020oscillatory}, the author proposed a model with long-range attractive and short-range repulsive force in the spatial component. The attractive force is the same as in Eq.~\eqref{eq.2.15}, but the spatial repulsive force is chosen as the Gaussian function $\text{I}_{\text{rep}}(\textbf{x}_{j}-\textbf{x}_{i}) = \frac{\textbf{x}_{j}-\textbf{x}_{i}}{\sigma} e^{-|\textbf{x}_{j}-\textbf{x}_{i}|^2/\sigma}$, where $\sigma$ determines the range of interaction. The phase interaction is the same as Eq.~\eqref{eq.2.16}. The combined effect of long-range attractive and short-range repulsive interactions brings about new collective states. Most notable of them is the tendency to form concentric circular patterns along with exhibiting oscillatory behaviors. A particularly fascinating oscillatory behavior emerges for the parameters $K = -0.1$ and $\sigma = 0.4$, where the entire system exhibits a collective oscillation. Figures~\ref{Fig.503}(a) and \ref{Fig.503}(b) illustrate this through a spatial plot in the $X$-$Y$ plane and a scatter plot in the $(\phi, \theta)$ plane, respectively. In this regime, the swarmalators form two elliptical clusters with a wave-like phase distribution. These ellipses oscillate in sync, with their major and minor axes alternating in length in a coordinated manner. The time series of the system's center of positions, given by
\begin{equation}
    R_x(t) = \frac{1}{N} \sum_i x_i(t), \quad R_y(t) = \frac{1}{N} \sum_i y_i(t),
\end{equation}
is shown in Fig.~\ref{Fig.503}(c). After a prolonged transient phase, both $R_x$ and $R_y$ begin to oscillate with some identical low frequency, superimposed with some identical higher frequency. The resulting trajectory of the center of mass is depicted in Fig.~\ref{Fig.503}(d).

{\it Heaviside coupling function}: Short-range interaction is also used to study the phase dynamics of the swarmalators. In Ref.~\cite{hong2018active}, the author use a Heaviside function defined by 
\begin{equation}
    H(x) = \begin{cases}
                1 ~~~~ x\ge 0, \\
                0 ~~~~ x<0,
            \end{cases}
\end{equation}
to incorporate the effect of spatial dynamics on the phases. More specifically, the function $G(\textbf{x}_{j}-\textbf{x}_{i})$, which has been chosen in the form $\frac{1}{|\textbf{x}_{j}-\textbf{x}_{i}|^{\gamma}}$ so far, is now defined as
\begin{equation}
    G(\textbf{x}_{j}-\textbf{x}_{i}) = \left( 1 - \frac{|\textbf{x}_{j}-\textbf{x}_{i}|^2}{\sigma}\right) H(\sigma - |\textbf{x}_{j}-\textbf{x}_{i}|^2),
\end{equation}
where $\sigma$ controls the range of the interaction. To make the model more realistic, a white noise term $\xi_i(t)$ with zero mean and strength $T$ characterized by 
\begin{equation}
    \langle \xi_i(t) ~ \xi_j(t')\rangle = 2 T \delta_{ij} ~ \delta(t-t')
\end{equation}
is added to the phase equation. Simulations show that even when the phase coupling strength $K$ is positive, the system exhibits active phase wave and async state along with the anticipated sync state, depending on the values of the other system parameters $J$, $\sigma$, and $T$. Notably, the active phase wave state is found primarily for small values of $\sigma$, as a large $\sigma$ corresponds to $G(\textbf{x}_{j}-\textbf{x}_{i}) \approx 1$ where only the sync and async states prevail. For small noise strength $T$, the sync state is more prevalent, and async state occurs for larger $T$ values. A finite-size scaling analysis of the order parameter $R$ reveals that the system shows the mean-field transition nature similar to the mean-field XY model for large values of $\sigma$.

%\subsubsection{Coupling mechanisms}

\subsubsection{Swarmalators with external forcing}

Imagine a group of fireflies in a forest flashing in unison. If a flashing LED is placed at the center of the group, it influences the flashing rhythms of the individuals, leading them to synchronize with the LED. Similarly, external perturbations can introduce new dynamics in a swarmalator system. To investigate this, Lizarraga et al.~\cite{lizarraga2020synchronization} studied the effect of a periodic external force that directly acts on the swarmalators' phases. The stimulus is positioned at the center of the swarmalators' initial configuration. Under this forcing, the phase dynamics are governed by
\begin{equation}
	\label{eq.24}
	\dot{\theta}_{i} = \frac{K}{N}\sum_{\substack{j = 1\\j \neq i}}^{N} \frac{\sin(\theta_{j}-\theta_{i})}{|\textbf{x}_{j}-\textbf{x}_{i}|} + F \frac{\cos(\Omega t - \theta_{i})}{|\textbf{x}_{0}-\textbf{x}_{i}|}, 
\end{equation}
where $F$, $\Omega$, and $\textbf{x}_0$ denote the amplitude, frequency, and position of the external stimulus, respectively. The spatial dynamics remain the same as in Eq.~\eqref{eq.2.15}.

\begin{figure}[hpt]
	\centerline{
		\includegraphics[width= 0.8\columnwidth]{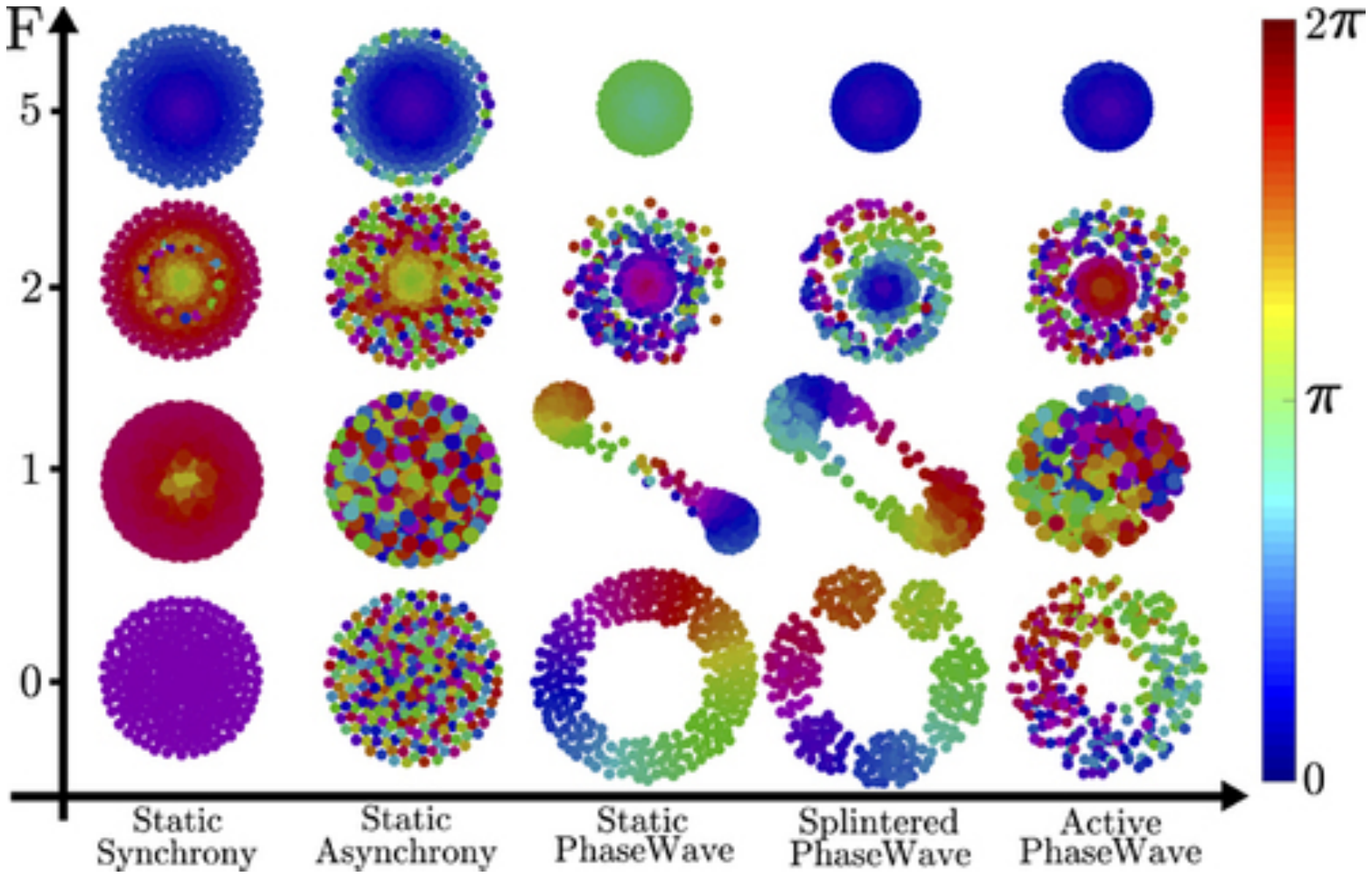}}
	\caption{State transitions using Eq.~\eqref{eq.24} as a function of the force amplitude $F$  at $\Omega = 3\pi/2$ are shown for each of the five non-forced states. The following parameter values are used: $J = 0.1$, $K = 1$ for the static synchrony state; $J = 0.1$, $K = -1$ for the static asynchrony state; $J = 1$, $K = 0$ for the static phase wave; $J = 1$, $K = -0.1$ for the splintered phase wave; and $J = 1$, $K = -0.75$ for the active phase wave. \\ Source: Reprinted figure with permission from Ref.~\cite{lizarraga2020synchronization}.}
	\label{Fig.3}
\end{figure}

As the amplitude $F$ increases, swarmalators near the stimulus begin to synchronize with the external frequency $\Omega$. All five states from the unforced model -- static sync, static async, static phase wave, splintered phase wave, and active phase wave -- undergo transitions from partial to full synchronization with increasing $F$. Complete phase synchronization occurs beyond a critical threshold of $F$, which is independent of $\Omega$. In the static sync state, swarmalators near the stimulus synchronize first, forming a small cluster around it. This cluster expands with increasing $F$ until it encompasses the entire system (see first column of Fig.~\ref{Fig.3}). A similar pattern emerges in the static async state, with the only difference being that the initial phases are desynchronized in the absence of forcing. The static phase wave and splintered phase wave states exhibit more complex dynamics. For $F = 0.5$, swarmalators near the stimulus rotate clockwise around it, while those farther away move counterclockwise. As $F$ increases to $1$, two clusters form and rotate slowly around the source (see third and fourth columns of Fig.~\ref{Fig.3}). At higher forcing strengths, the clusters merge and synchronize their phases, forming a single central cluster whose size grows with $F$. However, the fully synchronized region is smaller in this case compared to the static sync and static async states due to the higher value of $J$ in the phase wave states. In the active phase wave state, the transition under forcing resembles that of the splintered phase wave, but without splitting into distinct groups. These various phase transitions under periodic forcing are illustrated in Fig.~\ref{Fig.3}. This model has also been studied with phase lag in Ref.~\cite{sharma2026influence}.

\textcolor{black}{An alternative way is to introduce an external position-level forcing by adding a linear “pseudo-force” that attracts each swarmalator toward a prescribed point, which (despite not being a physical force in first-order dynamics) provides a simple control knob for spatial confinement~\cite{hughes2025crystal}. As the pseudo-force strength increases, the system undergoes transitions from splintered and active phase-wave states to a static antisynchronized state, with a concurrent reduction in the number of splintered groups and a marked increase in positional ordering. To quantify this “crystallization,” the authors propose a Fourier-transform-based order parameter for the spatial density, enabling a compact diagnostic of triangular crystal order and its change across transitions.}

\subsubsection{Time delayed interactions}
\begin{figure}[hpt]
	\centerline{
		\includegraphics[width = 0.95\columnwidth]{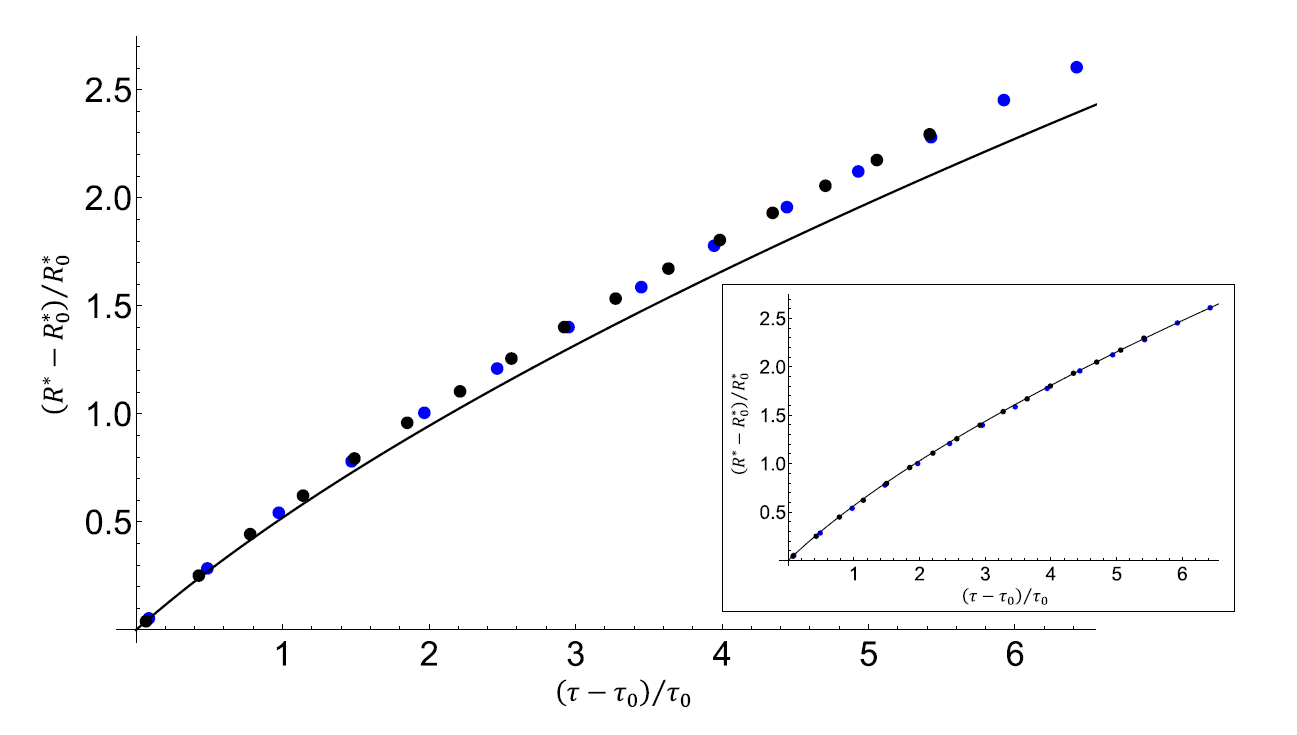}}
	\caption{Dots indicate simulation results with Eqs.~\eqref{delay1}-\eqref{delay2}, while the solid line represents the theoretical prediction. The discrepancy between the two disappears when the theoretical $y$-values are scaled by a factor of approximately 1.09, as illustrated in the inset. The parameters used are $(J = 1, K = -0.7)$. \\ Source: Reprinted figure with permission from Ref. \cite{blum2024swarmalators}.}
	\label{delay-2d-fig}
\end{figure}

Delays are a fundamental yet understudied aspect of many natural and technological systems. In the context of swarmalators, they arise in various forms--for example, through fluid-mediated interactions in microswimmers or as intrinsic features of gene expression during embryonic development~\cite{petrungaro2019information}. Robotic swarms also encounter delays due to digital communication lags, affecting coordination and information sharing. These delays can influence either the internal state, the spatial state, or both. Theoretical studies often focus on the effect of delays in the internal state within the framework of the original swarmalator model. Theoretical studies often focus on the effect of delays in the internal state within the framework of the original swarmalator model~\cite{blum2024swarmalators,kumpeerakij2025aging}. In physical terms, phase represents an internal state, such as the phase of a gene expression cycle. This type of internal information is typically communicated through chemical signals, a process significantly slower than the mechanisms used to convey spatial position. It is assumed that at time $t$, the $i$th swarmalator responds to the phase of the $j$th swarmalator as it was $\tau$ units of time earlier, i.e., at time $t - \tau$. Therefore, the instance of the model considered here is achieved just by substituting $\theta_j(t)$ by $\theta_j(t-\tau)$ in Eqs.~\eqref{eq.2.15}-\eqref{eq.2.16},
\begin{align}
	\dot{\textbf{x}}_{i} &= \frac{1}{N} \sum_{\substack{j = 1\\j \neq i}}^{N}\bigg[\frac{\textbf{x}_{j}-\textbf{x}_{i}}{|\textbf{x}_{j}-\textbf{x}_{i}|} \left(1+J \cos\left(\theta_{j}(t-\tau)-\theta_{i}(t)\right)\right) - \frac{\textbf{x}_{j}-\textbf{x}_{i}}{|\textbf{x}_{j}-\textbf{x}_{i}|^2}\bigg], \label{delay1}\\
	\dot{\theta}_{i} &= \frac{K}{N}\sum_{\substack{j = 1\\j \neq i}}^{N} \frac{\sin\left(\theta_{j}(t-\tau)-\theta_{i}(t)\right)}{|\textbf{x}_{j}-\textbf{x}_{i}|} 
	\label{delay2}.
\end{align}
The most important control parameter here is the delay time $\tau$ along with the other system parameters $J$ and $K$. The primary focus is given to the region of the $(J, K)$ parameter space that, in the absence of delay, corresponds to the active phase wave state described in Ref.~\cite{o2017oscillators}. Remember that, in this regime, swarmalators form an annulus-shaped cluster and exhibit circular motion around its center--some rotating clockwise, others counterclockwise--while their internal phases evolve continuously. It is within this region that the introduction of delay leads to the emergence of notable collective behaviors. Regardless of the value of $\tau$, swarmalators show transient breathing oscillations (when one measures the average speed of the system) at initial times that arise from rearrangement of spatial positions or from the expansion and contraction of the circular disc-like spatial structure that they form in the 2D plane. After this initial transient, the system enters into a coherent, synchronized collective motion characterized by decaying oscillations of the average speed and disc radius. A critical value of the delay, denoted by $\tau = \tau_c$, separates two distinct long-term behaviors. For $\tau > \tau_c$, the system transitions from a breathing transient to a ``quasistationary pseudocrystal" state, marked by a slow, creeping motion of the particles. In contrast, for $\tau < \tau_c$, the breathing transient evolves into a ``boiling state", characterized by convective, boiling-like particle motion near the surface. At large times, the oscillations of the system decay and the phases advance uniformly with rate $\Omega$. Thus we assume $\tilde{R}(t) = R^{\ast} + \delta \tilde{R} (t)$, and $\theta_i(t) = \Omega t + \delta \theta_i (t)$ where $\tilde{R}(t)$ is the radius of the disc at time $t$ and $R^{\ast}$ is the radius at $t \rightarrow \infty$, and $\delta \tilde{R} (t)$ and $\delta \theta_i (t)$ are decaying functions. With these assumptions, one can analytically solve the system by analyzing the density profile in the steady state and arrive at the following pair of equations~\cite{blum2024swarmalators},
\begin{align}
    \dot{\theta} &= \frac{5K}{\pi \tilde{R}(t)} \sin \left[\theta(t-\tau) - \theta(t)\right],\\
    \dot{\tilde{R}} &= \frac{5}{2\pi} \left[ -\left(1+J\cos\left[\theta(t-\tau) - \theta(t)\right]\right) + \frac{2\pi}{5 \tilde{R}}\right].
\end{align}
These can be solved in the $t \rightarrow \infty$ limit which yield
\begin{equation}
    R^{\ast} = -\frac{5K}{\pi} \frac{\sin(\Omega \tau)}{\Omega}, ~~ R^* = \frac{2\pi}{5} \frac{1}{1+J\cos(\Omega \tau)}. \label{delay-radius}
\end{equation}
Combining these two, one gets
\begin{equation}
    1 + J\cos (\Omega t) =-\frac{2\pi^2}{25} \frac{\Omega}{K \sin(\Omega \tau)},
\end{equation}
the solution of which gives the value of $\Omega$ as a function of $\tau$ and substituting it into Eq.~\eqref{delay-radius} gives the analytical values of the disc radius. It is also noted that for $\tau < \tau_0 = -\frac{2\pi^2}{25} \frac{1}{K(1+J)}$ and $\Omega=0$, one can obtain $R^{\ast}_0 = \frac{2\pi}{5(1+J)}$. The disc radius is plotted against delay in dimensionless units in Fig.~\ref{delay-2d-fig} and the result is compared with the simulation data (dots).

\subsubsection{Effect of phase lag}
A notable extension of the model involves the incorporation of a frustration parameter, or phase lag, within the phase interaction function. In the context of the Kuramoto model, the influence of such a parameter has been extensively examined under various scenarios~\cite{sakaguchi1986soluble,omel2012nonuniversal}, uncovering a wide range of dynamics including the emergence of chimera states~\cite{martens2016chimera,parastesh2021chimeras}--patterns in which synchronized and desynchronized oscillations coexist. These chimera states have also been identified in neuronal networks, where they are driven by the combined effects of different coupling mechanisms, such as electrical, chemical, and ephaptic synapses~\cite{majhi2019chimera}. To investigate the effect of phase lag in the swarmalator dynamics, the following model is discussed~\cite{SENTHAMIZHAN2025116164},
\begin{align}
	\dot{\textbf{x}}_{i} &= \frac{1}{N} \sum_{\substack{j = 1\\j \neq i}}^{N}\bigg[\frac{\textbf{x}_{j}-\textbf{x}_{i}}{|\textbf{x}_{j}-\textbf{x}_{i}|} (1+J \cos(\theta_{j}-\theta_{i}-\alpha_x)) - \frac{\textbf{x}_{j}-\textbf{x}_{i}}{|\textbf{x}_{j}-\textbf{x}_{i}|^2}\bigg], \label{phaselag-1}\\
	\dot{\theta}_{i} &= \frac{K}{N}\sum_{\substack{j = 1\\j \neq i}}^{N} \frac{\sin(\theta_{j}-\theta_{i}-\alpha_{\theta})}{|\textbf{x}_{j}-\textbf{x}_{i}|} 
	\label{phaselag-2},
\end{align}
where $\alpha_x$ and $\alpha_{\theta}$, both less than $\pi/2$, represent the frustration parameters in the phase interactions. For different combinations of $J$ and $K$ with $\alpha_x = \alpha_{\theta} < \pi/2$, the system displays a variety of distinct dynamical behaviors. These include static configurations such as the static asynchronous state and static chimera, along with active states like the active phase wave, active synchronized state, and active chimera. The model also gives rise to globally dynamic patterns, including global translational motion and synchronized global translational motion. It is also notable that independently varying $\alpha_x$ and $\alpha_{\theta}$ in the vicinity of $\pi/2$ does not result in the emergence of additional states within the $J-K$ parameter space.

\subsubsection{Higher-order harmonics}
\begin{figure}[hpt]
	\centerline{
		\includegraphics[width = 0.9\columnwidth]{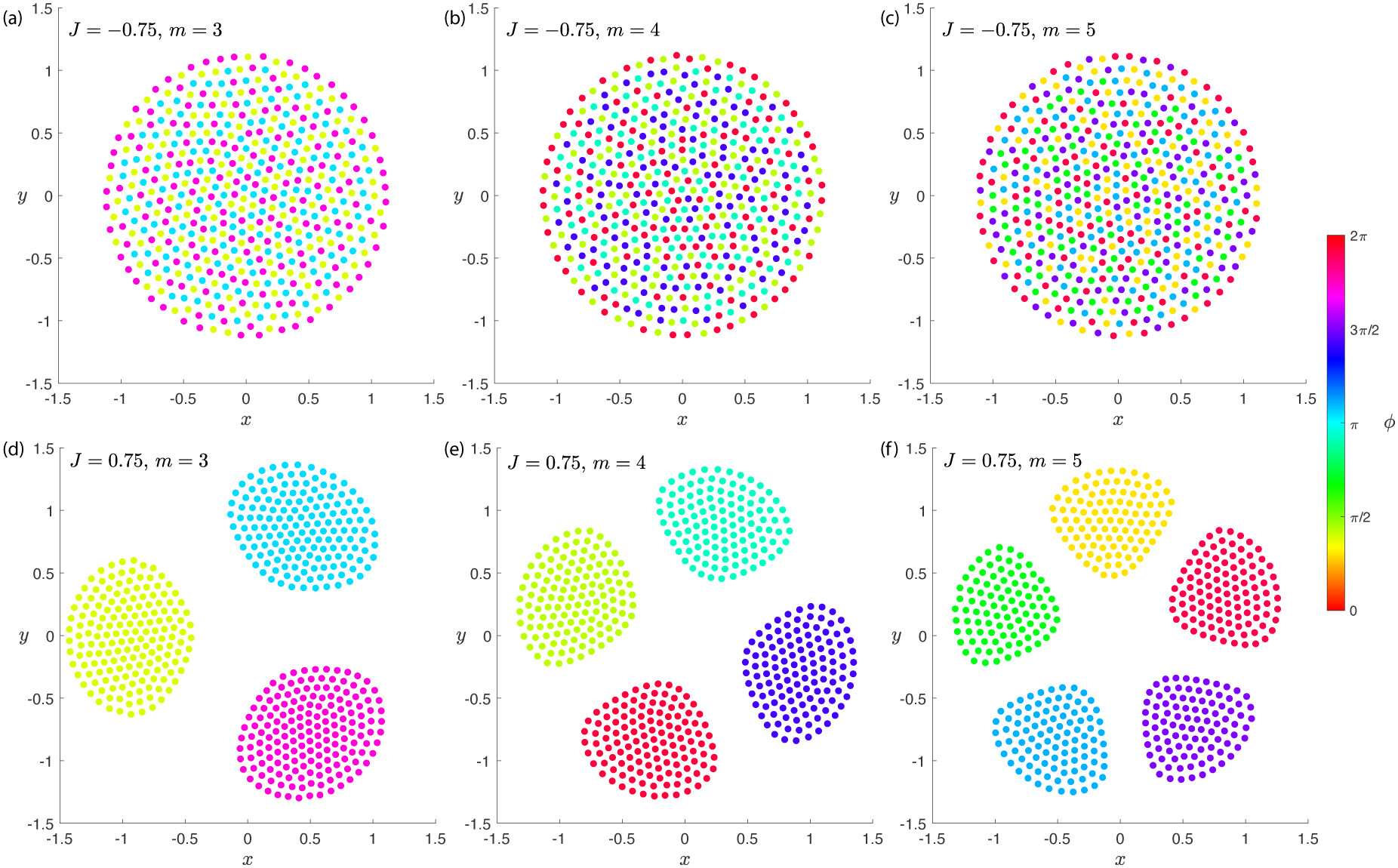}}
	\caption{Static configurations of the higher harmonic swarmalator system governed by Eqs.~\eqref{eq.2.15} and \eqref{harmonic-fn} with $K = 1$ and $N = 500$. Top row is for $J = -0.75$ with (a) $m = 3$, (b) $m = 4$, and (c) $m = 5$. Bottom row is for $J = 0.75$ with (d) $m = 3$, (e) $m = 4$, and (f) $m = 5$. \\ Source: Reprinted figure with permission from Ref. \cite{smith2024swarmalators}.}
	\label{harmonic}
\end{figure}

As an approach to exploring more general coupling mechanisms, the original swarmalator model (given by Eqs.~\eqref{eq.2.15}-\eqref{eq.2.16}) is extended to incorporate higher harmonic terms in the phase dynamics. Given that pairwise interactions are typically antisymmetric and that phase variables are $2\pi$-periodic, the coupling functions can be represented as Fourier sine series. Attention is given to truncated versions of these series, retaining only the most dominant harmonics. Specifically, the analysis focuses first on scenarios where the first and second harmonics are equally dominant, followed by cases where a single higher harmonic dominates the interaction. The phase dynamics with $m$-th order harmonic is given by the following equation
\begin{equation}
    \dot{\theta}_{i} = \frac{K}{N}\sum_{\substack{j = 1\\j \neq i}}^{N} \frac{\sin\big(m(\theta_{j}-\theta_{i})\big)}{|\textbf{x}_{j}-\textbf{x}_{i}|}. \label{harmonic-fn}
\end{equation}
When the phase coupling strength $K$ is positive, for $J < 0$, swarmalators organize into a single spatial cluster containing $m$ distinct phases. In contrast, when $J > 0$, they form $m$ separate spatial clusters, each associated with a unique phase. Examples of these occurrences for both the cases with $m=3,4,5$ are shown in Fig.~\ref{harmonic}. For $m=2$ (not shown here in Fig.~\ref{harmonic}), two disjoint spatial clusters are formed that have a phase difference $\pi$ between them. The distance between the center of the positions of these clusters is also calculated from the governing dynamics as $1/(1-J)$~\cite{smith2024swarmalators}. For $J<0$, only a single disc is formed where swarmalators with $\pi$ phase difference arrange themselves by mixing their positions inside the disc. The amount of mixing increases when the magnitude of $J$ decreases (i.e., becomes more negative). All the collective states that were observed in the original model are also realized here for certain values of $J$ and $K$. In Ref.~\cite{senthamizhan2024data}, a machine learning approach based on convolutional neural networks has been adopted to analyze this model, specifically utilizing the U-Net architecture, to identify distinct dynamical states and to determine the critical boundaries separating them. One can also consider a combination of harmonics in Eq.~\eqref{harmonic-fn} rather than the single $m$-th order harmonics. One particular example of such a phase dynamics is the following
\begin{equation}
    \dot{\theta}_{i} = \frac{1}{N}\sum_{\substack{j = 1\\j \neq i}}^{N} \frac{\Big(K_1\sin(\theta_{j}-\theta_{i})+K_2\sin\big(2(\theta_{j}-\theta_{i})\big)\Big)}{|\textbf{x}_{j}-\textbf{x}_{i}|}. \label{harmonic-fn-1}
\end{equation}
Smith~\cite{smith2024swarmalators} studied this model and reported the formation of a single cluster or two disjoint clusters and analyzed their stability in the $K_1$-$K_2$ plane. A special case is also demonstrated where two clusters situated at some distance apart are joined by a few trapped swarmalators that waver between these clusters back and forth, either periodically or chaotically. These trapped swarmalators are termed ``vacillators". We refer the readers to Ref.~\cite{smith2024swarmalators} where a reduced vacillator system is analytically studied.

\subsubsection{Distributed velocities and frequencies}
\begin{figure}[hpt]
	\centerline{
		\includegraphics[width = 0.9\columnwidth]{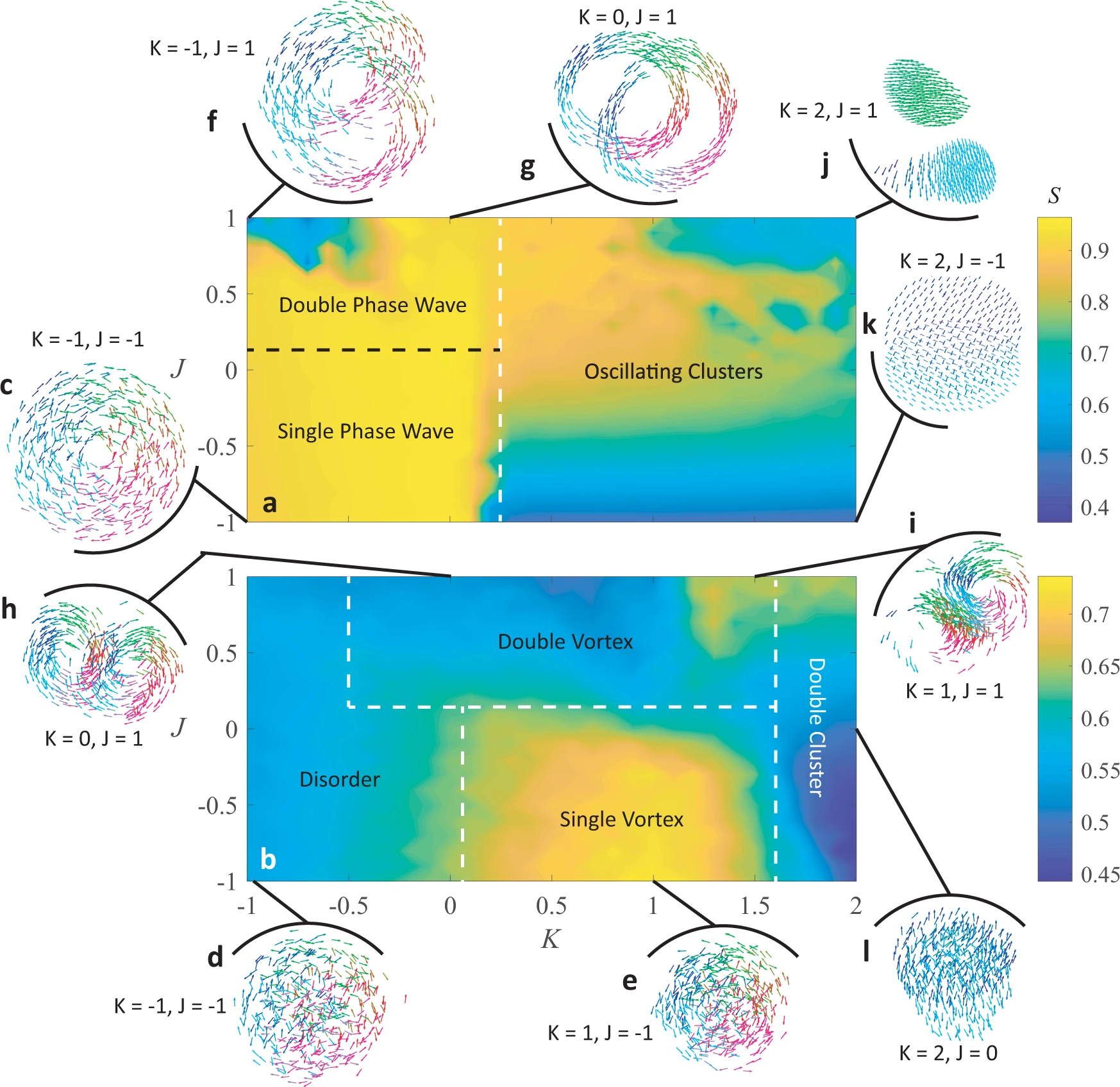}}
	\caption{Frequency-coupled chiral swarmalators given by Eqs.~\eqref{nonid-1}-\eqref{nonid-2}. Heat maps of $S$ across $K-J$ parameter space are shown for natural frequency distributions (a) exactly half of the swarmalators have $\omega_i=1$ and the other half have $\omega_i=-1$, and (b) exactly half of the swarmalators have their natural frequency randomly selected from one uniform distribution $\omega_i \in U[1,3]$ and the second half have their natural frequency selected from another uniform distribution $\omega_i \in U[-3,-1]$. (c) Phase wave. (d, e) Disordered vortices. (f, g) Double phase waves. (h,i) Double vortex. (j) Dense revolving clusters. (k, l) Revolving clusters. \\ Source: Reprinted figure with permission from Ref. \cite{ceron2023diverse}.}
	\label{nonid-natcom}
\end{figure}

The foundational models of swarmalators often assume idealized conditions such as identical natural frequencies, uniform velocities, non-chiral motion, and global coupling. While these assumptions offer clarity and analytical tractability, they limit the applicability of the model to real-world systems. More recent developments have moved beyond these constraints by incorporating heterogeneity, chirality, and locality into the framework~\cite{ceron2023diverse,ansarinasab2024thespatial,kongni2024expected}. In particular, allowing swarmalators to possess non-identical natural frequencies reflects the diversity commonly observed in biological and artificial agents, where intrinsic variability and noise are unavoidable. Systems such as microrobotic swarms often display behavioral differences due to imperfections in fabrication or calibration. The following discussion focuses on the impact of nonidentical properties--specifically variability in intrinsic frequencies and velocities--on the collective dynamics, offering insights into how such heterogeneity influences synchronization, pattern formation, and phase coherence within swarmalator populations. The following model is considered
\begin{align}
	\dot{\textbf{x}}_{i} &= \textbf{v}_{i} + \frac{1}{N} \sum_{\substack{j = 1\\j \neq i}}^{N}\bigg[\frac{\textbf{x}_{j}-\textbf{x}_{i}}{|\textbf{x}_{j}-\textbf{x}_{i}|} (1+J \cos(\theta_{j}-\theta_{i}- Q_x)) - \frac{\textbf{x}_{j}-\textbf{x}_{i}}{|\textbf{x}_{j}-\textbf{x}_{i}|^2}\bigg], \label{nonid-1}\\
	\dot{\theta}_{i} &= \omega_i + \frac{K}{N}\sum_{\substack{j = 1\\j \neq i}}^{N} \frac{\sin(\theta_{j}-\theta_{i}-Q_{\theta})}{|\textbf{x}_{j}-\textbf{x}_{i}|} 
	\label{nonid-2},
\end{align}
where $\omega_i$'s are chosen independently and the velocities are chosen as $\textbf{v}_{i} = c_i \bf{n}_i$ that depend on agent’s natural frequency ($\omega_i$) and its radius of revolution in real space ($R_i$) by choosing $c_i = \omega_i R_i$. $\bf{n}_i$ is a vector pointing in the direction orthogonal to the angle denoted by $\theta_i$, within the global coordinate system. The model can be extended to scenarios in which an agent’s internal state (phase) is directly associated with its orientation along a circular orbit. When $|\textbf{v}_i| = 0$, the agent remains stationary, and its phase serves solely as an internal state. In contrast, for $|\textbf{v}_i| \neq 0$, the agent undergoes periodic circular motion in physical space with a defined radius of revolution, and its phase corresponds to the angular position along its orbit within a global coordinate frame. This formulation of $\textbf{v}_i$ allows for modeling swarmalator systems in which each agent's phase either represents a purely internal state ($c_i = 0$), or is linked to its spatial orientation ($c_i = -1$ or $c_i = 1$). Additional phase offset terms, $Q_x$ and $Q_{\theta}$, introduced in Eqs.~\eqref{nonid-1}-\eqref{nonid-2}, facilitate a form of `frequency coupling'. These terms are designed to enhance the interaction between agents possessing natural frequencies of opposite sign, thus enabling more realistic emergent dynamics that mirror those observed in hydrodynamically and mechanically coupled systems~\cite{riedel2005self}. Depending on the values of $c_i$, the phase offset terms $Q_x$ and $Q_{\theta}$, three distinct scenarios are identified as,
\begin{itemize}
    \item {\it Non-chiral swarmalators}: No frequency coupling, i.e., $c_i=0$ for all agents, $Q_x = 0$, and $Q_{\theta}=0$.
    \item {\it Revolving swarmalators}: When $c_i \ne 0$ for all agents and $Q_x=Q_{\theta}=0$;
    \item {\it Frequency-coupled chiral swarmalators}: In this case, $c_i \ne 0$, $Q_x \ne 0$, and $Q_{\theta} \ne 0$. $Q_x$ and $Q_{\theta}$ are defined as, $Q_x = \frac{\pi}{2} \left| \frac{\omega_j}{|\omega_j|} - \frac{\omega_i}{|\omega_i|}\right|$, $Q_{\theta} = \frac{\pi}{4} \left| \frac{\omega_j}{|\omega_j|} - \frac{\omega_i}{|\omega_i|}\right|$, therefore, the sign difference between the natural frequencies of two agents plays an important role in how they will affect each other’s motion and phase coupling.
\end{itemize}
A plethora of new collective states are found in each of these three cases. The behavior of the order parameter $S$ in the third case for two different types of frequency distribution are shown in Fig.~\ref{nonid-natcom}. 

In the configuration discussed above, the velocities of the swarmalators are either zero or coupled to their natural frequencies. Ansarinasab et al.~\cite{ansarinasab2024thespatial} considered independent velocities and frequencies of the swarmalators where they are randomly chosen from various kinds of distributions such as the uniform, Gaussian, Lorentzian. Their study reveals that the uniform distribution displays a phase transition curve with a notably more abrupt, explosive nature. Varying the phase coupling strength unfolds the emergence of three distinct spatial patterns: drifting periodic, irregular, and static configurations.

\subsubsection{Coupling mechanisms and different network structures}
Swarmalator dynamics can be significantly influenced by the nature of coupling and the structure of the underlying interaction network. While early models primarily assumed all-to-all (global) coupling, real-world systems often feature more complex interaction rules, including local, distance-dependent, or heterogeneous network topologies. Additionally, coupling mechanisms can vary from purely attractive or repulsive to mixed or adaptive forms, each introducing new behaviors and collective states. This section explores how different coupling schemes and network architectures shape the collective dynamics of swarmalators.

{\it Coupling disorder}: The preceding discussions have focused on the long-term behavior of swarmalators under purely attractive or repulsive phase coupling. However, in many real-world systems, coupling interactions are more nuanced. For instance, such complexity is evident in neural network models \cite{hopfield1982neural} and in the vocal interactions of Japanese tree frogs \cite{aihara2008mathematical}. Motivated by these observations, recent investigations have explored swarmalator dynamics under mixed coupling scenarios. In Ref.~\cite{hong2021coupling}, the phase coupling strength $K_{ij}$ between the $i$th and $j$th swarmalators is randomly assigned from a bimodal distribution:
\begin{equation}
	h(K_{ij}) = p \delta(K_{ij}-K_a) + (1-p) \delta(K_{ij}- K_r), \label{double-delta}
\end{equation}
where $K_a$ and $K_r$ denote attractive and repulsive coupling strengths, respectively, and $p$ is the probability of an attractive interaction. The governing dynamics follow Eqs.~\eqref{eq.2.15} and \eqref{eq.2.16}, with the modification that $K_{ij}$ values are randomly drawn while maintaining the symmetry $K_{ij} = K_{ji}$. A control parameter can be defined as $Q = -K_r/K_a > 0$. Using an annealed approximation of the quenched disorder, it is shown that the system exhibits a phase transition from a desynchronized to a synchronized state at the critical probability $p_c = Q/(1+Q)$. 

Within the incoherent regime $0 < p < p_c$, nonstationary behaviors such as splintered phase waves and active phase waves emerge for appropriate values of $J$, even when the majority of interactions are attractive ($p > 0$). To detect these dynamic states, the mean velocity $\bar{v}$ is calculated as
\begin{equation}
    \bar{v} = \frac{1}{N} \sum_{i=1}^{N} \sqrt{\dot{x}_i^2 + \dot{y}_i^2}.
\end{equation}
The order parameters $S$, $R$, $U$, and $\bar{v}$ are used to characterize different states and monitor transitions. The introduction of mixed randomness gives rise to a variety of deformed spatial-phase patterns, resembling chimera states observed in oscillator populations with identical natural frequencies \cite{kuramoto2002coexistence, abrams2004chimera}. These patterns can be interpreted through the lens of annealed averages of the mixed couplings.

Interactions between agents in a group are sometimes non-reciprocal, i.e., the coupling strengths are asymmetric; $J_{ij} \ne J_{ji}, K_{ij} \ne K_{ji}$. Non-reciprocal interactions can be considered into the system through a leader-follower approach where a few individuals (leaders) exert stronger coupling on others, or by an autonomous decision-making approach where some units are primarily influenced by their own dynamics without much knowledge of others. These types of non-reciprocal coupling are seen to produce several complex states, including the chimera and the antiphase state~\cite{yu2025collective}. 
\textcolor{black}{In place of the non-reciprocal pairwise coupling strength, each swarmalator can also be assigned an intrinsic coupling strength $K_i$, drawn from the same distribution as in Eq.~\eqref{double-delta}~\cite{yu2025swarmalator}.}

{\it Finite-cutoff coupling range}: The previous sections examined swarmalator dynamics under global spatial and phase interactions. However, in most real-world multi-agent systems, interactions are typically local, limited to neighboring agents. This observation has motivated investigations into swarmalator models where interactions—either spatial or phase-based—are restricted to local neighborhoods~\cite{lee2021collective,schilcher2021swarmalators}. As a step in this direction, Lee et al.\ \cite{lee2021collective} analyzed the steady-state configurations of swarmalators subjected to a finite interaction range in the spatial dynamics. In this framework, swarmalators interact only with others located within a finite distance $r$. The spatial evolution of the $i$th swarmalator is then governed by
\begin{equation} 
\label{eq.25} 
\dot{\textbf{x}}_{i} = \frac{1}{N_i(r)} \sum_{j \in \Lambda_i(r)}\bigg[ \frac{\textbf{x}_{j}-\textbf{x}_{i}}{|\textbf{x}_j-\textbf{x}_{i}|^{\alpha}} (1+J \cos(\theta_{j}-\theta_{i})) - \frac{\textbf{x}_{j}-\textbf{x}_{i}}{|\textbf{x}_j-\textbf{x}_{i}|^2}\bigg],
\end{equation}
where $\Lambda_i(r)$ denotes the set of swarmalators located within a radius $r$ of the $i$th swarmalator (excluding itself), and $N_i(r)$ is the number of such neighbors. The phase dynamics remains unchanged from Eq.\ \eqref{eq.2.16}. A linear attraction kernel with $\alpha = 0$ is employed for analytical simplicity, although similar results persist for $\alpha = 1$.

\begin{figure}[hpt]
	\centerline{
		\includegraphics[width = 0.75 \linewidth]{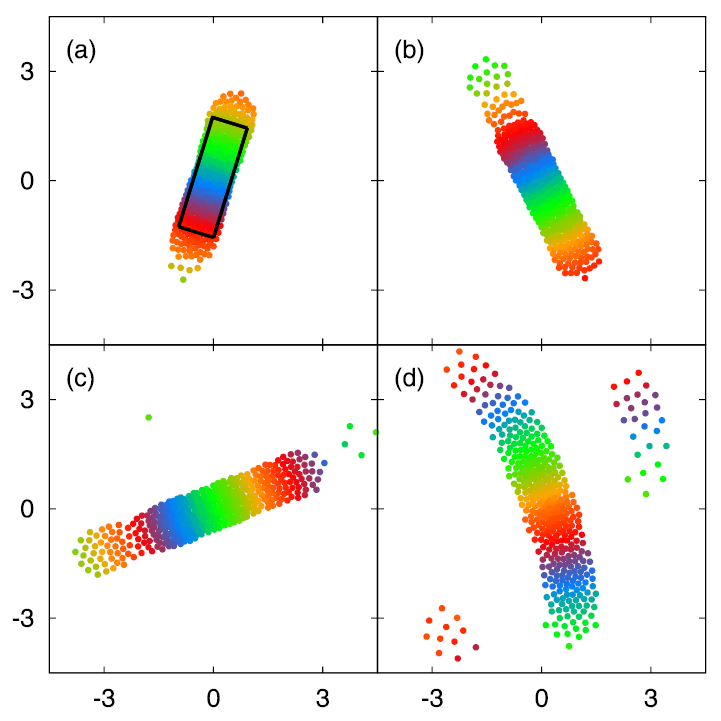}}
	\caption{Various bar-like patterns observed for $1.2 \leq r \leq 1.8$ in a system of $N = 400$ swarmalators with $K = 0$ and $J = 1$. Figures (a)--(d) correspond to $r = 1.8$, $1.6$, $1.4$, and $1.2$, respectively. In figure (a), the black rectangle has a length of $\pi$ and a width of $1$. \\ Source: Reprinted figure with permission from Ref. \cite{lee2021collective}.}
	\label{cutoff}
\end{figure}

In the limit $r \rightarrow \infty$, the system recovers previously observed steady states under global coupling—namely, the static sync, static async, and static phase wave states. When the interaction range $r$ is finite, novel behaviors emerge. For values of $r$ exceeding the steady-state diameter $D_{\infty}$ (measured in the $r \rightarrow \infty$ limit), the system exhibits multiple identical replicas of the steady states. The number of such replicas is sensitive to initial conditions. This phenomenon can be explained as follows: if a group of swarmalators occupies a circular region of radius $r$ and remains separated from others by at least distance $r$, it effectively evolves as an independent unit. Each group mirrors the dynamics of the globally coupled system, resulting in a cluster of diameter $D_{\infty}$ and a minimum separation of $D_{\infty}$ between groups. For $r < D_{\infty}$, distorted versions of static sync and static async states appear, characterized by non-uniform swarmalator density. Within the intermediate range $1 < r < 1.8$, bar-shaped spatial structures emerge from the static phase wave state as shown in Fig.~\ref{cutoff}.

{\it Time varying attractive-repulsive coupling}: Studies on swarmalators presented thus far have predominantly assumed static network structures, where the coupling among the units remains constant over time. An alternative approach involves incorporating time-varying interactions into the system \cite{ghosh2022synchronized}. In this context, Sar et al.\ \cite{sar2022swarmalators} proposed a model of swarmalators featuring time-dependent competitive phase interactions and investigated the resulting long-term behaviors. Each swarmalator moves within a two-dimensional plane, interacting with neighbors inside a uniform circular zone referred to as the {\it vision radius} $r$. Phase coupling between swarmalators is attractive if one lies within the other's vision radius; otherwise, the interaction becomes repulsive. As the agents move in space, the coupling matrix evolves continuously based on their spatial configuration.

The time-dependent competitive phase dynamics is governed by the following equation:
\begin{equation}	
	\label{eq.27}
	\dot{\theta}_{i} =\frac{K_a}{N_{i}(r)}\sum_{\substack{j = 1\\j \neq i}}^{N} \text{A}_{ij}\frac{\sin(\theta_{j}-\theta_{i})}{|\textbf{x}_{j}-\textbf{x}_{i}|} + \\ \frac{K_r}{N-1-N_{i}(r)}\sum_{\substack{j = 1\\j \neq i}}^{N} \text{B}_{ij}\frac{\sin(\theta_{j}-\theta_{i})}{|\textbf{x}_{j}-\textbf{x}_{i}|},
\end{equation}
where $K_a$ and $K_r$ denote the attractive and repulsive coupling strengths, respectively. The matrices $\text{A}$ and $\text{B}$ represent the attractive and repulsive coupling schemes, defined as
\begin{equation*}
	\text{A}_{ij} =
	\begin{cases}
		1 & \text{if $j \in \Lambda_{i}(r)$}\\
		0 & \text{otherwise}
	\end{cases}	;\hspace{5pt} \text{B}_{ij} =
	\begin{cases}
		1 & \text{if $j \not\in \Lambda_{i}(r) \cup \{i\}$}\\
		0 & \text{otherwise}
	\end{cases}
\end{equation*}
Here, $N_i(r)$ and $\Lambda_i(r)$ are as defined in Eq.\ \eqref{eq.25}. The spatial dynamics remains the same as described in Eq.\ \eqref{eq.2.15}.

\begin{figure}[hpt]
	\centerline{
		\includegraphics[width = 0.6 \linewidth]{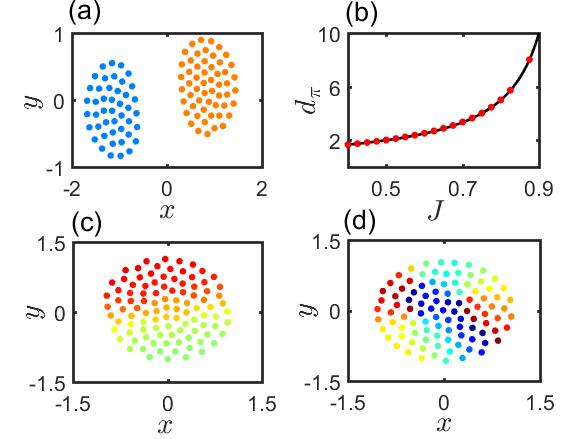}}
	\caption{(a) Snapshot of static $\pi$ state for $(J,r,K_a,K_r)= (0.5,0.8,0.5,-0.5)$. (b) Distance between the center of positions of the clusters in static $\pi$ state. Red dots are numerically calculated values. Black curve is theoretical prediction given by $d_{\pi} = 1/(1-J)$. Snapshots of the mixed phase wave state for (c) $(J,r,K_a,K_r)= (0.1,1.38,0.5,-0.5)$ and (d) $(J,r,K_a,K_r)= (0.1,0.36,0.1,-1.0)$ at $t=1000$ time unit. Simulations are done with $N=100$ swarmalators. \\ Source: Reprinted figure with permission from Ref. \cite{sar2022dynamics}.}
	\label{att-rep}
\end{figure}

In the limiting cases of vision radius $r$, the system replicates all steady states found in the globally coupled version governed by Eqs.\ \eqref{eq.2.15} and \eqref{eq.2.16}. Numerical simulations reveal that for suitable choices of $K_a$, $K_r$, $J$, and $r$, the swarmalators organize into a stationary configuration consisting of two distinct spatial clusters. Each cluster exhibits internal phase synchrony, with a phase difference of $\pi$ between them--this configuration is referred to as the {\it static $\pi$} state (see Fig.\ \ref{att-rep}(a)). Analytical treatment shows that the cluster centers maintain a constant spatial separation of $d_{\pi} =1/(1-J)$, illustrated in Fig.\ \ref{att-rep}(b). For lower values of $J$, swarmalators group in space with others of similar phases, yet do not form clearly separated clusters. In this state, full phase synchrony is absent, although spatial proximity correlates with small phase differences that increase with distance. This behavior defines an active dynamical state known as the {\it mixed phase wave}. Snapshots of this state under various parameter values are shown in Figs.\ \ref{att-rep}(c) and \ref{att-rep}(d).

\begin{figure}[hpt]
	\centerline{
		\includegraphics[width = 0.75 \linewidth]{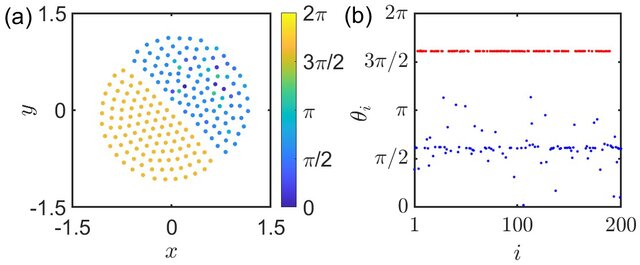}}
	\caption{Chimera State found in the system given by Eqs.~\eqref{eqcom1}-\eqref{eqcom2}. In this configuration, one community exhibits complete phase synchronization, while the other remains asynchronous. The simulation is performed with $K_3 = -0.4$, and $(\mathrm{dt}, t, N) = (0.01, 5000, 200)$. (a) A snapshot at $t = 5000$ time units illustrates the emergence of the chimera state. (b) The phases of the swarmalators are plotted against their indices, with red and blue dots representing the first and second communities, respectively. Phase synchronization is observed in the first community (red), while desynchronization characterizes the second community (blue). \\ Source: Reprinted figure with permission from Ref. \cite{ghosh2023antiphase}.}
	\label{anti-chim}
\end{figure}

{\it Different network structures}: In the discussions so far, swarmalator interactions have been primarily modeled using either global or local coupling frameworks. However, many natural and engineered systems are better represented by complex network topologies, where connections among agents are neither entirely global nor strictly local, but are shaped by underlying structural rules. Introducing network-based interactions into swarmalator models allows the exploration of collective dynamics under more realistic constraints, revealing how the interplay between spatial organization and temporal synchronization is influenced by network architecture. One such framework is the community of swarmalators where they are distributed in multiple groups and there are inter-community as well as intra-community interactions among them. Let $C_i$ denote the set of indices of swarmalators belonging to the $i$th group. Then $\sum_{i=1}^{p}|C_i| = N$, the total number of swarmalators in the population and $\cup_{i=1}^{p} C_i = \{1,2,\cdots,N\}$ where $p$ denotes the number of communities of groups. Assuming without loss of generality that the $i$th swarmalator belongs to the $n$th group, one can write the governing equations as,
\begin{align}
    \Dot{\mathbf{x}}_i &= \sum_{m = 1}^{p}\dfrac{1}{|C_m|}\sum_{j \in C_m\setminus \{i \}}\Bigg[\dfrac{\mathbf{x}_j-\mathbf{x}_i}{|\mathbf{x}_j-\mathbf{x}_i|}\big(1 + J_{n,m} \cos(\theta_j - \theta_i)\big)- \dfrac{\mathbf{x}_j-\mathbf{x}_i}{|\mathbf{x}_j-\mathbf{x}_i|^2}\Bigg],
    \label{eqcom1}\\
    \Dot{\theta}_i &= \sum_{m = 1}^{p}\dfrac{K_{n,m}}{|C_m|}\sum_{j \in C_m\setminus \{i \}} \dfrac{\sin(\theta_j-\theta_i)}{|\mathbf{x}_j-\mathbf{x}_i|}.
    \label{eqcom2}
\end{align}
One retrieves Eqs.~\eqref{eq.2.15}-\eqref{eq.2.16} with $p=1$. The two-community interaction with $p=2$ is studied by Ghosh et al.~\cite{ghosh2023antiphase}. A particular case has been considered where the spatial coupling strengths are set at $J_{1,1}=J_{2,2}=J_{1,2}=J_{2,1} = 0.1$, and the phase couplings are set as $K_{1,1}=-0.1, K_{2,2}=-0.2$. The inter-community phase coupling strengths $K_{1,2}$ and $K_{2,1}$ are chosen symmetrically such that $K_{1,2} = K_{2,1} \equiv K_3$, say, and it has been varied to investigate the emerging states of the system. Six different collective states are reported of which the chimera state and the antiphase sync state have to do with the community structure of the underlying network~\cite{ghosh2023antiphase}. The chimera state is further identified as breathing chimera where the phase coherence order parameter corresponding to the desynchronized group exhibit irregular oscillations over time. The occurrence of the chimera state is presented in Fig.~\ref{anti-chim}. Apart from the community network structure discussed here, Kongni et al.~\cite{kongni2023phase} studied a population of swarmalators with multiplex network structure where in each layer swarmalators move in the 2D plane. The state variables $\theta$ of a particular layer are attractively or repulsively coupled to the phases of the other layer depending on a predefined vision range like the one defined earlier in Eq.~\eqref{eq.27}.

\subsubsection{Non-Kuramoto internal dynamics}
\label{sec.3.1.8}
\begin{figure}[hpt]
	\centerline{
		\includegraphics[width = 0.75 \linewidth]{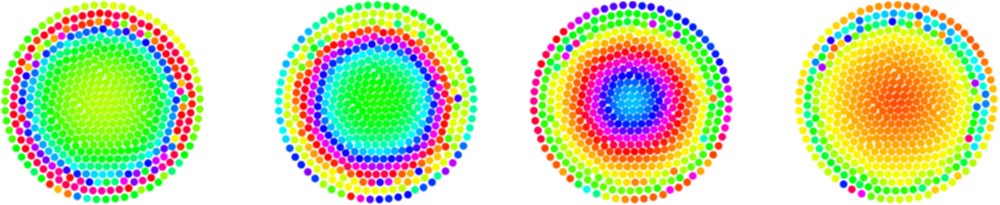}}
	\caption{Snapshots at different time of the radial wave for parameters $J = 0$ and $K = 0.55$ illustrating the traveling wave along the radial direction. \\ Source: Reprinted figure with permission from Ref. \cite{yadav2025collective}.}
	\label{pulse-2d}
\end{figure}

Most of the studies on swarmalators considered so far rely on the Kuramoto-type phase evolution to describe the internal dynamics of the agents. While the Kuramoto model provides a foundational framework for understanding synchronization phenomena, it assumes simple sinusoidal coupling and uniform oscillatory behavior. However, real-world systems often exhibit richer and more complex internal dynamics that cannot be captured by the Kuramoto model alone. To explore a broader spectrum of behaviors, recent investigations have considered alternative formulations for the internal dynamics, incorporating features such as pulse-like interactions by using the Winfree model~\cite{winfree1967biological}. The phase dynamics involving the pulse-like interactions among the agents is given by the following equation
\begin{equation}
    \Dot{\theta}_i = \omega_i+ R(\theta_i) \frac{K}{N}  \sum\limits_{i\neq j}^N\frac{P(\theta_j)  }{|\mathbf{x}_j-\mathbf{x}_i|},
\label{eq-pulse}
\end{equation}
where $P(\theta)$ and $R(\theta)$ are the pulse and response functions, respectively, which are of the form
\begin{equation}
    P(\theta) = a_n (1+\cos \theta)^n; \; \; \; \; R(\theta) = \left[q(1-\cos\theta) - \sin \theta\right]. \label{pulse}
\end{equation}
A diverse array of collective states is observed ranging from the bump state where a static synchronized core of swarmalators are surrounded by desynchronized ones with phase drift, to the radial wave state where the swarmalators are synchronized along a fixed radius but change their phases with varying radius~\cite{yadav2025collective}. See Fig.~\ref{pulse-2d} for time evolution of the radial wave state. Also see Table I of Ref.~\cite{yadav2025collective} for comparison of the collective states found using the Kuramoto phase dynamics and the Winfree dynamics.

{\it Amplitude oscillator based model}: Most studies on swarmalators focus on internal dynamics driven by phase oscillators, typically modeled using the Kuramoto or the Winfree frameworks. However, internal dynamics can also be governed by amplitude-based chaotic oscillators, introducing richer behavior. Synchronization of such chaotic oscillators, despite their sensitive dependence on initial conditions, has been a subject of significant interest. One such model has been proposed by Ghosh et al.~\cite{ghosh2024amplitude} where they consider the internal dynamics of the swarmalators as the chaotic R{\"o}ssler oscillator,
\begin{align}
    \Dot{\mathbf{x}}_i &= \dfrac{1}{N}\sum_{j\neq i}^{N}\big(\mathbf{x}_j-\mathbf{x}_i\big)  \bigg[\big(A+J_1 e^{-E_{ij}}\big)-\dfrac{\big(B-J_2 e^{-E_{ij}}\big)}{|\mathbf{x}_j-\mathbf{x}_i|^{2}}\bigg],
    \label{eq-amp1}\\
    \Dot{\mathbf{\xi}}_i &= f(\mathbf{\xi}_i) + \dfrac{K}{N} \sum_{j\neq i}^{N} H (\mathbf{\xi}_i,\mathbf{\xi}_j,\mathbf{x}_i,\mathbf{x}_j).
    \label{eq-amp2}
\end{align}
Here, $\xi_i = ({\xi_x}_i, {\xi_y}_i,{\xi_z}_i) \in \mathbb{R}^3$ is the state variable of the $i$th R{\"o}ssler oscillator and $f(\xi_i)$ is the uncoupled dynamics corresponding to it. $E_{ij}$ is the difference between the $i$th and $j$th oscillators given by $E_{ij} = |\xi_j - \xi_i|$. The function $H$ stands for the effect of spatial configuration of internal dynamics and for instance can be chosen as the following,
\begin{equation}
\label{eq-amp3}   H(\mathbf{\xi}_i,\mathbf{\xi}_j,\mathbf{x}_i,\mathbf{x}_j)=  \left[0, \frac{{\xi_y}_j-{\xi_y}_i}{|\mathbf{x}_j-\mathbf{x}_i|^\gamma}, 0\right ]^{\mathcal{T}}.
\end{equation}
The beauty of the model is that the counterparts of all five states of the original model are realized here as well.

%\color{black}

\subsection{Swarmalators in one dimension (1D)}
The 2D swarmalator model and its several offspring offer a rich framework to study the interplay between spatial movement and internal phase dynamics, leading to a variety of complex collective behaviors. However, the power-law attraction and repulsion often complicate the analytical understanding of these emergent phenomena. To gain deeper insights into the underlying mechanisms and simplify the mathematical analysis, a one-dimensional (1D) version of the swarmalator model has been proposed by O'Keeffe et al.~\cite{o2022collective}. The model is governed by a pair of equations:
\begin{align}
\dot{x}_i &= \omega_i +  \frac{J}{N} \sum_j \sin(x_j - x_i) \cos(\theta_j - \theta_i), \label{1d-x}\\
\dot{\theta}_i &= \nu_i + \frac{K}{N} \sum_j \sin(\theta_j - \theta_i) \cos(x_j - x_i) \label{1d-theta},
\end{align}
where $x_i\in \mathbb{S}^1$ denotes the spatial position of the $i$th swarmalator on the 1D ring and $\theta_i \in \mathbb{S}^1$ is its internal phase as earlier. The reduction of this model from the 2D model involves a series of algebraic and trigonometric calculation which are shown in Sec.~\ref{sec.4.3}. This reduced setting retains the essential features of coupling between space and phase while allowing for clearer interpretation of the resulting collective states. Equation~\eqref{1d-theta} describes synchronization dynamics that are modulated by spatial proximity. The standard Kuramoto sine term promotes phase synchronization by reducing pairwise phase differences between swarmalators, while the cosine-modulated coupling term, $K_{ij} = K \cos(x_j - x_i)$, ensures that the strength of interaction increases for nearby individuals, making the synchronization dependent on spatial separation. In contrast, Eq.~\eqref{1d-x} governs the spatial dynamics influenced by phase similarity. Here, the sine term encourages spatial aggregation by minimizing pairwise distances, and the coupling term $J_{ij} = J \cos(\theta_j - \theta_i)$ amplifies attraction between swarmalators with similar phases. Collectively, Eqs.~\eqref{1d-x} and~\eqref{1d-theta} can also be interpreted as representing coupled dynamics on the unit torus, where both phase and position interactions are~interdependent.
\begin{figure}[hpt]
    \centering
    \includegraphics[width=\columnwidth]{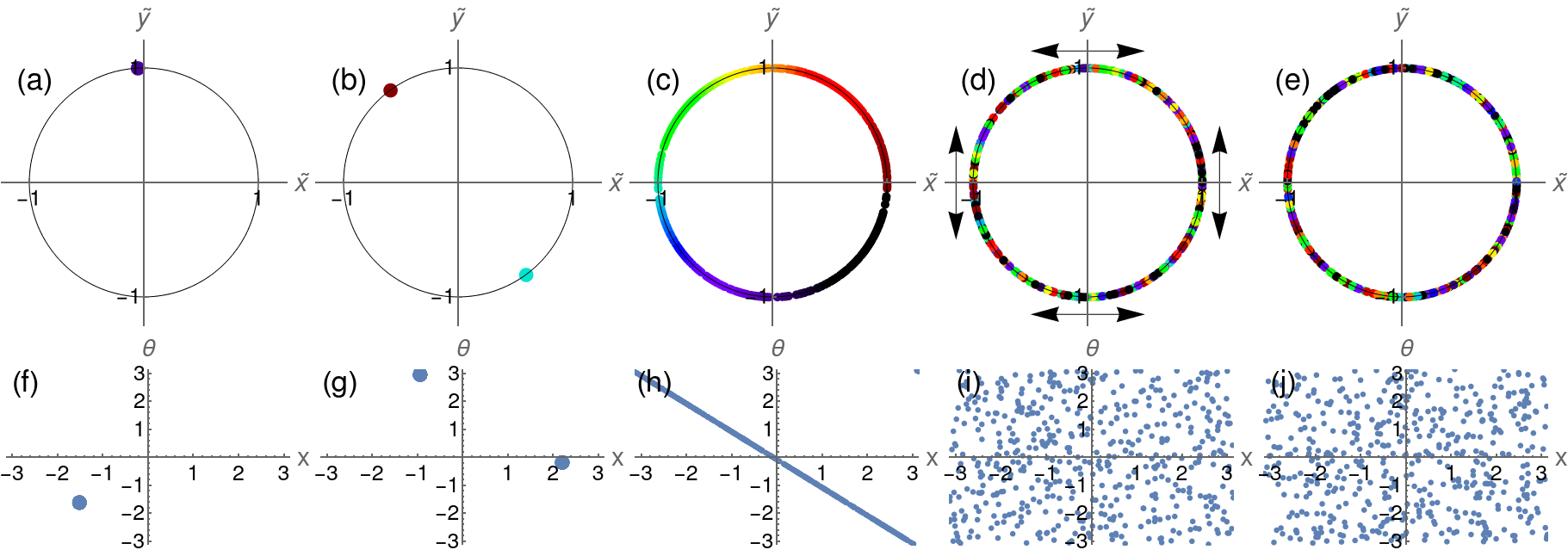}
    \caption{Collective states of the 1D ring model (Eqs.~\eqref{1d-x}-\eqref{1d-theta}). (a, f) Static sync for $(J, K) = (1, 1)$. (b, g) Static $\pi$ state for $(J, K) = (1, 1)$. (c, h) Static phase wave for $(J, K) = (1, -0.5)$. (d, i) Active async for $(J, K) = (1, -1.05)$. Here the swarmalators jiggle about in $(x, \theta)$, as indicated by the double-ended arrows, with no global space-phase order, as indicated by the scatterplot in (i). The amount of motion or jiggling depends on the population size $N$. (e, j) Static async for $(J, K) = (1, -2)$.} 
    \label{1d-model}
\end{figure}

\begin{figure}[hpt]
    \centering
    \includegraphics[width=0.6\columnwidth]{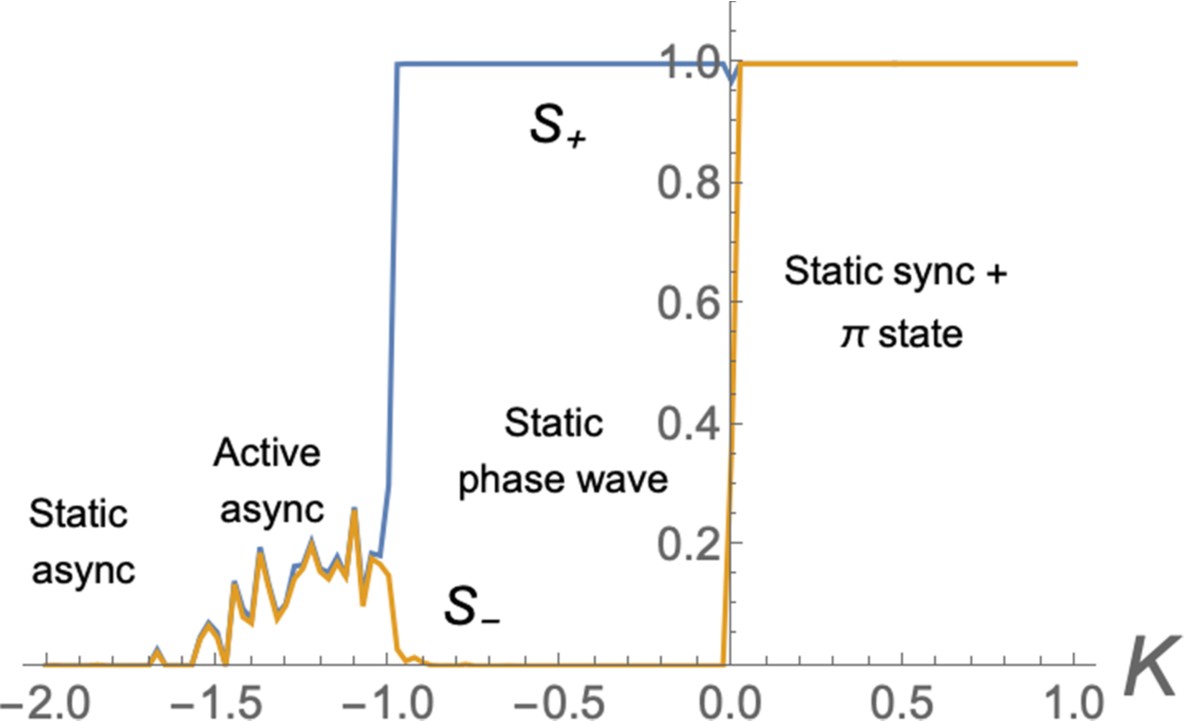}
    \caption{The order parameters $S_{\pm}$ were evaluated for the ring swarmalator model (Eqs.~\eqref{1d-x}-\eqref{1d-theta}) with $\nu_i = \omega_i = 0$ and $J = 1$. Without loss of generality, the condition $S_+ > S_-$ was imposed by interchanging $(S_+, S_-)$ with $(S_-, S_+)$ when necessary. The system described by Eqs.~\eqref{1d-x} and~\eqref{1d-theta} was integrated using a fourth-order Runge-Kutta (RK4) method with time step and total duration set to $(dt, T) = (0.1, 500)$. To eliminate transient behavior, the initial $90\%$ of the time series was discarded, and the mean of the final $10\%$ was used for analysis. A system size of $N = 10$ swarmalators was chosen to clearly demonstrate the characteristics of the active asynchronous state, as the fluctuations in $S_{\pm}$ that define this regime tend to vanish for larger values of $N$. \\ Source: Reprinted figure with permission from Ref. \cite{o2022collective}.} 
    \label{1d-spm-fig}
\end{figure}

{\it Identical swarmalators:} The terms $\omega_i$ and $\nu_i$ can be set to zero by a change of frame without loss of generality. One can convert the trigonometric functions to complex exponentials and derive the following equations:
\begin{align}
    \dot{x}_i &= \frac{J}{2} \left( S_+ \sin \left[ \Phi_+ - (x_i+\theta_i)\right] + S_- \sin\left[ \Phi_- - (x_i-\theta_i\right]\right), \\
    \dot{\theta}_i &= \frac{K}{2} \left( S_+ \sin \left[ \Phi_+ - (x_i+\theta_i)\right] - S_- \sin\left[ \Phi_- - (x_i-\theta_i\right]\right),
\end{align}
where the order parameters $S_{\pm}$ are defined as
\begin{equation}
    W_{\pm} = S_{\pm} e^{\mbox{i} \Phi_{\pm}} = \frac{1}{N} \sum_{j} e^{\mbox{i} (x_j \pm \theta_j)}. \label{1d-spm}
\end{equation}
These order parameters measure the system's total amount of space-phase order. The nature of ordering in the system can be understood by examining its limiting cases. In the case of static sync, where all swarmalators occupy the same position and phase, i.e., $(x_i, \theta_i) = (x^*, \theta^*)$, the order parameters attain their maximum value, $S_{\pm} = 1$, as directly follows from substitution into Eq.~\eqref{1d-spm}. Conversely, in a static async state, where spatial positions $x_i$ are completely uncorrelated with phases $\theta_i$, the order parameters reach their minimum value, $S_{\pm} = 0$. Between these two extremes lies a nontrivial scenario where position and phase are perfectly correlated via a linear relationship, $x_i = \pm \theta_i + c$, for some constant $c$, leading to $(S_+, S_-) = (0,1)$ or $(S_+, S_-) = (1,0)$ depending on the sign of the correlation. This is the 1D analogue of the static phase wave state. Simulations show that the identical swarmalator system settles into five collective states that are depicted in Fig.~\ref{1d-model}. Initial positions and phases were drawn from $[-\pi, \pi]$ in all states except the static sync state (shown in Fig.~\ref{1d-model} (a,f)), which were drawn from $[0, \pi]$. Fig.~\ref{1d-spm-fig} shows that the $S_{\pm}(K)$ curves can distinguish between all but the static sync and static $\pi$ states that only differ in their basins of attractions.

The 1D ring model is mathematically tractable. The bifurcation points can be found using the Jacobian analysis of the fixed points in the finite $N$ limit or using the continuity equation in the $N \rightarrow \infty$ limit. These methods are discussed in Ref.~\cite{o2022collective}. The nonidentical swarmalators with distributed velocities and frequencies are also described in Sec.~\ref{sec.4.3}.

{\it Different coupling mechanisms}: The ring model has been studied with disordered coupling strengths by choosing $K$ from a double delta distribution (like the one used in Eq.~\eqref{double-delta}) with peaks at $K_p>0$ and $K_n<0$ and fixing $J=1$~\cite{o2022swarmalators,hao2023attractive}. The combination of attractive and repulsive couplings leads to several unsteady states. One of them is the buckled phase wave state where the static phase wave destabilizes into noisy, unsteady phase waves with nonzero velocity. In Ref.~\cite{sar2025effects}, a function of the form $G(x_j - x_i) = \left(\frac{1+\cos(x_j-x_i)}{2}\right)^p$ is introduced in both spatial and phase dynamics to investigate the effect of spatial proximity or coupling range on the dynamics of swarmalators. Novel type of phase waves of the form $\theta_i = \pm q x_i + c$, named the $q$-wave, are found where $q$ depend on the exponent $p$ of the pulse function $G$. For example, $p=2$ produces $1$-wave, $2$-wave, and $3$-wave.

{\it External forcing, noise, and phase lag}: The addition of an external periodic forcing term of the form $F \sin(\Omega t - \theta_i)$ to the phase dynamics impacts the overall behavior of the system and phase-locked state and the existence of chimera state have been reported~\cite{anwar2024forced}. An analytical study of the model with thermal noise has been carried out by Hong et al.~\cite{hong2023swarmalators}. The effect of phase lag or frustration has also been investigated by Lizárraga et al.~\cite{lizarraga2023synchronization}, where  many disordered states are found, some of which are similar to turbulence generated in a flattened media.

\begin{figure}[hpt]
    \centering
    \includegraphics[width=0.75\columnwidth]{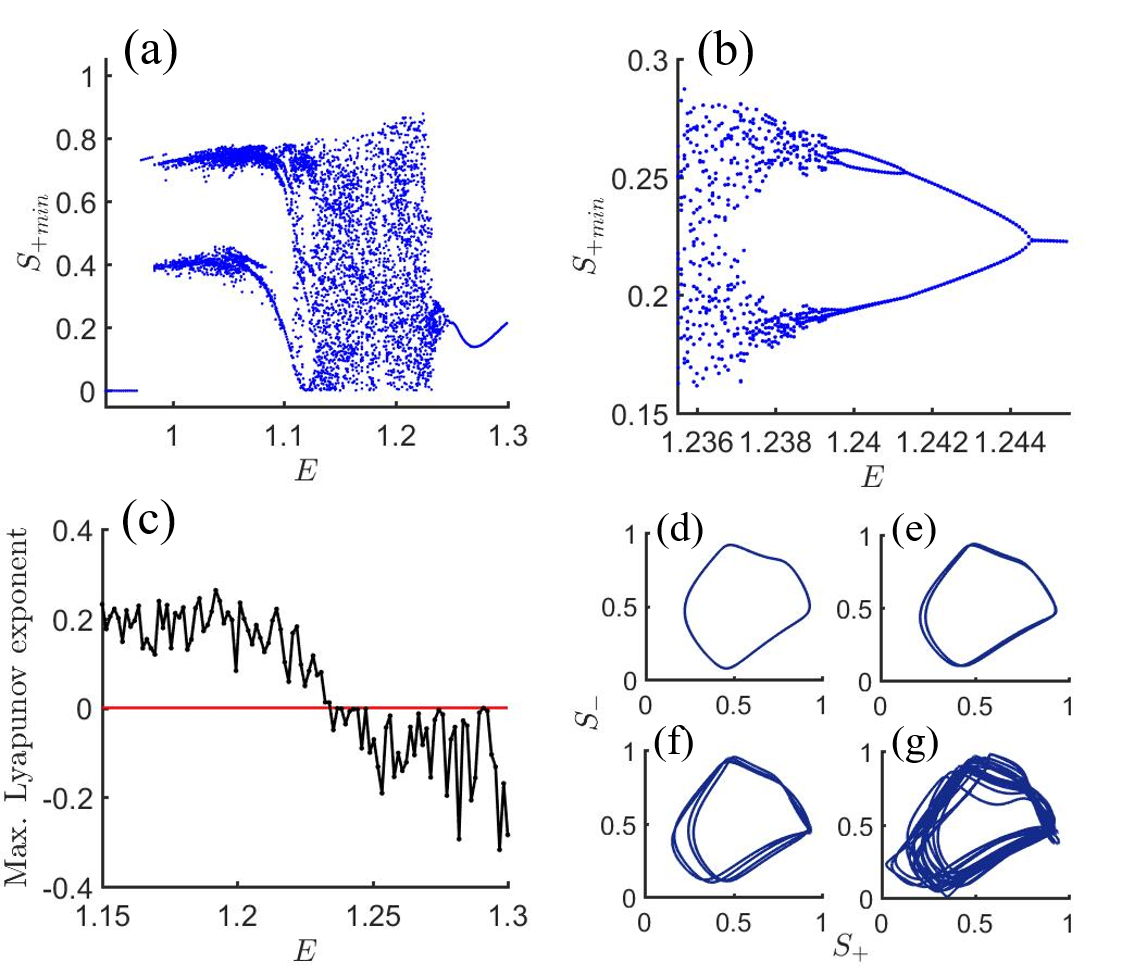}
    \caption{Route to Chaos with varying driving strength $E$ in model Eqs.~\eqref{1d-pin1}-\eqref{1d-pin2}. (a) Bifurcation diagram of $S_+$ as a function of $E$. (b) A magnified view highlighting the period-doubling route to chaos. (c) Corresponding Lyapunov exponents confirming the onset of chaotic dynamics. (d)–(g) Illustration of the period-doubling sequence in the $(S_+,S_-)$ plane for $E=1.248,1.243,1.241,1.23$, respectively. \\ Source: Reprinted figure with permission from Ref. \cite{sar2023pinning}.} 
    \label{1d-pin}
\end{figure}

{\it Random pinning}: This segment describes the dynamics of swarmalators in the presence of random pinning, a concept well established in nonlinear dynamics and statistical physics. Random pinning describes the tendency of a system to become anchored to local inhomogeneities, requiring external forcing to induce motion. A classical example is found in charge density waves, where phase oscillators are used to model how the phase of the wave interacts with the underlying medium. Beyond a critical threshold of forcing, these pinned phases begin to depin, giving rise to rich and complex collective behavior. In terms of swarmalators, this framework offers a new perspective for analyzing how swarmalator systems respond to spatial and phase disorder. The swarmalator model with pinning and external driving is
\begin{align}
    \dot{x_i} &= E -b \sin(x_i - \alpha_i) + \frac{J}{N} \sum_j \sin(x_j - x_i) \cos(\theta_j - \theta_i), \label{1d-pin1} \\
    \dot{\theta_i} &= E -  b \sin(\theta_i - \beta_i) + \frac{K}{N} \sum_j \sin(\theta_j - \theta_i ) \cos(x_j - x_i ), \label{1d-pin2}
\end{align}
where $E$ and $b$ are the driving and pinning strengths, respectively, and $\alpha_i, \beta_i$ are the pinning sites. Sar et al.~\cite{sar2023pinning} studied this model with symmetric coupling strengths $J=K$ and symmetric linearly spaced pinning sites $\alpha_i=\beta_i=2 \pi i/N$. Low dimensional chaos in terms of the order parameters $S_{\pm}$ has been observed where it is achieved via an inverse period doubling as shown in Fig.~\ref{1d-pin}. Asymmetric pinning strengths and coupling sites are also considered in a later work~\cite{sar2023swarmalators}.

{\it Higher-order interactions}: Recent developments in physics and related disciplines have highlighted the growing importance of interactions that go beyond simple pairwise connections~\cite{boccaletti2023structure,battiston2021physics}. In particular, three-body and four-body interactions have emerged as crucial mechanisms driving complex collective phenomena. Consequently, the study of network dynamics has expanded to incorporate higher-order structures that more accurately represent the multifaceted nature of real-world interactions. These complex relationships are often formalized through simplicial complexes, which consist of hierarchical building blocks such as 1-simplices (edges), 2-simplices (filled triangles), and higher-dimensional simplices--each capturing different layers of interaction within the network~\cite{majhi2022dynamics}. Building on this motivation, a swarmalator model is introduced that integrates both pairwise and higher-order, specifically three-body interactions in the form
\begin{align}
    \dot{x}_{i} &= v_{i}+\dfrac{J_1}{N} \sum\limits_{j=1}^{N} \sin(x_{j}-x_{i}) \cos(\theta_{j}-\theta_{i}) +\dfrac{J_2}{N^2} \sum\limits_{j=1}^{N} \sum\limits_{k=1}^{N} \sin(2x_{j}-x_{k}-x_{i}) \cos(2\theta_{j}-\theta_{k}-\theta_{i}),\\
    \dot{\theta}_{i} &= \omega_{i}+\dfrac{K_1}{N} \sum\limits_{j=1}^{N} \sin(\theta_{j}-\theta_{i}) \cos(x_{j}-x_{i}) +\dfrac{K_2}{N^2} \sum\limits_{j=1}^{N} \sum\limits_{k=1}^{N} \sin(2\theta_{j}-\theta_{k}-\theta_{i}) \cos(2x_{j}-x_{k}-x_{i}),
\end{align}
where $J_1,K_1$ are the pairwise coupling strengths, and $J_2,K_2$ are the higher-order coupling strengths. Although no new states are found with the addition of the higher-order couplings, it is the explosive or first-order phase transitions and bistability between those states that are the significance of the inclusion of higher-order interactions into the swarmalator system~\cite{anwar2024collective}.

%\subsubsection{Higher-order interactions}
%Collective dynamics of swarmalators with higher-order interactions~\cite{anwar2024collective}

%\subsubsection{Sakaguchi swarmalators}
%Synchronization of Sakaguchi swarmalators~\cite{lizarraga2023synchronization}

%\subsubsection{Non-Kuramoto phase dynamics}
{\it Winfree-type pulse coupling}:
Analogous to Sec.~\ref{sec.3.1.8} where the 2D swarmalator model had been embedded with non-Kuramoto phase dynamics, here also the 1D model can be studied in the light of Winfree-type pulse coupling. The phase dynamics is then governed by the equation
\begin{equation}
    \dot{\theta}_i = \omega_i + R(\theta_i)  \frac{K}{N} \sum_{j=1}^{N} P(\theta_j) \cos(x_j - x_i),
\end{equation}
where $R(\theta)$ and $P(\theta)$ are defined earlier in Eq.~\eqref{pulse}. Very recently, this model has been studied by Ghosh et al.~\cite{ghosh2025dynamics} with identical swarmalators. Also, this model is extended to 2D and new emerging states, like mixed states are observed \cite{acharya2025a}.

The 1D model is a simple model in the sense that many of the emerging states are solvable, yet a complex one that exhibits a diverse array of collective states. It is understood that whenever the coupling strengths of the model are positive and large, the system goes to a sync state where one can perceive coordinated motion or rhythmic oscillation. However, in certain contexts, synchronization can be detrimental to a system’s function, making its regulation or suppression essential. In a recent study, Sar et al.~\cite{sar2025strategy} proposed a control strategy based on Hamiltonian control theory
to suppress synchronization in a system of swarmalators confined to a 1D ring.

%Dynamics of pulsating swarmalators on a ring~\cite{ghosh2025dynamics}

%\subsubsection{Strategy to control synchronization}
%Strategy to control synchronized dynamics in swarmalator systems~\cite{sar2025strategy}

\subsection{Swarmalators in three and higher dimensions}
\begin{figure}[hpt]
    \centering
    \includegraphics[width=0.75\columnwidth]{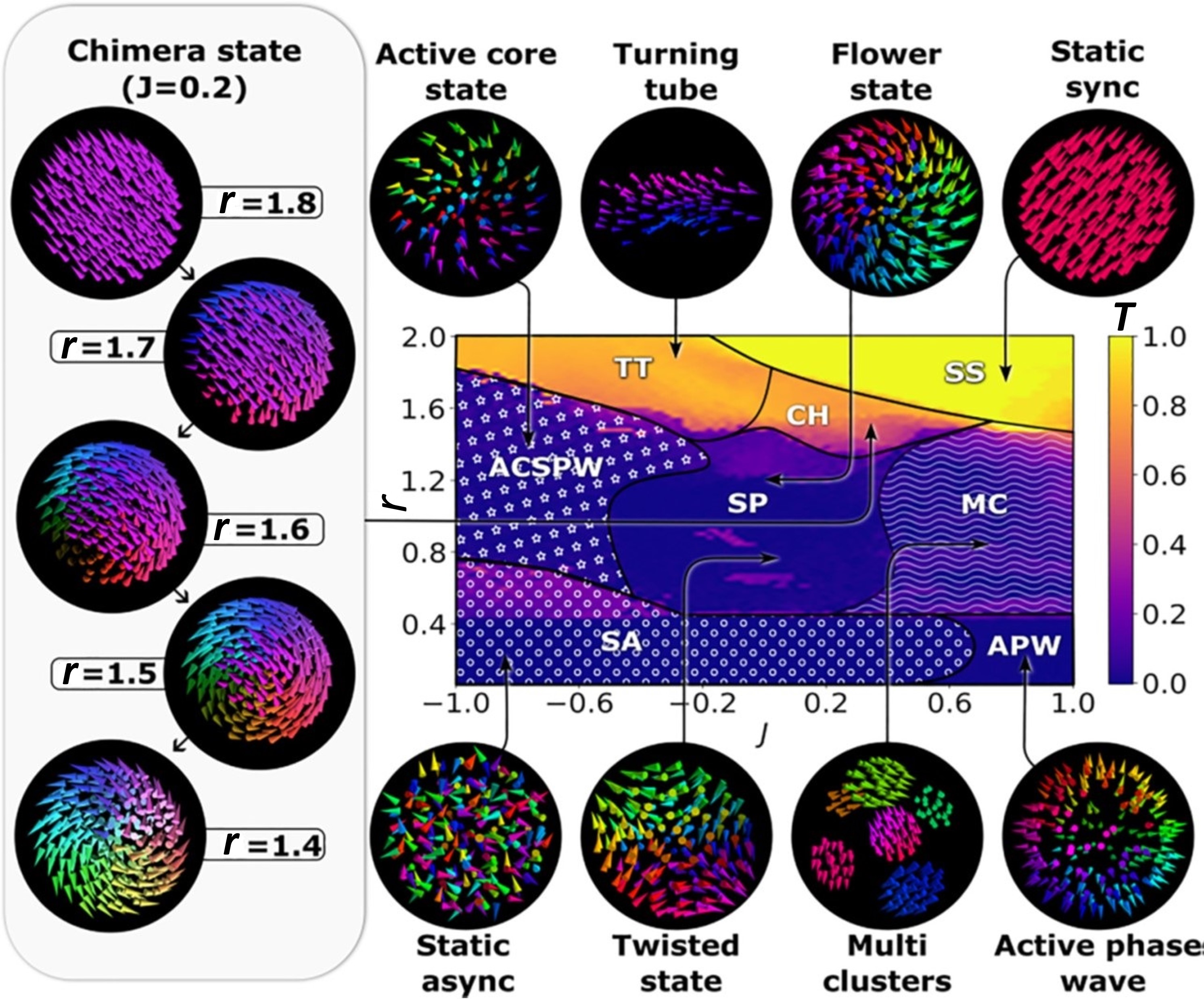}
    \caption{The synchronization order parameter $T$ is presented in the two-parameter $(J, r)$ phase diagram for competing attractive and repulsive interactions, with $K_a = K_r = 0.5$. The left panel illustrates the progression of a chimera state as the vision radius $r$ increases from $1.4$ to $1.8$. Snapshots of the swarmalators, highlighting various emergent collective states, are displayed in the top and bottom panels. For simulations, Eqs.~\eqref{3d-x}-\eqref{3d-theta} are used. \\ Source: Reprinted figure with permission from Ref. \cite{yadav2024exotic}.} 
    \label{3d-model}
\end{figure}

To more accurately capture the self-organizing spatiotemporal patterns and uncover the underlying mechanisms of natural collectives, a generalized $D$-dimensional swarmalator model is introduced, incorporating both $D$-dimensional spatial and orientation vectors. In this framework, the alignment of orientation vectors characterizes the intrinsic dynamics, enhancing the model’s predictive capability for real-world collective behaviors. The governing equations of this generalized $D$-dimensional model are given by~\cite{yadav2024exotic},
\begin{align}
	\dot{\textbf{x}}_{i} &= \textbf{v}_{i} + \frac{1}{N-1} \sum_{\substack{j = 1\\j \neq i}}^{N}\bigg[\frac{\textbf{x}_{j}-\textbf{x}_{i}}{|\textbf{x}_{j}-\textbf{x}_{i}|^{\alpha}} (1+J (\sigma_i \cdot \sigma_j)) - \frac{\textbf{x}_{j}-\textbf{x}_{i}}{|\textbf{x}_{j}-\textbf{x}_{i}|^{\beta}}\bigg] + \xi_i^{\textbf{x}}(t), \label{3d-x}\\
	\dot{\sigma}_{i} &= W_i \sigma_i + \sum_{\substack{j = 1\\j \neq i}}^{N} K_{ij} \left[\frac{\sigma_j - (\sigma_j \cdot \sigma_i) \sigma_i}{|\textbf{x}_{j}-\textbf{x}_{i}|^{\gamma}} \right] + \xi_i^{\sigma}(t),
	\label{3d-theta}
\end{align}
where $\textbf{x}_{i}$ is the $D$-dimensional position vector of the $i$th swarmalator and $\sigma_i$ is the orientation vector on the $D$-dimensional unit hypersphere that characterizes the internal dynamics. The significance of the exponents $\alpha$, $\beta$, and $\gamma$ is discussed earlier in Sec.~\ref{sec.3.1.1}. $W_i$ is the $D \times D$ anti-symmetric angular velocity matrix corresponding to the $i$th swarmalator. The coupling strength $K_{ij}$ is chosen as
\begin{equation}
    K_{ij} = \begin{cases}
        \frac{K_a}{N_i(r)} \hspace{40 pt} \text{for} \;\;|\textbf{x}_{j}-\textbf{x}_{i}| \le r\\
        \frac{K_r}{N-1-N_i(r)} \hspace{19 pt} \text{for} \;\;|\textbf{x}_{j}-\textbf{x}_{i}| > r,
    \end{cases}
\end{equation}
in a manner similar to Eq.~\eqref{eq.27} where $r$ denotes the vision radius of the swarmalators. $\xi^\textbf{x}_i(t)$ and $\xi^\sigma_i(t)$ represent Gaussian white noise processes with zero mean and strengths $d_{\textbf{x}_k}$ and $d_{\sigma_k}$, respectively. These noise terms are characterized by $\langle \xi^{\textbf{x}_k}_i(t) \; | \;\xi^{\textbf{x}_k}_j(t') \rangle = 2d_{\textbf{x}_k} \delta(t - t')$ and $\langle \xi^{\sigma_k}_i(t) \;| \;\xi^{\sigma_k}_j(t') \rangle = 2d_{\sigma_k} \delta(t - t')$, where $k = 1, 2,  \ldots, D$. Since $\sigma_i$ is affected by noise, it must be renormalized at each time step to maintain unit length. The synchronization order parameter $T$ measures the collective alignment of the orientation vectors of the swarmalators and is defined as the magnitude of their mean orientation vector by
\begin{equation}
    T = \frac{1}{N} \Bigg| \sum_{j=1}^N \sigma_j\Bigg|.
\end{equation}
A case study for $D=3$ with $\textbf{v}_{i} = W_i=0$ for all $i$, $d_{\textbf{x}_k}=d_{\sigma_k}=0$ featuring $N=100$ swarmalators was carried out in Ref.~\cite{yadav2024exotic} and several collective states had been reported, as shown in Fig.~\ref{3d-model}.

%==============================================================

\section{Theory}  \label{sec.4}
\subsection{Overview: the 2D swarmalator model}
\begin{figure}[hpt]
    \centering
    \includegraphics[width=0.6\columnwidth]{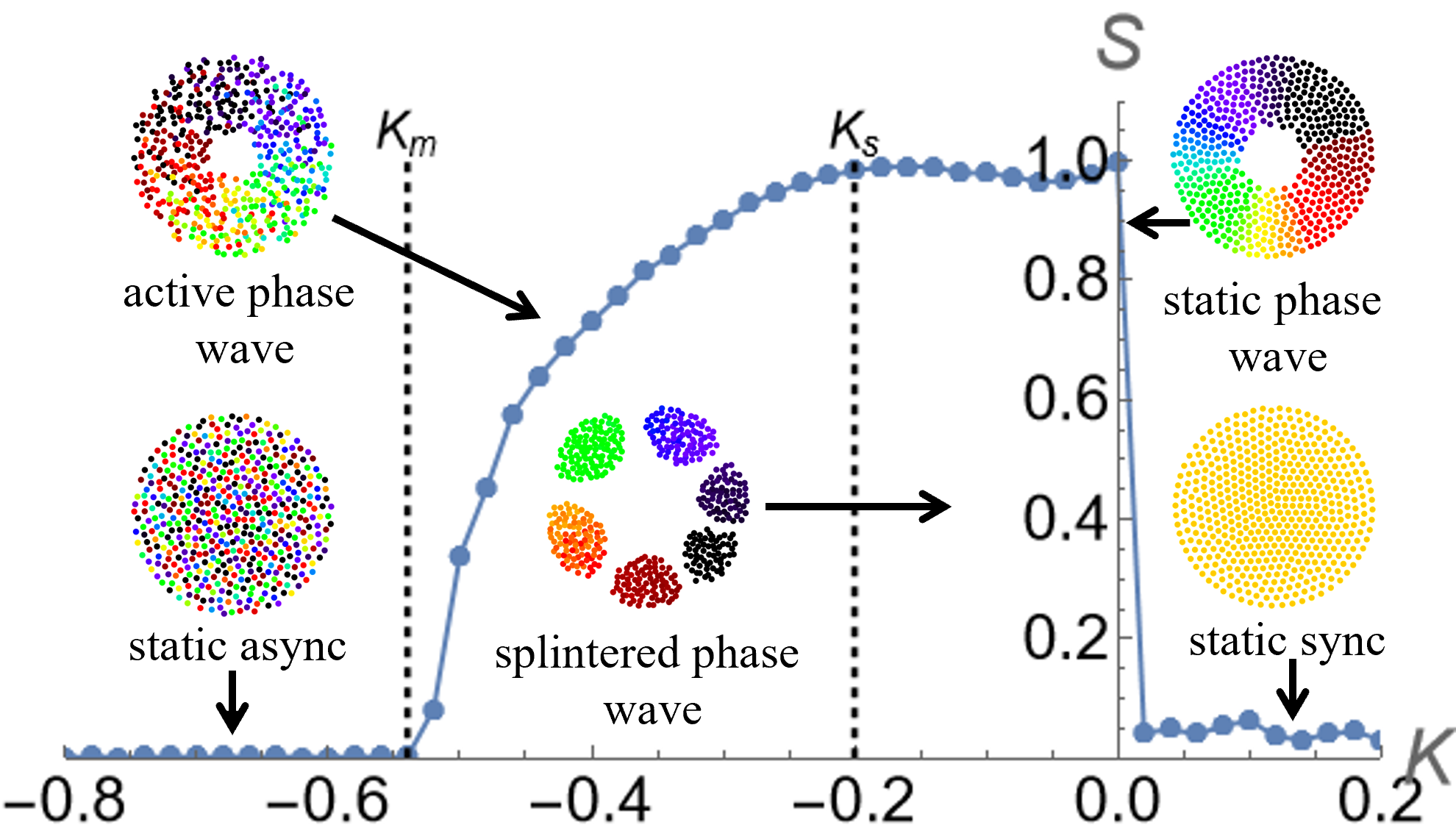}
    \caption{Rainbow order parameter $S(K)$ of the 2D swarmalator model introduced in \cite{o2017oscillators} (Eqs.~\eqref{eq.2.15}-\eqref{eq.2.16}). Insets show the collective states of the model where swarmalators are represented as colored dots in the $(x,y)$ plane where the color denotes the swarmalators phase. Static async: $(J,K) = (0.5,-0.8)$, active phase wave $(J,K) = (0.5,-0.4)$, splintered phase wave $(J,K) = (0.5,-0.1)$, static phase wave $(J,K) = (0.5,0)$, static sync $(J,K) = (0.5,0.2)$. In the three static states, swarmalators do not move in space or phase. In the splintered phase wave, each colored chunk is a vortex: the swarmalators oscillate in both space and phase. In the active phase wave, the oscillations are excited into rotations. The swarmalators split into two groups with counter-rotate in $x$ and $\theta$. \\ Source: Reprinted figure with permission from Ref. \cite{o2017oscillators}.} 
    \label{2d-model}
\end{figure}

Theories for swarmalators are hard to develop. Their governing equations are both nonlinear and high dimensional whose analysis is inherently difficult. Even the original 2D swarmalator model, which is radically simplified, is not well understood. Here we talk through our efforts to trying to understand this model theoretically, which we see as a first case-study towards a general theory of swarmalators.

Figure~\ref{2d-model} replots the model's five collective states along with its bifurcation structure by solving the governing Eqs. \eqref{eq.2.15} and \eqref{eq.2.16}, as encoded by its order parameter curve $S(K)$. For $K < K_m$,  a crystal-like async state arises, where the individual swarmalators are frozen into a circular disk with no motion or overall phase order (as seen by the uniform distribution of colors). Beyond a critical melting point $K_m$, the state melts into an active phase wave state where the individual units run in a space-phase vortex. At a later $K_s$, the vortex splints its a ring of mini-vortices called the splintered phase wave. For $K>0$, a synchronous disk is formed, and for the special case $K=0$ an another phase wave which is now static. These states mimic several real world swarmalator states such as the vortex arrays of sperm \cite{riedel2005self} and the synchronous disks of Quincke rollers \cite{reyes2023magnetic}.

Like the old puzzles about the Kuramoto model \cite{strogatz2000kuramoto}, the collective behaviors of the swarmalator model cry out for explanation. A good theory should provide expressions for the melting point $K_m$, the splitting point $K_s$ and ideally expressions for the $S(K)$. Expressions for the density $\rho(x,\theta)$ in each state would be a bonus.

Over the past ten years or so, we have been trying to build such a theory. We have made some progress on the density front, having managed to construct $\rho(x,\theta)$ in the three static states (async, sync, phase wave) using a self consistency method \cite{o2017oscillators}. But the $\rho(x,\theta)$ for the two active states (active phase wave, splintered phase wave) as well as the critical points $K_m$, $K_s$ and order parameter curve $S(K)$ curve remain out of reach. We outline some of the walls, we have run into when trying to pin these quantities down. Our goal is to show how swarmalator theory is different to oscillator theory, and hopefully to entice you into solving some of the problems that continue to puzzle us.

\subsection{The search for the melting point $K_m$}
Let us focus on the melting point $K_m$, the point at which static async melts into the active phase wave (Fig.~\ref{2d-model}). How could you find this $K_m$? We confess we originally thought it would be easy. We know how to find the critical point $K_c$ at which the asynchronous state in the Kuramoto model destabilizes. You consider perturbations around the async density $\rho(\theta, \omega, t ) = 1/(2 \pi) + \epsilon \rho_1(\theta, \omega, t)$ and substitute into the continuity equation $\dot{\rho} + \partial_{\theta} (v \rho) = 0$, where the velocity in this state is simply $v(\theta, \omega) = \omega$ (in the async state oscillators move at their natural frequencies $\omega$ which are drawn from some distribution). Then you pull out an eigenvalue equation and solve for $K_c$. Following the same recipe for swarmalator model looked straightforward -- in fact it looked \textit{easier}, since in the swarmalator model the individual swarmalators are identical (same frequencies which can be set to zero) and static ($v=0$ which kills one of the terms in the continuity equation). 

Not so. The compact support of the async state (it being a disk of fixed width) is tricky to analyze. Linearizing around its density $\rho(x,\theta) = (2\pi)^{-1} H(R-r)$, where $H(.)$ is Heaviside's step function and $R$ is the radius of the disk, leads to nonstandard eigenvalues equations in the form
\begin{align}
    \lambda b(x) + H(R-r) \int f(y-x) b(y) dy + \delta(R-r) \int g(y-x) b(y) dy = 0 \label{hard_eval} 
\end{align}
Here $f,g$ are complex kernels and $b(x)$ is the eigenfunction. The convolutions are already hard to deal with. But the generalized delta and Heaviside functions out front make things even worse. They pull us into a more advanced functional analytic setting, which one doesn't typically encounter in studies of regular oscillators \cite{strogatz1991stability}. As a result, no one has been able to solve Eq.~\eqref{hard_eval} for $\lambda$. It is one of the big open problems in the swaramalator field.

You might think you could skip these issues with nonstandard eigenvalues and find $K_m$ using Kuramoto's famous self-consistency trick \cite{kuramoto1975self}. But that path has its own blocker. Recall the essence of the approach is to write the density in the supercritical sync state $\rho(\theta, \omega; R)$ in terms of the order parameter $R = | \langle e^{\mbox{i} \theta} \rangle|$ using the continuity equation and then invoke self consistency. We summarize by the three equations that are needed, %The key part is that the 1d continuity equation in the steady state is easy to solve: $\partial_{\theta}(v \rho) = 0 \rightarrow \rho = C / v$ for some constant $C$.  
\begin{align}
    R &= \Big| \int e^{\mbox{i} \theta} \rho(\theta, \omega) g(\omega) d \theta d\omega \Big|, \hspace{1 cm}  \partial_{\theta}(v(\theta ; R) \rho(\theta, \omega, t)) = 0, \hspace{1 cm} v(\theta, \omega; R) = \omega - K R \sin \theta . \label{c1} \\ 
   S &= \max_{\pm} \Big| \int_{\Gamma(r, \phi, \theta)} e^{\mbox{i} ( \phi \pm \theta) } \rho(S_+, S_-, r,\phi,\theta)_{apw} r dr d \phi d \theta  \Big|, \hspace{0.5 cm} \nabla (v(r,\phi,\theta; S_{\pm}) \rho(r, \phi, \theta)) = 0. \label{c2}
\end{align}
Equation~\eqref{c2} gives the analogous equations for the swarmalator order parameter $S$ in Eq. \eqref{smaxmin} (but without the expressions for the velocity; it is too long to display here). You can see its much more complicated: a 3D countour integral over the (unknown) support $\Gamma(r,\phi,\theta)$ of the active phase wave whose (unknown) density $\rho_{apw}$ depends on both of the rainbow order parameters $S_{\pm}$. Moreover, the 3D steady state continuity equation $\nabla (v(r,\phi,\theta; S_{\pm}) \rho(r, \phi, \theta))$ is \textit{far} harder to solve than its 1D equivalent $\partial_{\theta}(v(\theta ; R) \rho(\theta, \omega, t)) = 0$. The latter simplifies to $\rho = C/v$ for some constant $C$, the former to a monstrous 3d integro-differential equation of form $ v_r \, \partial_r \rho + (v_\phi / r) \, \partial_\phi \rho + v_\theta \, \partial_\theta \rho + \rho \bigl( (1 / r) \, \partial_r (r v_r) + (1 / r) \, \partial_\phi v_\phi + \partial_\theta v_\theta \bigr) = 0$. 

The summary is this: implementing the two go-to methods to find criticality thresholds in systems of oscillators, linearization and self-consistency, in swarmalator systems is much more difficult. One cannot blindly adapt the tools of coupled oscillator theory to build a swarmalator theory.

Blocked by these mathematical walls, we retreat a little and consider a simpler model which captures the essence of the 2D model but was hopefully easier to analyze. Casting an eye back to Fig.~\ref{2d-model}, you can see the swarmalator states are radially symmetric, consisting of synchronous disks and annular phase waves. The interesting space-phase interactions are all in the azimuthal direction $\phi = \tan^{-1}(x,y)$. This suggests a 1D model where the spatial motion is confined to the 1D periodic domain, which plays the role of $\phi$, might capture the key physics.

%%%%%%%%%%%%%%%%%%%%%%%%%%%%%%%%%%%%%
\subsection{The 1D swarmalator model on the ring $x_i \in \mathbb{S}^1$} \label{sec.4.3}
How to build such a model? We wanted to be as principled as possible, so we wrote the 2D model in polar coordinates $(\dot{x},\dot{y}, \dot{\theta}) \rightarrow (\dot{r},\dot{\phi}, \dot{\theta})$ and peeled off the $(\dot{\phi}, \dot{\theta})$ equations. This is easiest in a slightly different parameterization of the 2D swarmalator model
\begin{align}
&\dot{\mathbf{x}}_i = \frac{1}{N} \sum_{ j \neq i}^N \Bigg[ (\mathbf{x}_j - \mathbf{x}_i) \Big( 1 + J \cos(\theta_j - \theta_i)  \Big) -   \frac{\mathbf{x}_j - \mathbf{x}_i}{ | \mathbf{x}_j - \mathbf{x}_i|^2}\Bigg], \label{linear_parabolic1} \\ 
& \dot{\theta_i} = \frac{K}{N} \sum_{j \neq i}^N \sin(\theta_j - \theta_i ) \Big(1 - \frac{| \mathbf{x}_j - \mathbf{x}_i|^2}{\sigma^2} \Big) H(\sigma - |\mathbf{x}_j - \mathbf{x}_i|), \label{linear_parabolic2}
\end{align}
\noindent
where $I_{att}(x) = x$,  $I_{rep}(x) = x / |x|^2$ and $G(x) = (1 - |x|^2/ \sigma^2) H({\sigma - |x|})$. So $\sigma$ is a hard-cutoff coupling range. The benefit of this model is that in polar coordinates, its dynamics take especially simple form,
\begin{align}
\dot{r_i} &= H_{r}(r_i, \phi_i) +  \Bigg[ S_+ \cos \Big( \Psi_+ - (\phi_i+\theta_i) \Big)  + S_- \cos \Big( \Psi_{-} -  (\phi_i- \theta_i) \Big) \Bigg], \\
\dot{\phi_i}&=  H_{\phi}(r_i, \phi_i)  + \frac{J}{2 r_i} \underbrace{\Bigg[ S_+ \sin \Big( \Psi_+ - (\phi_i+\theta_i) \Big)  + S_- \sin \Big( \Psi_{-} -  (\phi_i- \theta_i) \Big) \Bigg],}_{1d \; model} \label{e2} \\
\dot{\theta_i} &=  K \Big(1- \frac{r_i^2}{\sigma^2} \Big) R_0 \sin (\Phi_0 - \theta_i) - \frac{K}{\sigma^2} R_1 \sin(\Phi_1 - \theta_i) + \frac{K r_i}{\sigma^2} \underbrace{\Bigg[ S_+ \sin \Big( \Psi_+ - (\phi_i+\theta_i) \Big) - S_- \sin \Big( \Psi_- - (\phi_i- \theta_i) \Big) \Bigg].}_{1d \; model} \label{e3}
\end{align}
\noindent
You can see the equations decouple into baseline functions $H_r(r_i, \phi_i), H_{\phi}(r_i, \phi_i)$, which depend only on the spatial coordinates, and Kuramoto-like terms $S_{\pm} \sin(\phi_{\pm} + (\phi \pm \theta))$ which encode the desired space-phase interaction. Here,
\begin{align}
H_r(r_i, \phi_i) &=  \frac{1}{N} \sum_{j} \Big( r_j \cos(\phi_j - \phi_i) - r_i \Big) ( 1 - d_{ij}^{-2} ), \quad
H_{\phi}(r_i, \phi_i) =  \frac{1}{N} \sum_{j} \frac{r_j}{r_i} \sin(\phi_j - \phi_i) ( 1 - d_{ij}^{-2} ), \\
Z_0 = R_0 e^{\mbox{i} \Psi_0} &=  \frac{1}{N} \sum_{j}  e^{\mbox{i} \theta_j}, \quad
\hat{Z}_0 = \hat{R}_0 e^{\mbox{i} \hat{\Psi}_0} =  \frac{1}{N} \sum_{j \in N_i}  e^{\mbox{i} \theta_j}, 
\; \; \; Z_2 = R_2 e^{\mbox{i} \Psi_2} =  \frac{1}{N} \sum_{j} r_j^2 e^{\mbox{i} \theta_j}
\; \; \; \hat{Z}_2 = \hat{R}_2 e^{\mbox{i} \hat{\Psi}_2} =  \frac{1}{N} \sum_{j \in N_i} r_j^2 e^{\mbox{i} \theta_j} \\
\tilde{W}_{\pm} = \tilde{S}_{\pm} e^{\mbox{i} \Psi_{\pm}} &=  \frac{1}{N} \sum_{j} r_j e^{\mbox{i} (\phi_j \pm \theta_j)}, \quad
\hat{W}_{\pm} = \hat{S}_{\pm} e^{\mbox{i} \hat{\Psi}_{\pm}} =  \frac{1}{N} \sum_{j \in N_i} r_j e^{\mbox{i} (\phi_j \pm \theta_j)}.
\end{align}
\noindent
The underbraced terms in Eq.~\eqref{e2},~\eqref{e3} contain the physics we want to isolate. We thus defined the \textit{1d swarmalator model}:
\begin{align}
\dot{x}_i &= \tilde{\omega}_i +  \frac{\tilde{J}}{N} \sum_j \sin(x_j - x_i) \cos(\theta_j - \theta_i) \label{w1}, \\
\dot{\theta}_i &= \tilde{\nu}_i + \frac{\tilde{K}}{N} \sum_j \sin(\theta_j - \theta_i) \cos(x_j - x_i) \label{w2}.
\end{align}
When you switch to $(\xi, \eta) = (x+\theta, x-\theta)$ coordinates, imitates the $\phi,\theta$ interactions in the 2D model as
\begin{align}
\dot{\xi}_i = \omega_i+  K r \sin(\phi - \xi) + J s \sin(\psi - \eta),  \\
\dot{\eta}_i =  \nu_i +  J r \sin(\phi - \xi) + K s \sin(\psi - \eta),  
\end{align}
where,
\begin{align}
& r e^{\mbox{i} \phi}, \; \; s e^{\mbox{i} \psi} = \langle e^{\mbox{i} \xi} \rangle, \; \; \langle e^{\mbox{i} \eta} \rangle,  \\
& (\omega, \nu) = (\tilde{\omega} + \tilde{\nu}, \tilde{\omega} - \tilde{\nu}), \\
& (J,K) = ( (\tilde{J}+\tilde{K})/2, (\tilde{J}-\tilde{K})/2).
\end{align}
This 1D model has lovely structure. It is a pair of coupled Kuramoto models and also has perfect space-phase symmetry: the mapping $(x,\theta), (\xi, \eta) \rightarrow (\theta, x), (\eta, \xi)$ leaves the equations unchanged. These properties make it one of the few models of swarmalators / mobile oscillators that is tractable. It has inspired a burst of theoretical studies of swarmalators~\cite{o2022collective, yoon2022sync, hong2023swarmalators, o2022swarmalators, hao2023attractive, o2025stability, sar2024solvable, sar2023pinning, anwar2024collective, anwar2024forced, o2025global, anwar2025forced} and as we now discuss apes many features of the 2D model, as intended.
\begin{figure}[h!]
    \centering
    \includegraphics[width=\textwidth]{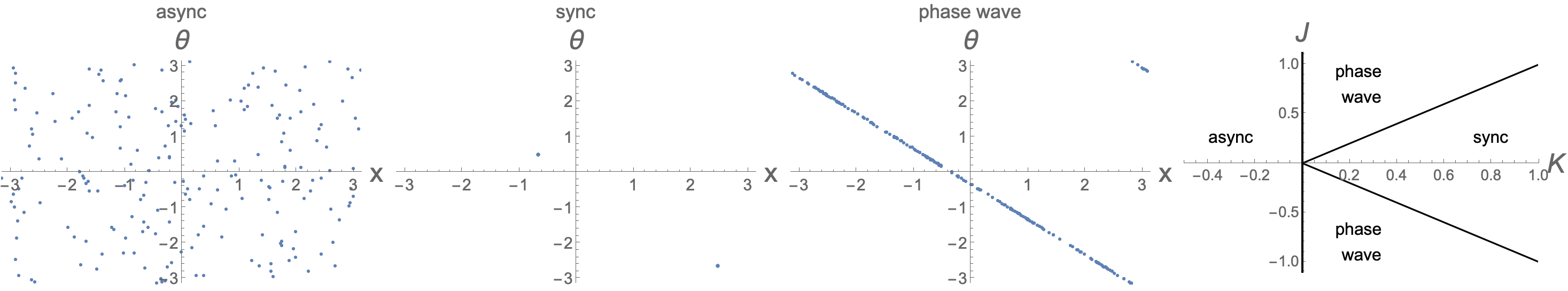}
    \caption{Three left panels: collective states of 1D swarmalator model Eqs.~\eqref{w1}-\eqref{w2}. Right most panel: phase diagram in $(J,K)$ space.}
    \label{fig:1d_model_states}
\end{figure}

\textit{Identical swarmalators}. Figure~\ref{fig:1d_model_states} shows that for identical swarmalators the model produces 1D analogues of the async, sync, and phase wave states. The async state is now fully, not compactly supported, allowing us to derive the 1D analogues of the melting point $K_m$.  The stabilities of the phase wave and sync states have also been derived with stadard finite-$N$ linearization allowing us to draw the bifurcation diagram. The global stability of the sync state has also been proved \cite{o2025global}. 

On top of that, analyzing the stability of the phase wave in the continuum limit has given us a tractable version of the nonstandard eigenvalue Eq.~\eqref{hard_eval}. Linearizing around its density described by $\rho(\xi,\eta) = \delta(\eta) / 2 \pi$ -- which is similar that 2D async state which $\rho(x,\theta) = (2 \pi)^{-1} H(R-r)$ -- leads to an eigenvalue equation
\begin{align}
   \lambda b &= K b \cos \eta + \sin \eta \left( K b_{\eta}  + J b_{\xi} \right) - \frac{s_1}{2 \pi} K \sin \phi_1 \delta'(\eta) +\frac{r_1}{2 \pi} \Big( K \delta(\eta) \cos(\xi - \phi_1) + J \sin(\xi - \phi_1) \delta'(\eta) \Big) \label{ee}
\end{align}
where $b = b(\xi,\eta)$ is the eigenfunction. Like our target Eq.~\eqref{hard_eval}, this contains generalized functions, but without the complication of the convolutions. As we show in \cite{o2025stability}, this lets us derive the  critical eigenvector
\begin{align}
    b(\xi,\eta) = \frac{1}{2\pi} \Big( a_0(\eta) + \cos(\xi) a_1(\eta) + \sin(\xi) b_1(\eta) \Big),
\end{align}
where the Fourier coefficients are delta series
\begin{align}
    & a_0(\eta) = \sum_{n=1} c_{n} \delta^{(n)}(\eta), \hspace{1cm}  a_1(\eta) = \sum_{n=0} d_{n} \delta^{(n)}(\eta), \hspace{1cm} b_1(\eta) = \sum_{n=0} e_{n} \delta^{(n)}(\eta).
\end{align}
Inserting the ansatz into the eigenvalue equation and simplifying yields the eigenvalues
\begin{align}
    & \lambda_{0} = 0, \\
    & \lambda_{1} = -n K, \hspace{0.5 cm} n > 2, \; \; (n \in \mathbb{Z}^+), \\
    & \lambda_{2} =  \frac{1}{4} \Big( -K \pm \sqrt{9 K^2 - 8 J^2} \Big).
\end{align}
which match the eigenvalues derived from the finite-$N$ analyzes \cite{o2022collective}.

This is the first swarmalator state with compact support whose stability can be computed exactly. It thus serves its purpose as a warm up problem for the 2D model.

\begin{figure}[t!]
    \centering
    \includegraphics[width=0.8\textwidth]{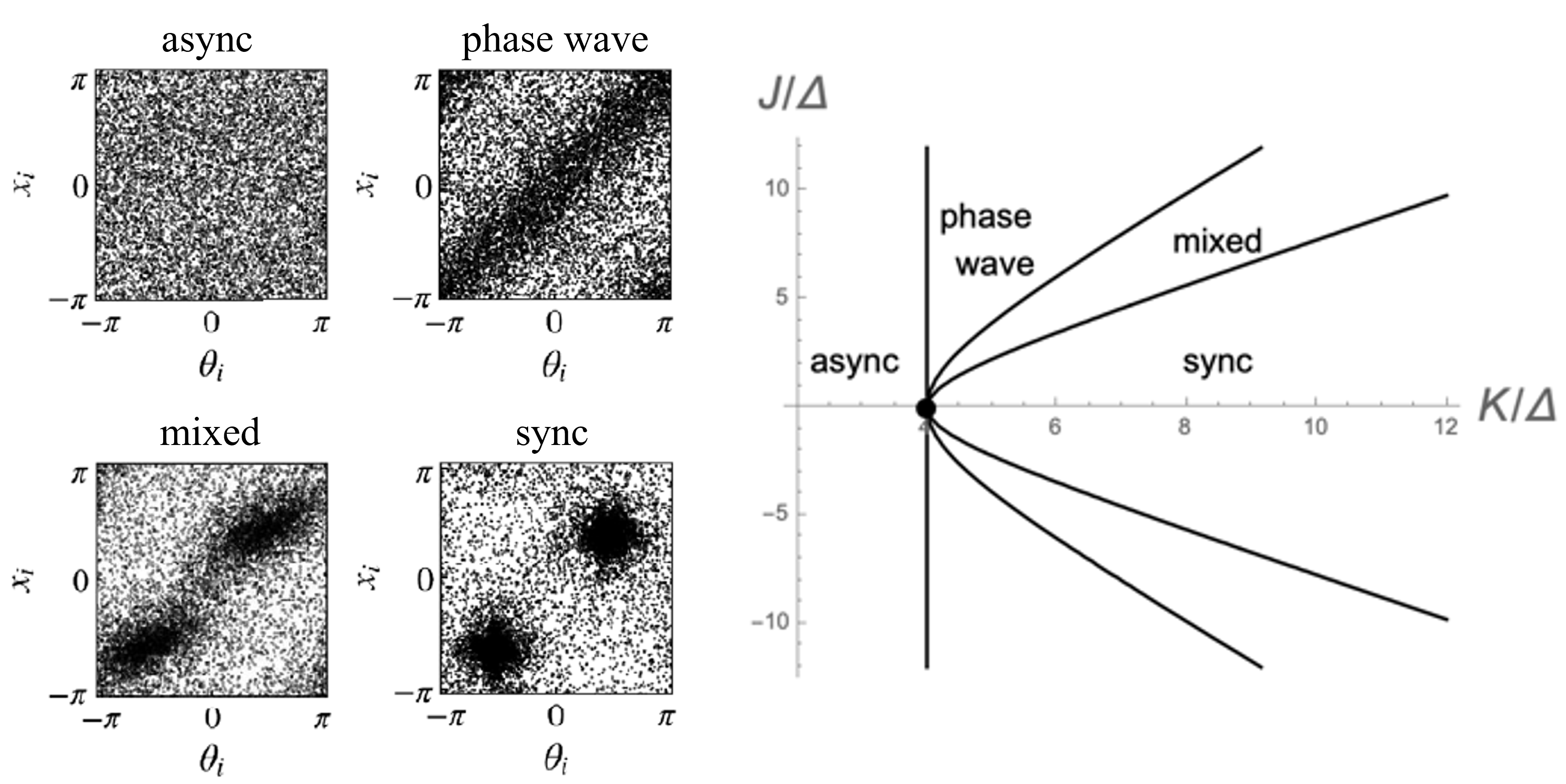}
    \caption{Left, collective states of the 1D swarmalator model with natural frequencies drawn from Lorentzians.  Right, bifurcation diagram where $\Delta$ is the width of the common width of the Lorentzians. \\ Source: Reprinted figure with permission from Ref. \cite{yoon2022sync}.}
    \label{fig:1d_model_states_nonidentical}
\end{figure}

\textit{Non identical swarmalators}. Figure~\ref{fig:1d_model_states_nonidentical}(a) shows that when the swarmalators are nonidentical blurry analogies of async, sync and phase waves states arise. A new mixed state is also found, which blends sync and the phase wave. Figure~\ref{fig:1d_model_states_nonidentical}(b)  shows the bifurcation structure which is interesting: there is a tetracritical point where all four states coincide. 

These behaviors of the 1D model were the perfect playground to try and port the tools from oscillators studies to swarmalators. It allowed us to adapt Kuramoto's self consistency and also derive a quasi-OA ansatz analysis to derive expressions for $r(J,K), s(J, K)$ in all but the mixed state \cite{yoon2022sync} as,
\begin{align}
    &  \Big(r_{pw}, \; s_{pw}\Big) = \Big(\sqrt{1 - \frac{4\Delta}{K} }, \; 0\Big), \label{s1} \\
    &  r_{sync} = s_{sync} = \sqrt{1 - \frac{4 J \Delta}{J^2-K^2} }, \label{s2} \\
    &  r_{mixed}, s_{mixed} = ?
\end{align}
We used these to derive \textit{existence} conditions for each state, but their \textit{stabilities} are open problems (expecting for the async state whose stability we have derived). One ought to be able to adapt the eigenvalue analysis of the phase wave in the previous section here. That's a nice low hanging fruit for anyone interested.

%%%%%%%%%%%%%%%%%%%%%%%%%%%%%%%%%%%%%
\subsection{The 2D swarmalator model with periodic boundary conditions, $x, y \in \mathbb{S}^1$}
With the 1D model under our belt so to speak, it is natural to try and generalize the model to 2D, where now both the $x$ and $y$ motion are confined to the unit circle $x,y \in \mathbb{S}^1$. We thus considered the model
\begin{align}
    \dot{x}_i &= u_i + \frac{J'}{N} \sum_j \sin(x_j - x_i) \cos(\theta_j - \theta_i) \label{x1}, \\
    \dot{y}_i &= v_i + \frac{J'}{N} \sum_j \sin(y_j - y_i)\cos(\theta_j - \theta_i) \label{y1}, \\
    \dot{\theta}_i &= w_i + \frac{K'}{N} \sum_j \sin(\theta_j - \theta_i) \Big(\cos(x_j - x_i) + \cos(y_j - y_i) \Big), \label{t1} 
\end{align}
which you can think of as swarming happening on a plane with periodic boundary conditions. Notice we have neglected hard shell repulsion terms for the sake of simplicity.

2D analogues of the async, sync, and phase wave states are found for both identical and nonidentical swarmalators \cite{o2024solvable} and were analyzable with the same techniques with the 1D model. These were the first results about the bifurcations/stabilities of 2D swarmalators. The model is also studied with higher-order interactions and first-order phase transitions have been reported along with bistable regions~\cite{anwar2025two}.

A different generalization of the 1D model was also proposed~\cite{lizarraga2024order}, aiming to recover the states observed in the 2D model~\cite{o2017oscillators} while maintaining the advantages gained by considering periodic boundary conditions. Up to this point, modifications to the 1D model did not include explicit repulsion terms. These, responsible for the formation of states where particles spread out in space instead of clustering at a single point. Under this premise, the model was defined as 

	\begin{align}
		\begin{split}
			\dot{x}_i &= \frac{1}{N}\sum_{j= 1}^N \left\{J^-\sin (x_{ji})\cos(\theta_{ji}) - J^+\left[1- \cos (x_{ji})\right] \sin(\theta_{ji}) \right\},\\
			\dot{y}_i &= \frac{1}{N}\sum_{j= 1}^N \left\{J^-\sin (y_{ji})\cos(\theta_{ji}) - J^+\left[1- \cos (y_{ji})\right] \sin(\theta_{ji}) \right\},\\
			\dot{\theta}_i &= \frac{K'}{N}\sum_{j= 1}^N \sin(\theta_{ji})\left[2+ \cos(2x_{ji}) + \cos(2y_{ji})\right],
		\end{split}
		\label{eqS2:2d_general}
	\end{align}
where $(x_i, y_i, \theta_i)\in(\mathbb{S}^1, \mathbb{S}^1, \mathbb{S}^1)$, and differences between these are written in a compact form (e.g., $\theta_{ji} = \theta_j - \theta_i$). The phase coupling strength is denoted by $K'$, and $J^{\pm}$  scale the spatial interaction terms. Specifically, $J^+$ strengthens quasi-repulsive interactions between individuals. The prefix ``quasi" is used because no formal metric was employed to define distances on the torus. 

\begin{figure}
    \centering
    \includegraphics[width=\textwidth]{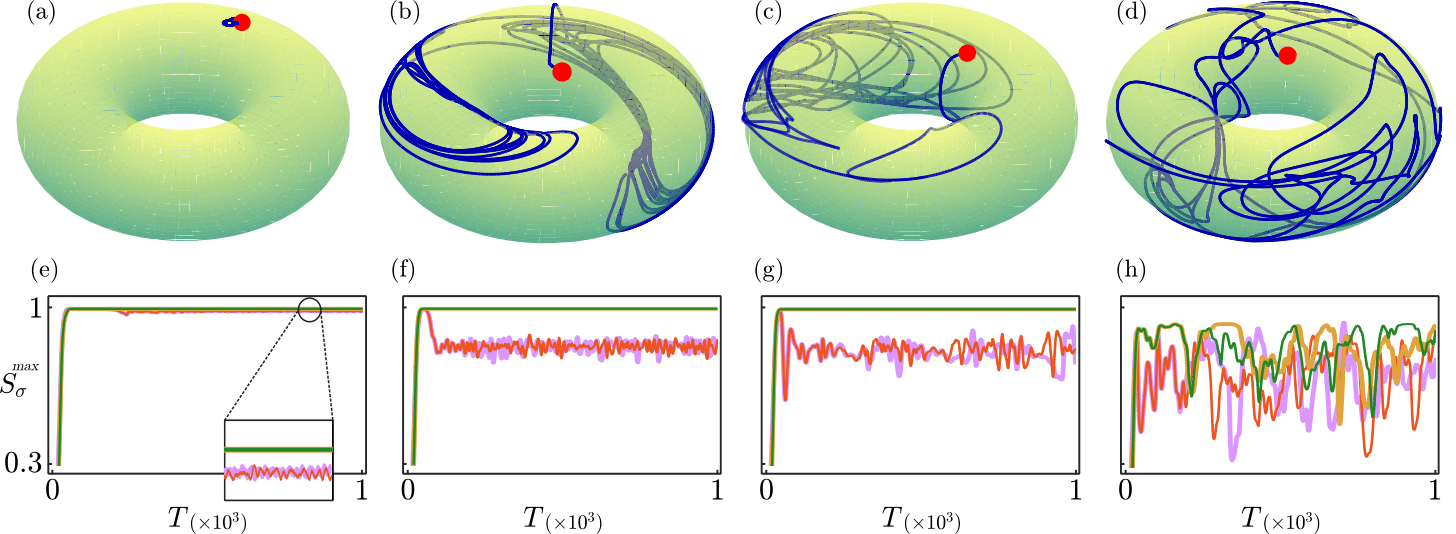}
    \caption{Top row: trajectories (blue) followed by single particles in the $x-\theta$ torus. The red circles are positioned at the start of each trajectory. Bottom row: evolution of $S^{max}_{(\theta\pm x)}$ and $S^{max}_{(x\pm y)}$ for the original initial conditions (purple and gold, respectively) and the perturbed initial conditions (orange and green, respectively). Parameters $(K', J^+, J^-)$ are set as (a, e) $(-0.5, -0.25, 0.25)$, (b, f) $(-0.5,-0.5 , 0)$, (c, g) $(-0.5, -0.75, -0.25)$, and (d, h) $(-0.5, -1.75, -1.25)$. The small rectangle in (a) shows a magnification of the circled lines.}
    \label{fig:torus_OCDT}
\end{figure}

Once quasi-repulsion is introduced, as in Eq.~\eqref{eqS2:2d_general}, spatially distributed states begin to show up. Moreover, since periodic boundary conditions are considered, stability conditions can be derived explicitly. A striking effect spotted as the difference between attraction and repulsion strengths grows and becomes negative is the emergence of active states that exhibit signatures of chaos. An instance of this behavior is shown in Fig.~\ref{fig:torus_OCDT}, which presents a bi-dimensional projection of the trajectory followed by a single particle in four different scenarios. The correlations are measured by $S_{(\theta\pm x)}^{max} = \max\left[S_{(\theta+x)}, S_{(\theta-x)} \right]$ and $S_{(x\pm y)}^{max} = \max\left[S_{(x+y)}, S_{(x-y)}\right]$, considering a generalized form of the Kuramoto order parameter given by
\begin{equation}\label{s-sigma}
	S_\sigma e^{\iu\phi_{\sigma}} = \frac{1}{N}\sum_{j = 1}^N e^{\iu\sigma_j},
\end{equation}
where $\sigma$ can represent any individual variable $x$, $y$, $\theta$, or a linear combination of these.

%%%%%%%%%%%%%%%%%%%%%%%%%%%%%%%%%%%%%%%%%%%%%%%%%
\subsection{The 1D swarmalators model on the line $x_i \in \mathbb{R}^1$}
Finally, one could also define a 1D model where motion is confined to a line $x \in \mathbb{R}^1$ instead of a ring $x \in \mathbb{S}^1$. We do this by peeling off the radial and phase components $(\dot{r}, \dot{\theta})$ of the governing ordinary differential equations of the 2D model (recall we pulled off the azimuthal $\dot{\phi}, \dot{\theta}$ for our ring model). This inspired the following model
\begin{align}
    \dot{x_i} &=  \frac{1}{N} \sum_j^N \Big[(x_j - x_i)( 1 + J \cos(\theta_j - \theta_i)) - sign(x_j-x_i)\Big], \label{eom-x-line} \\
    \dot{\theta_i} &= \omega_i + \frac{K}{N}  \sum_j^N \sin(\theta_j - \theta_i ) \left( 1 - \frac{(x_j - x_i)^2}{\sigma^2} \right).  \label{eom-theta-line}
\end{align}
\noindent
Here we assume $\sigma$ to be large enough so that the parabolic kernel is always positive. The sign function might seem odd, but is the natural replacement of the $x/|x|^2$ repulsion term in 1D. Combined with the linear attraction term, it leads to states of uniform density in the uncoupled limit $J=0$. The parabolic term might seem strange too, but allows the model to be written in terms of global order parameters like so
\begin{align}
\dot{x_i} &= -x_i(1 + J R_0 \cos (\Psi_0 - \theta_i)) + J R_1 \cos(\Psi_1 - \theta_i) - G(x), \\ 
\dot{\theta_i} &= \omega_i +  K R_0 \sin(\psi_0 - \theta_i) - \frac{K}{\sigma^2} R_2 \sin(\Psi_2-\theta_i) + 2 x_i R_1 \sin(\Psi_1 - \theta) - x_i^2 R_0 \sin(\Psi_0 -\theta),
\end{align}
where the order parameters are
\begin{align}
    Z_n &= R_n e^{i \Psi_n} = N^{-1} \sum_j x_j^n e^{i \theta_j}, \\
    G(x) &= N^{-1} \sum_j sign(x_j - x_i).
\end{align}
We ran some simulations of this model and found as well as analogues of the sync and phase waves states. There were also signs of chaos in the $K<0$ regime that remind us of algal or bacterial turbulence \cite{dunkel2013fluid,baruah2025emergent}. But other than that, the model is wide open. A natural place to start would be to test the potential connection to the aforementioned turbulent systems by compute probability density functions of the density and velocity and looking for heavy-tailedness.
\par Table~\ref{layland} collects these theoretical findings and shows we are approaching our targets of the melting point $K_m$, splitting point $K_s$, and $S(K)$ curve of the 1D and 2D models.

%%%%%%%%%%%%%%%%%%%%%%%%%%%%%%%%%%%%%%%%%%%%%%%%
%\subsection{Summary}
\begin{table*}[t!]
\centering
\rowcolors{1}{blue!10}{white}
\begin{tabular}{|c|c|c|c|c|c|c|c|c|} 
\hline 
\rowcolor{gray!50} 
Model & $\rho_{\text{async}}$ & $\rho_{\text{sync}}$ & $\rho_{\text{pw}}$ & $K_m $ & $S(K)$  & $\rho_{\text{active pw}}$ & $K_s$ & $\rho_{\text{splintered pw}}$ \\ 
\hline 
1D model $x \in \mathbb{S}^1$ & \cellcolor{blue!20}\checkmark \cite{o2022collective,yoon2022sync} & \cellcolor{blue!20}\checkmark \cite{o2022collective,yoon2022sync} & \cellcolor{blue!20}\checkmark \cite{o2022collective,yoon2022sync} & \cellcolor{blue!20}\checkmark \cite{o2022collective,yoon2022sync} & \cellcolor{blue!20}\checkmark \cite{yoon2022sync} & \cellcolor{blue!20}\checkmark \cite{yoon2022sync} & \cellcolor{gray!50}NA & \cellcolor{gray!50}NA \\ 

\hline 
2D model $x,y \in \mathbb{S}^1$ & \cellcolor{blue!20}\checkmark \cite{o2024solvable} & \cellcolor{blue!20}\checkmark \cite{o2024solvable} & \cellcolor{blue!20}\checkmark \cite{o2024solvable} & \cellcolor{blue!20}\checkmark \cite{o2024solvable} & \cellcolor{blue!20}\checkmark \cite{o2024solvable} & \cellcolor{blue!20}\checkmark \cite{o2024solvable} & \cellcolor{gray!50}NA & \cellcolor{gray!50}NA \\ 
\hline 
2D model $x \in \mathbb{R}^2$ & \cellcolor{blue!20}\checkmark \cite{o2017oscillators} & \cellcolor{blue!20}\checkmark \cite{o2017oscillators} & \cellcolor{blue!20}\checkmark \cite{o2017oscillators} & \cellcolor{red!20}? & \cellcolor{red!20}? & \cellcolor{red!20}? & \cellcolor{red!20}? & \cellcolor{red!20}? \\ 
\hline 
1D model $x \in \mathbb{R}^1$ & \cellcolor{red!20}? & \cellcolor{red!20}? & \cellcolor{red!20}? & \cellcolor{red!20}? & \cellcolor{red!20}? & \cellcolor{red!20}? & \cellcolor{red!20}? & \cellcolor{red!20}? \\ 
\hline 
\end{tabular}
\caption{Lay of the land. The models definition in the first column are defined by Eqs.~(\ref{w1},~\ref{w2}), Eqs.~(\ref{x1},~\ref{y1},~\ref{t1}), Eqs.~(\ref{linear_parabolic1},~\ref{linear_parabolic2}), and Eqs.~(\ref{eom-theta-line},~\ref{eom-x-line}), respectively. Here, a $\checkmark$ indicates the emergence of a particular
state in the corresponding model, while question marks (?) denote
its absence and NA represents not-applicable.} 
\label{layland}
\end{table*}

%---------------------------------------------------------
\section{Predator-swarmalator models}  \label{sec.5}
The dynamics of swarmalator systems can undergo significant transformations when exposed to external agents that mimic predatory behavior. The presence of such agents often triggers adaptive responses among the swarmalators including increased cooperation, strategic positioning, or the development of defensive mechanisms. These induced behaviors can result in emergent phenomena, reshaping the spatial and phase coherence within the system. Exploring such predator-induced interactions offers insights into the robustness and adaptability of swarmalator collectives under environmental changes.

The objective is to analyze the behavior of the swarmalator system described by Eqs.~\eqref{eq.2.15} and~\eqref{eq.2.16} in the presence of predator-like agents that significantly influence the system’s long-term dynamics. Similar to swarmalators, these agents possess both spatial and phase dynamics. These agents are referred to as \textit{contrarians} due to the contrasting nature of their interactions compared to regular swarmalators. Let $q$ denote the number of contrarians, and let $z_i$ and $\psi_i$ represent their positions and phases, respectively, for $i = 1, 2, \ldots, q$. The number of contrarians is assumed to be much smaller than the number of swarmalators, i.e., $q \ll N$, a condition under which the model’s outcomes remain unaffected. The governing equations for the swarmalators under this setup is as follows:
\begin{align}
    \Dot{\mathbf{x}}_i &= \dfrac{1}{N}\sum_{\substack{j = 1\\j \neq i}}^{N}\Bigg[(\mathbf{x}_j-\mathbf{x}_i) \big(1 + J \cos(\theta_j - \theta_i)\big)- \dfrac{\mathbf{x}_j-\mathbf{x}_i}{|\mathbf{x}_j-\mathbf{x}_i|^2}\Bigg] + \dfrac{1}{q}\sum_{j =1}^{q}\Bigg[ (\mathbf{z}_j-\mathbf{x}_i) -  \dfrac{\mathbf{z}_j-\mathbf{x}_i}{|\mathbf{z}_j-\mathbf{x}_i|^2}\Bigg],
    \label{eq-predator1}\\
    \Dot{\theta}_i &= \dfrac{K_1}{N}\sum_{\substack{j = 1\\j \neq i}}^{N} \dfrac{\sin(\theta_j-\theta_i)}{|\mathbf{x}_j-\mathbf{x}_i|} + \dfrac{K_2}{q}\sum_{j =1}^{q} \dfrac{\sin(\psi_j-\theta_i)}{|\mathbf{z}_j-\mathbf{x}_i|}.
    \label{eq-predator2}
\end{align}
The phases of the swarmalators are coupled to those of the contrarians with a strength $K_2$, and this interaction is modulated by the spatial distance between them.

\begin{figure}[h!]
    \centering
    \includegraphics[width=0.75\textwidth]{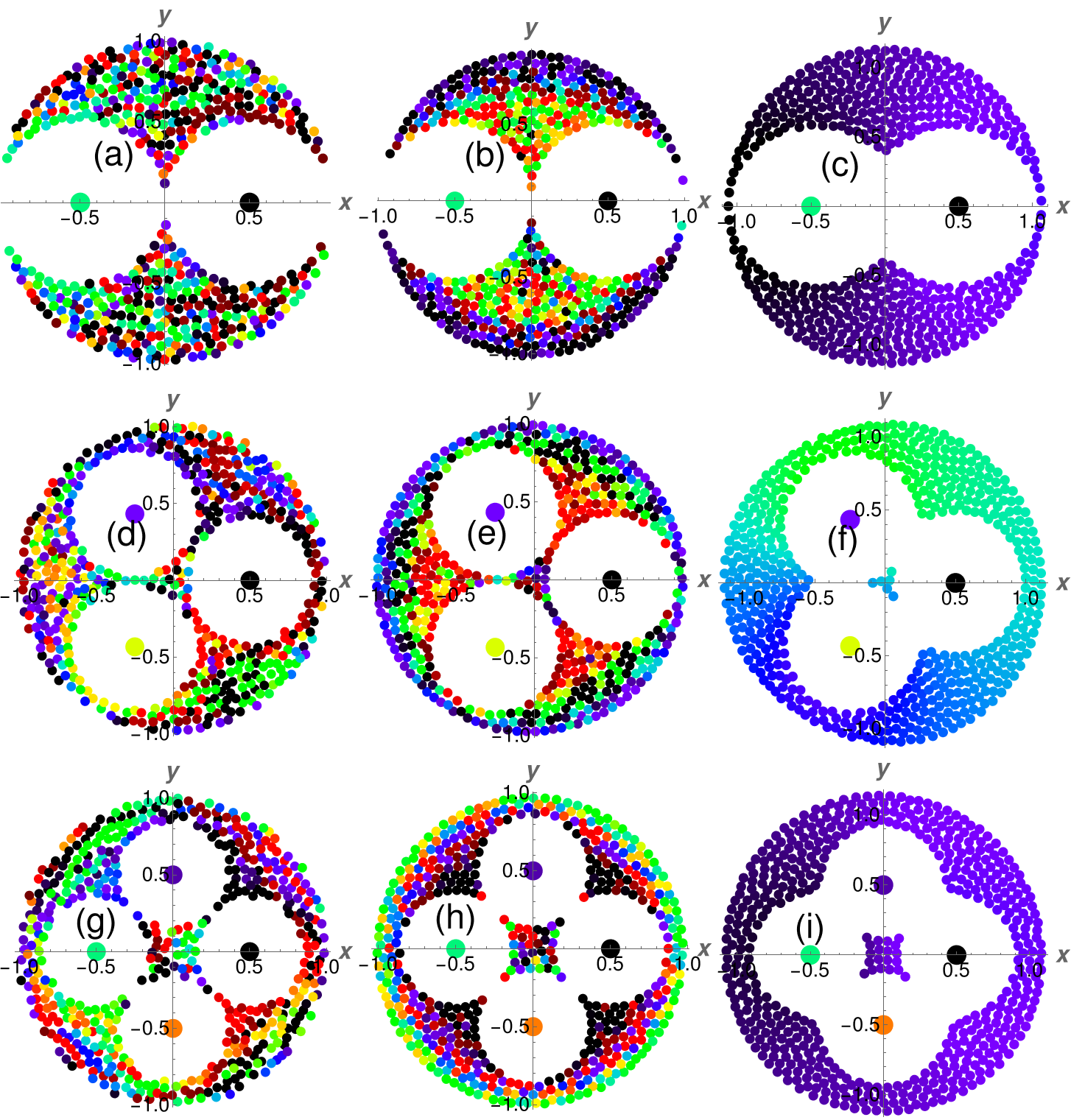}
    \caption{The emergent behavior of swarmalators influenced by multiple contrarians is illustrated, where both swarmalators and contrarians are color-coded based on their phases. Simulations are performed using Eqs.~\eqref{eq-predator1}-\eqref{eq-predator2} with $N = 500$ swarmalators and $q = 2, 3, 4$ contrarians, whose initial positions and phases are assigned following Eq.~\eqref{contrarian} with $r_i = 0.5$. The top row panels (a)–(c) correspond to $q = 2$, the middle row (d)–(f) to $q = 3$, and the bottom row (g)–(i) to $q = 4$. Each column represents a different coupling scheme: $K_1 = -K_2 = -0.5$ in the first, $K_1 = -K_2 = 0$ in the second, and $K_1 = -K_2 = 0.5$ in the third. The interaction strength $J$ is fixed at $-0.5$ for all configurations. \\ Source: Reprinted figure with permission from Ref. \cite{sar2025dynamics}.}
    \label{predators}
\end{figure}

{\it Static predators}: For the sake of simplicity, first assume that the contrarians are stationary in space and their phases do not change. Their static configuration can be chosen as~\cite{sar2025dynamics},
\begin{equation}
    \mathbf{z}_i = \left(r_i\cos \dfrac{2 \pi (i-1)}{q},r_i\sin \dfrac{2 \pi (i-1)}{q}\right),\hspace{1 pt}
    \psi_i = \dfrac{2 \pi (i-1)}{q},
    \label{contrarian}
\end{equation}
for $i=1,2,...,q$, where $r_i$ is the spatial distance of the $i$th contrarian from the origin. Various collective states where the contrarians are surrounded by the swarmalators or where the contrarians escape the swarmalators pack are found for different values of $J$, $K_1$, and $K_2$. The $q=1$ case is explored in a detailed way in Ref.~\cite{sar2025dynamics}. Some emerging states of the model for $q=2,3,4$ are illustrated in Fig.~\ref{predators}.

{\it Moving predators}: The contrarian or the predator can also move in space without being static all the time. Tendency of a predator is to hunt or minimize distance with other swarmalators. Accordingly, predator's spatial dynamics can be considered as,
\begin{equation}
    \Dot{\mathbf{z}}_j= \dfrac{C}{N}\sum_{k =1}^{N}  \dfrac{\mathbf{x}_k-\mathbf{z}_j}{|\mathbf{x}_k-\mathbf{z}_j|^{\beta}}, \label{eq-predator3}
\end{equation}
where $C$ is the hunting strength and $\beta$ is the exponent of the power-law interaction that can vary.

%\clearpage

%---------------------------------------------------------
\section{Applications and implications}\label{sec.6}
\begin{figure}[h!]
    \centering
    \includegraphics[width=0.75\textwidth]{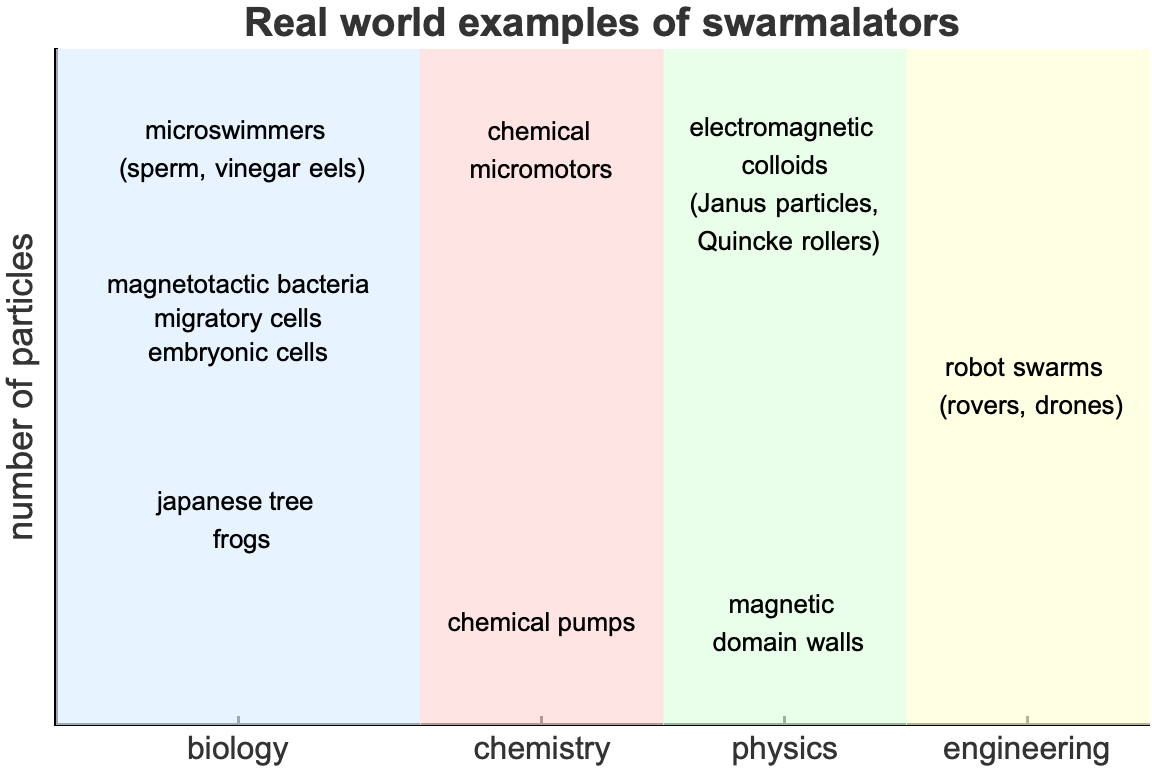}
    \caption{Real-world examples of swarmalators.}
    \label{fig:examples}
\end{figure}

The swarmalator approach offers a promising framework for understanding and explaining collective behavior in nature whenever spatial movement and temporal coordination are interconnected or dependent on one another. It has shown to be useful in the design of distributed technical systems, such as swarm robotics, where both spatial coordination and synchronized timing are crucial. Figure~\ref{fig:examples} illustrates the real world swarmalator systems in various disciplines, namely biology, chemistry, physics, and engineering. We discuss these examples in the following.

\subsection{Biology}

\begin{figure}[h!]
    \centering
    \includegraphics[width=0.95\textwidth]{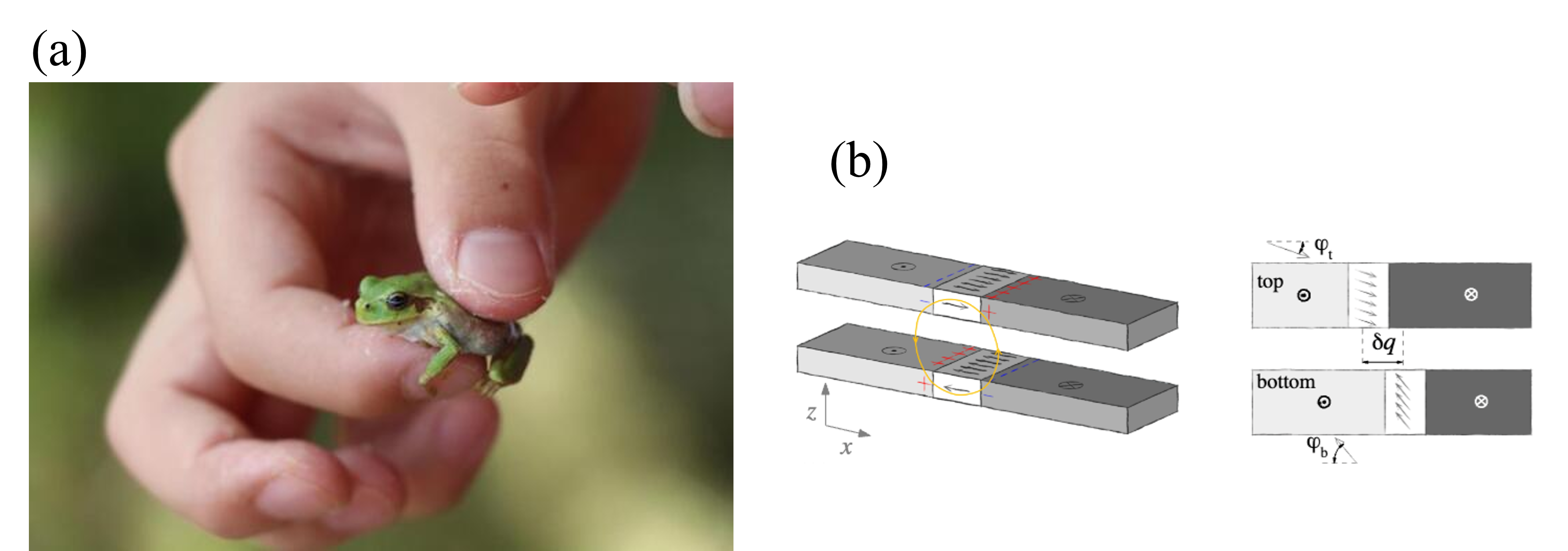}
    \caption{Some examples of real-world swarmalators. Left: Japanese Tree frog. Right: Magnetic domain walls.}
    \label{fig:two}
\end{figure}

\textit{Japanese tree frogs (Hyla japonica)} are beautiful creatures (see Fig.~\ref{fig:two}(a)) found in the wetlands of Japan, China, and Korea. During courtship rituals, males try to attract the attention of females by croaking periodically. It is fascinating that this croaking synchronizes; neighboring males want to croak $180$ degrees out of phase, so as not to `speak over one another' (more formally, to avoid acoustic interference). This sync mechanism influences how the frogs move around in space -- making them swarmalators -- and leads to intricate spatiotemporal patters. Field experiments show that frogs line up around the edges of paddy fields with phase difference $\theta_i - \theta_{i-1} = \pi$ \cite{aihara2014spatio}.

\textit{Microswimmers} like sperm and vinegar eels tend to synchronize the beating of their tails due to hydrodynamic coupling~\cite{riedel2005self, quillen2022fluid}. This synchrony couples to the swimmers' movements, establishing a swarmalation feedback loop. This manifests in emergent behavior like chains, waves, and vortex arrays \cite{riedel2005self, quillen2021metachronal,moore2002exceptional,schoeller2020collective}.

\textit{Embryonic cells} have under certain conditions genetic oscillation that can be modeled as a phase~\cite{tsiairis2016self}. These oscillations couple to their cells' movements, giving rise to emergent states like radial phase waves~\cite{tsiairis2016self}, which can be captured by the swarmalator model~\cite{ceron2023diverse}.

\textit{Larger animals} like locusts~\cite{ariel2015locust} and herds of cattle and bison exhibit collective movement and synchronization as well \cite{grandin2022genetics,ramos2015collective}. These animals have been observed in milling states, which are used for hunting and defense. \textcolor{black}{A recent work extends the swarmalator framework by introducing gaze direction as a proxy for visual attention, yielding a directional swarmalator model that captures coupled crowd motion, rhythmic synchronization, and gaze alignment in dance-floor dynamics~\cite{toiviainen2025modeling}. The model is validated against motion-capture recordings (silent-disco experiments) and provides a flexible tool to study how auditory and visual couplings and interaction ranges shape emergent social patterns.}

%%%%%%%%%%%%%%%%%%%%%%%%%%%%%%%%%%%%%%%%%%%
\subsection{Physics}

\textit{Magnetic domain walls} are a prime example of a swarmalator from physics. Figure~\ref{fig:two}(b) shows a schematic of two walls in close proximity. Each wall can be described by the position of its center of mass $q_i$ and the average orientation in phase of its magnetic field $\phi$ (this reduced model is called the $q-\phi$ model \cite{slonczewski1972dynamics}). When subjected to a magnetic field, the phases $\phi$ begin to oscillate and synchronize, which in turn causes spatial motion~\cite{hrabec2018velocity}. The result is a rich variety of spatiotemporal dynamics, some of which are captured by the swarmalator model \cite{sar2024solvable}. Similar dynamics can be observed by other magnetic particles such as skyrmions~ \cite{zhang2018manipulation}.

\textit{Electromagnetic colloids}. Janus particles are micrometer-sized spheres with one hemisphere painted with Nickel to make it ferro-responsive. By shining colloids of such particles with magnetic fields, one can get them to swarmalate: their dipoles begin to oscillate and synchronize which in turn couples to their motion leading to a diverse collective behavior \cite{yan2012linking}. Quincke rollers follow a similar effect, where the forcing is now done by electric fields~\cite{reyes2023magnetic, zhang2023spontaneous}. These colloids have great potential for application in biomedicine and other fields~\cite{wang2018light}. 

\textcolor{black}{Very recently, a feedback-controlled active-colloid platform realized an experimentally tunable swarmalator system in which hydrodynamic flows couple oscillatory (phase) dynamics to translational motion~\cite{heuthe2025tunable}. By varying a single control parameter, the system transitions between synchronized clusters, rotating aggregates, and dispersive phases, revealing an additional synchronization-dependent lateral interaction that can promote collective rotation. In a complementary control-oriented direction, recent theory shows that an externally driven mobile pacemaker can steer swarmalators between distinct spatiotemporal patterns (e.g., spindle/ripple/trapping), enabling formation and switching of organized states by tuning the external drive~\cite{xu2024collective}.}

\subsection{Chemistry}

\textit{Chemical micromotors}. A prominent class of chemically active matter consists of auto-catalytic particles, or micromotors, which drive motion through \textit{in situ} reactions rather than external fields~\cite{sanchez2015chemically}. These particles typically catalyze reactions—such as the decomposition of hydrogen peroxide on a platinum surface—to produce osmotic gradients that propel them forward. Under certain reactive or illuminated conditions, their reaction rates can oscillate, generating a temporal phase that couples with their spatial dynamics. Recent work by Wang and co-workers has provided compelling experimental evidence of this behavior, showing that silver micromotors can self-organize into swarmalator-like states characterized by spontaneous synchronization and collective motion waves~\cite{zhou2020coordinating,chen2022unraveling}.

\textit{Chemical pumps}. Complementing these mobile systems, chemical pumps utilize similar catalytic reactions to drive oscillatory behavior in flexible, surface-anchored sheets~\cite{manna2021chemical}. A catalytic patch decomposes reactants to generate fluid flow, which interacts with the sheets to induce “fishtailing’’ or circulatory oscillations. While single sheets oscillate independently, pairs can synchronize their motion through hydrodynamic and steric interactions. These tunable dynamics resemble swarmalator systems and offer potential applications in soft robotics and analytical chemistry as self-regulating ``chemical clocks.’’

%%%%%%%%%%%%%%%%%%%%%%%%%%%%%%%%%%%%%%%%%%%
\subsection{Engineering: Robotics}

The concept of swarmalation promises to be useful in technological systems, especially in mobile robotics for distributed coordination in monitoring and surveillance missions \cite{barcis2019robots,barcis2020sandsbots,ceron2024reciprocal}. For example, it can be used to gather and synchronize multiple robots or drones around a point of interest that requires observation \cite{barcis2019robots}. In general, swarming offers an alternative or complement to traditional path planning algorithms and trajectory control in robotics. Its distributed nature makes it scalable, robust against failures of nodes and communication links, and adaptive to dynamic environments. Synchronization is crucial for coordinated actions, such as joint sensing from multiple perspectives, which is essential for applications like 3D reconstruction. The beauty of using swarmalation in robotics is that a group of robots or drones can simply be instructed to form a desired space-time pattern around a specific point in space, and the system emerges to the desired state  autonomously without having to program any movement paths. 

\begin{figure}[h!]
    \centering
    \includegraphics[width=0.5\textwidth]{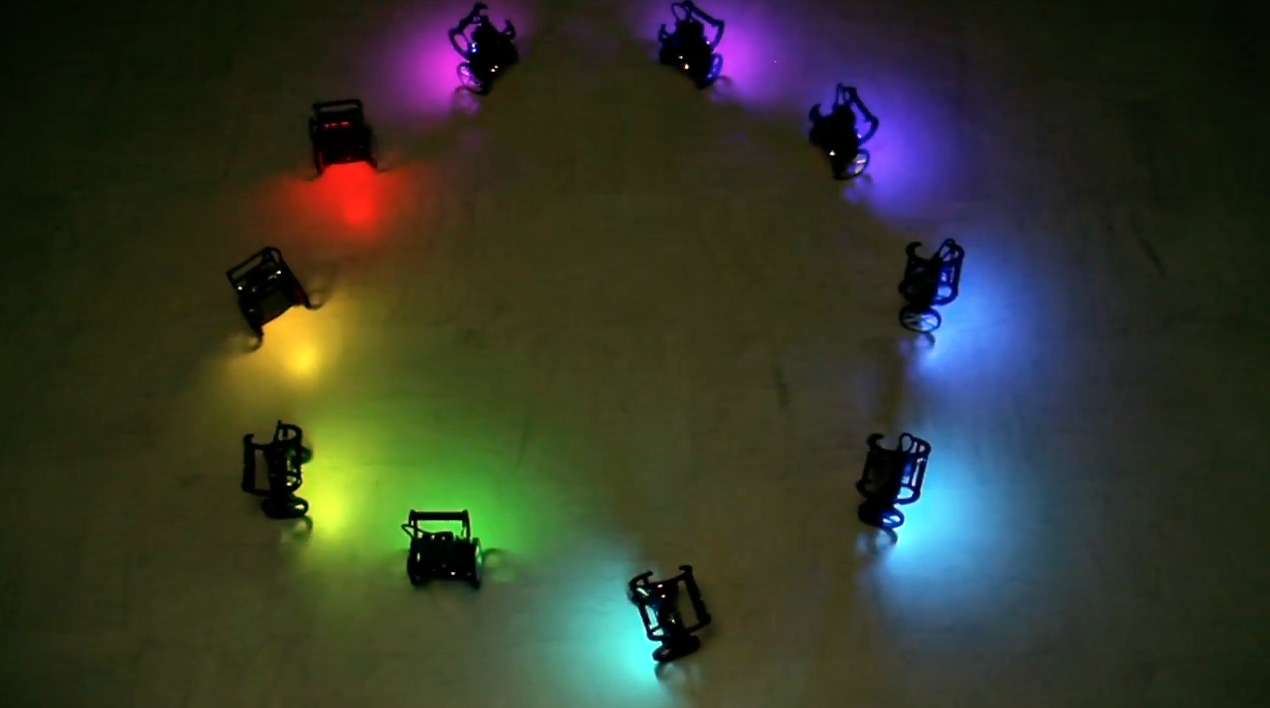}
    \caption{Robots forming a static phase wave. The colors indicate the phases. Photo by Agata and Michał Barciś for University of Klagenfurt.}
    \label{fig:robots-phase-wave}
\end{figure}

A first proof of concept for the application of the swarmalator concept in robot systems was presented in~\cite{barcis2019robots,barcis2020sandsbots}. From this work, some wheeled ground robots forming a static phase wave are illustrated in Fig.~\ref{fig:robots-phase-wave}, and a video of robots and quadcopters (Crazyflies) acting as swarmalators can be found on \url{www.youtube.com/watch?v=YCZREhgCF84}. Follow-up research on robotic swarmalators was done with Sphero BOLT robots \cite{beattie2025realizing}
and Crazyflies~\cite{quinn2025}. Also simpler ground robot systems like the vibration-driven bristle-bots (BBots) exhibit similar dynamics~\cite{giomi2013swarming}. They use angled bristles and motor oscillations as an internal phase, coupling with spatial movement to form collective patterns like clusters or streams. Multiple BBots self-organize through collisions and vibration-mediated interactions, mirroring swarmalator behavior where temporal oscillations drive spatial order. These mini\-malistic systems provide insights into active matter and potential microrobotics applications. The integration of swarmalation as a building block in an autonomous multidrone system was highlighted in~\cite{rinner2021multidrone}. 

The original swarmalator model \cite{o2017oscillators} is not directly applicable to robotics~\cite{barcis2019robots,schilcher2021swarmalators}. Substantial modifications are required to address the challenges and constraints of the real world. These include communication between the swarmalators to exchange their states, physical limitations on movement, and the need of each swarmalator to estimate its current state in terms of phase and location. \textcolor{black}{Stigmergy-mediated swarmalator mixtures suggest a robotics-relevant route to coordination via environmental encoding of phase cues, where obstacles/boundaries can be leveraged as control elements to steer segregation and collective rotation~\cite{horvath2025stigmergic}.}

\paragraph{Communication}
Several assumptions of the original model are too idealistic for wireless communication between robots. 
First, it is essential to change the coupling from time-continuous, global coupling to time-discrete, non-global coupling~\cite{barcis2019robots}. The coupling between network entities in the real world occurs through a communication channel constrained by bandwidth, latency, and potential information loss. While the original model assumes time-continuous interaction --- where swarmalators are permanently coupled obtaining instantaneous state information from others --- technical systems exchange information only at discrete points in time. For instance, wireless technologies like Wi-Fi and Bluetooth communicate via discrete data messages rather than via a continuous data flow. In discrete coupling, the question arises as to how frequently swarmalators should exchange their state information in order to balance pattern convergence with the efficient use of channel and energy resources~\cite{schilcher2025swarmalators,5912-s83l}. A recommendation for this state exchange period, as a function of the number of swarmalators and the parameter $J$, is given in~\cite{schilcher2025swarmalators}.
Second, if swarmalation is to be applied in large-scale systems, we can no longer assume a fully meshed network, where each node has a direct link to every other node to exchange state information. However, reducing the physical range of each node --- such that it is only coupled to a limited number of nearby nodes rather than all --- leads to a constellation where the system does not converge to one of the five basic states; the patterns are deformed or entirely different patterns emerge~\cite{schilcher2021swarmalators,lee2021collective}. This raises the question as to the optimal interaction range and asks for the design of adaptive schemes to tune this range over time.
Third, the system should withstand communication unreliability, since the exchange of state information may be subject to losses. Research indicates that the pattern formation shows a certain level of resilience to message loss. Specifically, a swarmalator system discretized with an Euler step size of $0.1$ (as in~\cite{o2017oscillators}), which performs a state exchange at each Euler step, is able to cope with more than $90\,\%$ loss without effects on the convergence probability~\cite{schilcher2021swarmalators}. This fact can be exploited to reduce communication overhead by increasing the state exchange interval beyond the Euler step size and randomizing the state exchange (called stochastic coupling~\cite{schilcher2021swarmalators}), thus saving energy and channel resources.

\paragraph{Movement}
Robotic swarmalators must account for the movement properties of real-world entities, which are constrained by the laws of physics and the robots' capabilities, such as speed, acceleration, movement direction, and turning radius \cite{barcis2019robots,schilcher2021swarmalators}. In this context, the velocity in the first system equation can be interpreted as the desired velocity of a given robot \cite{barcis2019robots}. Additionally, the orientation of a swarmalator and its dependence on velocity must be incorporated into the model. To prevent physical collisions, the repulsion function can be designed to include a safety margin around each~swarmalator.

\paragraph{State Estimation}
The system equations assume that each swarmalator $i$ knows the phase differences $\theta_j-\theta_i$ and the distance vectors $\mathbf{x}_j - \mathbf{x}_i$ to all other swarmalators $j$. 
Since the phase $\theta_i$ is an internal variable of an oscillator, it is inherently known to each node. Swarmalators can communicate their own phase to others or estimate the phases of others through local sensing. The situation is more challenging in the spatial domain: Each swarmalator has to employ a certain positioning method to determine its absolute position $\mathbf{x}_i$ or estimate its relative position in terms of distance $|\mathbf{x}_j - \mathbf{x}_i|$ and angle. 
The absolute position can be determined by Global Navigation Satellite Systems like GPS (meter accuracy, works only outdoors), ultrawideband technology (centimeter accuracy, requires calibration), and vision-based motion capture systems (millimeter accuracy, requires installation, calibration, and line of sight). Alternatively, the relative distances and angles can be estimated using ranging and angle of arrival techniques, employing  technologies like ultrawideband and acoustic sensing. This can be complemented by accelerometers and gyroscopes to measure motion and provide orientation and velocity information.
All these localization methods come with a certain level of inaccuracy, and the swarmalator system must be robust against these non-idealities. It has been shown that swarmalators tolerate localization errors as long as these errors remain within accepted bounds~\cite{schilcher2021swarmalators}. A~swarmalator model with additive noise in both system equations has been analyzed in \cite{hong2023swarmalators}.

%%%%%%%%%%%%%%%%%%%%%%%%%%%%%%%%%%%%%%%%%%%%%%%%%%%%%%%%%%%%%%%%%%%%%%%%%%%%%%%%%%%%%
\section{Conclusions and future directions} \label{sec.7}

Nonlinear oscillators can be found in a large variety of natural and artificial systems. In many cases, it is observed that large groups of such oscillators show partial to full synchronization. Classic examples include fireflies, pacemaker cells in the heart, neurons in the brain and metronomes placed on a platform.  In order to understand how such different types of oscillators display the same transition to synchronization, Winfree proposed to describe each oscillator by its phase only, disregarding changes in its amplitude even when perturbations were added. Inspired by these ideas, Kuramoto proposed this famous model of phase coupling that has become a paradigm in the area.

Parallel to these developments, the study of swarming took off with the work by Vicsek and his model of alignment. Similarly to synchronization, a large number of systems exhibit complex collective dynamical patterns, such as schools of fish, flocks of birds, bacteria and drones. Variations of the Vicsek model, such as the Cucker-Smale and Couzin models fueled the area and inspired several new theoretical and experimental works.

Contrary to pacemaker cells or neurons, that represent static oscillators, or flocks of birds that move in space but do not oscillate, several interesting systems exhibit both types of behavior simultaneously. Japanese tree frogs, for example, croak periodically to attract mates while moving is space. They arrange themselves so that neighbors are $\pi$ degrees out of phase. Microswimmers, like sperm and vinegar eels, tend to synchronize the beating of their tails due to hydrodynamic coupling. This synchrony couples to the swimmers' movements, leading to patterns like chains and vortex arrays. Particles that sync and swarm are swarmalators. In this article we have reviewed the phenomenology, theory and applications of these particles. The objectives of studying swarmalators extend beyond theoretical exploration and into practical applications. In ecological contexts, understanding how organisms synchronize their activities while maintaining certain spatial distributions could inform conservation strategies or the management of animal populations. In engineering, swarmalators can inspire the design of adaptive and self-organizing robotic systems, where robots can dynamically synchronize tasks while adjusting their positions to optimize performance. Thus, the study of swarmalators is not only about extending existing theories of swarming and synchronization but also about applying these insights to solve real-world problems across various domains.

Although swarmalators have been extensively studied in recent years, a number of important questions are yet to be fully answered. On the theoretical side, identifying critical transition points, such as the melting point $K_m$ and splitting point $K_s$, and deriving them analytically remains a major challenge. 
 The north star goal of our research efforts is to develop theory for swarmalators, using the 2D swarmalator model as a beginning stepping stone. Our goals are explicit formulas for the density $\rho(x,\theta)$ in each of its states, as well as its melting point $K_m$, splitting point $K_s$, and supercritical branch of the order parameter $S(K)$. These quantities are hard to attack directly, as discussed, so we took a divided conquer approach. We derived three simpler models which capture different pieces of the 2D model: the 1D ring model, the 1D line model, and the 2D model with periodic boundary conditions. 

The 1D ring and 2D periodic models were solved exactly, using standard linearization of fixed points, generalizations of Kuramoto's steady state analysis, and a quasi-OA ansatz. Moreover, the 1D model has a tractable version of the nonstandard eigenvalue equations. In this sense, the two models have served their purpose analytic warm ups for the harder 2D model. That said, the 1D model has value in its own right. It is a model for the many real world swarmalators which swarm in 1D such as bordertaxis microswimmers \cite{riedel2005self}. The 1D line model, on the other hand, is wide open for investigation, and might be related to active forms of turbulence as discussed. Also, investigation of swarmalators in three dimensions is not done in details, so exploring their collective dynamics and theoretical approaches could be an interesting research direction in near future. \textcolor{black}{Building on recent work that derives a hydrodynamic description for non-reciprocal swarmalators and identifies explicit 2D doubly-periodic travelling-wave states with a quantized topological index -- whose stability and emergent patterns depend sensitively on phase noise -- a natural extension is to develop a systematic theory for the long-time selection and robustness of these topological swarmalator states across broader model classes and higher dimensions~\cite{degond2022topological}.}

\textcolor{black}{Another key future direction is to develop topological swarmalator models in which coupling is defined by metric-free neighborhoods (e.g., fixed-number nearest neighbors or rank-based rules) rather than Euclidean distance. This setting raises open questions about how neighbor switching reshapes synchronization transitions, phase waves, and chimera-like states, and how robust these patterns remain under heterogeneity and noise. Progress will likely require reduced theories that explicitly couple evolving interaction graphs to phase dynamics, and data-driven methods to infer effective topological neighborhoods from spatiotemporal observations.}

Experimentally, realizing physical systems where phase and spatial dynamics are mutually coupled is still in its infancy. From an applications perspective, swarmalator-based control could enable richer coordination strategies in drone swarms or robotic systems. Exploring these directions may unlock new insights into collective behavior and inspire novel technologies. For applications in mobile robotics, aviation, or other technical areas, it will be crucial to reliably generate stable states representing a desired, specified pattern of locations, orientations, and synchronization. For example, in autonomous monitoring missions, swarmalators forming a ring-shaped constellation around a point of interest with all swarmalators oriented toward the center could play a role -- such as quad\-copters encircling a cellular tower and facing inward. Such a ring pattern could indeed be generated with the original swarmalator model. However, other applications are likely to require different constellation for which the model must be extended. \textcolor{black}{Looking ahead, integrating modern machine-learning–assisted discovery and control, for example, via data-driven system identification and reinforcement-learning or optimal-control approaches, could help learn interaction rules that robustly generate and stabilize user-specified spatial–phase formations under uncertainty.}

\section*{Acknowledgments} \label{6}
MAMA acknowledges financial support from FAPESP, grant 2021/14335-0, and CNPq, grant 303814/2023-3.

\section*{Declaration of generative AI and AI-assisted technologies in the manuscript preparation process}
During the preparation of this work the authors used ChatGPT in order to correct grammatical errors and restructure certain sentences. After using this tool, the authors reviewed and edited the content as needed and take full responsibility for the content of the published article.

\section*{Declaration of competing interest}
The authors declare that they have no known competing financial interests or personal relationships that could have
appeared to influence the work reported in this paper.

\section*{Author Contributions}
G. K. Sar: Writing – original draft (equal). K. O'Keeffe: Writing – original draft (equal). J. U. F. Lizárraga: Writing – original draft (equal). M. A. M. de Aguiar: Writing – original draft (equal). C. Bettstetter: Writing – original draft (Chapters 2, 6, and 7). D. Ghosh: Writing – original draft (equal).

%\section{References} \label{7}
%----------------------------------------------------------
\bibliographystyle{elsarticle-num}       % APS-like style for physics
\bibliography{references}

%\end{thebibliography}
\end{document}